\documentclass[preprint,11pt]{article}

\usepackage{bm}
\usepackage{epsfig}
\usepackage{color}
\usepackage{amssymb}
\usepackage{amsmath}

\newcommand {\lab}[1]{\label{eq:#1}}

\newcommand {\be}[1]{\begin{equation}{\lab{#1}}}
\newcommand {\ee}{\end{equation}}
\newcommand {\bea}{\begin{eqnarray}}
\newcommand {\eea}{\end{eqnarray}}

\begin{document}

\title{Low--dimensional $q$--tori in FPU lattices:\\
dynamics and localization properties}

\author{
\textbf{  H. Christodoulidi$^{a}$, C. Efthymiopoulos$^{b}$ },\\
$^{a}$  Universit\`a degli Studi di Padova,\\
Dipartimento di Matematica Pura e Applicata, \\
Via Trieste 63, 35121 - Padova, Italy\\
$^{b}$ Research Center for Astronomy and Applied Mathematics,\\
Academy of Athens, Greece }


\maketitle

\begin{abstract}
Recent studies on the Fermi-Pasta-Ulam (FPU) paradox, like the theory
of $q$--breathers and the metastability scenario, dealing mostly with 
the {\it energy localization} properties in the FPU space of normal 
modes ($q$--space), motivated our first work on {\it $q$--tori} in 
the FPU problem \cite{chrietal2010}. The $q$--tori are low-dimensional 
invariant tori hosting trajectories that present features relevant 
to the interpretation of FPU recurrences  as well as the energy 
localization in $q$--space. The present paper is a continuation of 
our work in \cite{chrietal2010}. Our new results are: We extend 
a method of analytical computation of $q$--tori, using 
Poincar\'{e}--Lindstedt series, from the $\beta$ to the $\alpha$--FPU 
and we reach significantly higher expansion orders using an improved 
computer--algebraic program. We probe numerically the convergence 
properties as well as the level of precision of our computed series.
We develop an additional algorithm in order to systematically locate 
values of the incommensurable frequencies used as an input in the 
PL series construction of  $q$--tori corresponding to progressively 
higher values of the energy. We generalize a proposition proved in 
\cite{chrietal2010} regarding the so-called `sequence of propagation' 
of an initial excitation in the PL series. We show by concrete 
examples how the latter interprets the localization patterns 
found in numerical simulations.  We focus, in particular, on  
various types of extensive initial excitations that lead to 
$q$--tori solutions with exponentially localized profiles. 
Finally, we discuss the relation between $q$--tori, $q$--breathers 
(viewed as one-dimensional $q$--tori), and the so--called 
`FPU--trajectories' invoked in the original study of the FPU 
problem. 
\end{abstract}

                             
\newpage
\section{Introduction}
\label{intro}
In the well known numerical experiment of Fermi, Pasta and Ulam in 
1955, reported in \cite{feretal1955}, a dynamical system consisting 
of $N$ nonlinearly coupled oscillators showed an integrable--like 
behavior, contradicting the ergodic hypothesis of Fermi.
This surprising result motivated numerable historical works, from
Korteweg -– de Vries, solitons and Toda to ergodic theory etc., 
that together with the KAM theorem discovered in the same period,
changed the perspective of statistical mechanics, ergodic  
and perturbation theory. 

The FPU recurrences in the energies of normal modes, observed in 
\cite{feretal1955} when exciting one or few low--frequency normal 
modes, lead to the conclusion that energy transfer between modes 
is practically frozen, with only few modes sharing the total energy, 
leaving the system far from equilibrium (see also the review 
\cite{lichetal2008}). A careful inspection of the averaged in time energy 
spectra shows an exponentially localized profile in $q$--space, 
characterized by a `plateau' in the low--frequency part of the 
energy spectrum, a so--called {\it natural packet} of modes 
\cite{beretal2004,beretal2005,gentaetal}, accompanied by an 
exponential tail of the energy distribution for the remaining 
modes. This state of the system is called `metastable state' 
\cite{bampon2006,fucetal1982,livetal1985,ponbam2005,dresden} 
and recent studies on its persistence and times to equipartition 
are given in \cite{benetal2011}.  

On the other hand, it was observed that there exist periodic 
orbits called {\it $q$--breathers}, introduced in the series of 
works \cite{flaetal2005}--\cite{flachtiz}, \cite{flaetal2006b}, 
\cite{kanetal2007}, which share many common 
features with trajectories rising by the excitation of just one 
normal mode (called `FPU--trajectories'). These Lyapunov orbits 
are continuations of the linear modes, which, at low energies, 
have a similar exponentially localized energy spectrum as the 
FPU--trajectories, characterized by an exponent that depends 
logarithmically on the system's parameters.

This latter remark initiated the idea of studying trajectories 
corresponding to the excitation of more than one linear modes, 
i.e. lying on {\it tori of low dimensionality} in the FPU phase 
space. These tori were named $q$--tori in \cite{chrietal2010}, 
and numerical evidence of their existence came out from 
the implementation of the method of Poincar\'{e} -- Lindstedt 
(PL) series. We should stress at this point that, while the 
existence of $q$--breathers is guaranteed by basic theorems 
on the continuation of periodic orbits, the corresponding 
demonstration for $q$--tori would require proving the 
convergence of the associated PL series. A rigorous proof 
appears at present hardly tractable. However, in the present 
paper we will provide numerical tests showing that our computed 
series exhibit the behavior of convergent PL series. In particular, 
we identify near-cancellations between terms of a big absolute 
size, leaving a small residual in the final series. Let us note 
that precisely this mechanism has been invoked in well known 
formal proofs of the convergence of Lindstedt series in simple 
models \cite{eli1997,gal1994}. Furthermore, as in 
\cite{chrietal2010} we employ the GALI indicator \cite{skoetal2007}, 
thus obtaining further evidence that our computed solutions lie on 
low-dimensional tori. Finally, we demonstrate that the $q$-–tori exhibit
a number of features not encountered in $q$--breathers. However, 
we emphasize that $q$--tori and $q$--breathers should not be 
regarded as competitive theories,  but rather as complementary 
interpretation tools for the FPU localization phenomena. 

We finally note that the study of FPU trajectories presenting 
energy localization by various means of classical perturbation 
theory is a known subject (for an early implementation of the 
simple Birkhoff normal form approach see \cite{delucetal1995}; 
see also references in \cite{gentaetal}). In particular, the 
problem of motions on or close to low-dimensional manifolds 
using the classical method of Birkhoff was studied in 
\cite{giomur2006}. As commented in \cite{chrietal2010}, 
this approach leads to Nekhoroshev-like estimates for the 
time of stability of motions. However, the corresponding 
estimates have a bad behavior as $N\rightarrow\infty$. 
For an alternative approach to the same problem using a 
`resonant' Birkhoff construction see \cite{gentaetal}. 
On the other hand, the fact that we find indications about 
cancellations in our PL series implies that the $q$--tori 
constructed by the present method should be recoverable also 
by some `indirect' (i.e. Kolmogorov-like normal form) approach. 
However, one can easily check that a recently proposed 
algorithm for the computation of low-dimensional tori via 
normal forms \cite{sansetal2011} has different divisors than 
in our construction. Thus, we leave the question of the existence 
of an appropriate Kolmogorov algorithm for the FPU $q$--tori 
as an open problem.  

The main results and the content of this paper are stated as follows: 
After introducing the FPU model in Section \ref{fpumodel}, we sketch 
the construction of $q$--tori by Poincar\'{e}--Lindstedt series in
Section \ref{plseries}. We also present our numerical indications 
regarding the convergence of the PL series with the help of a 
concrete example referring to a two-dimensional $q$--torus solution. 

In Section \ref{4toros}, we give an example of a 4-dimensional 
$q$--torus solution. Here, we focus on the energy localization 
properties of such a torus. Furthermore, we compare this object 
with a corresponding so-called `FPU--trajectory' (subsection 
\ref{fputra}), finding results analogous to the study in 
\cite{dresden}, but for groups of modes rather than one mode. 

Section \ref{sequence} deals with an extension of our proposition 
found in \cite{chrietal2010}, where, we prove the so--called propagation 
of {\it initial excitations} of modes for both the $\alpha$ 
and $\beta$ FPU models.  In the context of perturbation theory, an 
initial excitation means a particular selection of a subset of modes 
in $q$--space for which we consider an oscillation with non--zero 
amplitude at the zero order of perturbation theory. Then, this 
excitation {\it propagates} to new modes at subsequent orders, 
giving justification to the observed exponential 
localization profile of $q$--tori.  

In section \ref{numqtor} we give several numerical examples and 
localization profiles associated to $q$--tori, and we compare them 
with the predictions of the proposition developed in section 
\ref{sequence}. We examine, in particular: 

i) low--frequency packet excitations (subsection \ref{lowfreqmod}), 
paying emphasis on so-called {\it extensive} initial excitations, 
i.e. ones in which the number of initially excited consecutive modes varies 
proportionally to $N$. We predict, based on a leading term analysis 
of the associated PL series, that the form of their energy spectrum 
has an exponentially localized profile with a slope that depends 
{\it logarithmically } on the specific energy $\varepsilon=E/N$ 
and on the system's parameters. 

ii) `arbitrary' initial excitations (subsection \ref{patternqtor}), 
i.e. excitations of modes chosen arbitrarily within the whole 
spectrum, leading to the formation of a variety of localization 
patterns, whose form is predicted theoretically and confirmed 
by numerical experiments. 

iii) `generalized packet excitations' (subsection \ref{general}), 
i.e. excitations of extensive packets of modes 
in various arbitrarily chosen parts of the spectrum. In 
this case we study the formation of local exponential profiles 
far from the low-frequency part of the spectrum.   

iv) $q$--breathers, examined as a particular case of one-dimensional 
$q$--tori (subsection \ref{trajbreath}). In this case we examine 
how the $q$--breathers compare 
with the FPU--trajectories studied in the original FPU report, i.e. 
for  the lowest frequency mode excitation, but also for a high 
frequency mode excitation. In particular, we point out the 
similarity between $q$--breathers and FPU trajectories regarding 
their energy localization pattern, but also their different 
behavior related to the phenomenon of FPU recurrences at higher 
energies.   

Section \ref{concl} is a summary of our basic conclusions from the
present study and a discussion on future perspectives.

\section{The Fermi Pasta Ulam model}
\label{fpumodel}
The FPU Hamiltonian for a lattice of $N-1$ particles reads:
\begin{eqnarray}\label{fpuham}
H= {1\over 2}\sum_{k=1}^{N-1} y_k^2 + {  1   \over
2}\sum_{k=0}^{N-1}(x_{k+1}-x_k)^2 + {\alpha \over
3}\sum_{k=0}^{N-1}(x_{k+1}-x_k)^3
 + {\beta \over 4}
\sum_{k=0}^{N-1}(x_{k+1}-x_k)^4
\end{eqnarray}
where $x_k$ is the $k$--th particle's position with respect to
equilibrium and $y_k$ its canonically conjugate momentum. Fixed
boundary conditions are defined by setting $x_0=x_{N}=0$. The
cases $\alpha \neq 0$, $\beta = 0$, and $\alpha=0$, $\beta \neq 0$
are called FPU--$\alpha$ and FPU--$\beta$ model respectively.

The normal mode canonical variables $(Q_q,P_q)$ are introduced by
the linear canonical transformation
\begin{eqnarray}\label{lintra}
x_k &=&\sqrt{2\over N}\sum_{q=1}^{N-1} Q_q\sin\left({qk\pi\over
N}\right)\nonumber\\
 y_k&=&\sqrt{2\over N}\sum_{q=1}^{N-1}
P_q\sin\left({qk\pi\over N}\right).
\end{eqnarray}
Substitution of (\ref{lintra}) into (\ref{fpuham}) yields the
Hamiltonian in the normal mode space ($q$--space):
\begin{eqnarray}\label{fpuham2}
& &H={1\over 2 } \sum_{q=1}^{N-1} { (P_q^2+\Omega_q^2Q_q^2 ) }
+{\alpha\over 3\sqrt{2N}}\sum_{q,l,m=1}^{N-1} B_{qlm}
\Omega_q\Omega_l\Omega_m Q_qQ_lQ_m\\
& &+{\beta\over 8N}\sum_{q,l,m,n=1}^{N-1} C_{qlmn}
\Omega_q\Omega_l\Omega_m\Omega_n Q_qQ_lQ_mQ_n \nonumber
\end{eqnarray}
with normal mode frequencies
\begin{equation}\label{fpuspec}
\Omega_q=2\sin\left({q\pi\over2N}\right),~~~1\leq q\leq N-1~~~.
\end{equation}
The harmonic energy $E_q$ of each normal mode $q$ is given by
\begin{equation}\label{harmonicene}
E_q=  {1\over 2} (P_q^2+ \Omega_q ^2 Q_q^2)~~~.
\end{equation}
The coefficients $B_{qlm}$ and $C_{qlmn}$ are non--zero only for
particular combinations of the indices $q,l,m,n$, namely
\begin{eqnarray}\label{transa}
B_{qlm}&=&\sum_{\pm} (\delta_{q\pm l\pm m, 0} - \delta_{q\pm l\pm m,
2N})\nonumber \\
C_{qlmn}&=&\sum_{\pm} (\delta_{q\pm l\pm m \pm n, 0} -
\delta_{q\pm l\pm m \pm n, 2N})~~.
\end{eqnarray}
In the above expressions, all possible combinations of the $\pm$
signs must be taken into account. In the new canonical variables,
the equations of motion are:
\begin{eqnarray}\label{eqmo}
\ddot{Q}_q+\Omega_q^2Q_q=-{\alpha\over {\sqrt{2N}}}
\sum_{l,m=1}^{N-1} B_{qlm} \Omega_q\Omega_l\Omega_m
Q_lQ_m \nonumber\\
-{\beta\over 2N}\sum_{l,m,n=1}^{N-1} C_{qlmn}
\Omega_q\Omega_l\Omega_m\Omega_n Q_lQ_mQ_n~~.
\end{eqnarray}

\section{Construction of $q$--tori by Poincar\'{e}--Lindstedt series}
\label{plseries}

\subsection{PL Algorithm}
As in \cite{chrietal2010}, we will now construct quasi-periodic solutions 
lying on $q$--tori by implementing the method of Poincar\'{e}--Lindstedt 
series. The main steps of our constructive algorithm are the following:

As a starting point we consider first the trivial case $\alpha=\beta=0$. 
Let $${\cal D}_0\equiv\{q_1,q_2,\ldots,q_s\},~~\mbox{where}
~~1\leq q_i\leq N-1 ~~\mbox{with} ~~ q_i < q_j,~~\mbox{for}
~~i<j$$ be an arbitrary set of $s<N$ modes, called hereafter `seed
modes' (in analogy to \cite{flaetal2005}, where the case $s=1$
corresponding to $q$--breathers is considered). The set of
functions $Q_{q_i}(t)=A_{q_i}\cos(\Omega_{q_i}t+\phi_{q_i})$,
$i=1,\ldots,s$ and $Q_q(t)=0$ if $q\notin{\cal D}_0$, constitutes
a particular solution of the linear system. The resulting
trajectory lies on a $s$--dimensional torus, provided that the
frequencies $\Omega_{q_i}$ of Eq. (\ref{fpuspec}) satisfy no
commensurability relation. This turns out to be always the case if
$N-1$ is a prime number or 
$\log_2 N\in\mathbf{N}^*$ \cite{hem1959}. If, on the other hand,
the frequencies $\Omega_{q_i}$ satisfy $s'$ linearly independent
commensurability relations ($0<s'<s$), the trajectories lie on a
`resonant torus' of dimension $s-s'$, which is a sub--manifold of
the original $q$--torus of dimension $s$.

Passing now to the nontrivial case $a\neq 0$, or $\beta\neq 0$, we
aim to define quasi--periodic trajectories lying on
$s$--dimensional tori. To this end, let $\omega_{q_i}$,
$i=1,\ldots,s$ be a set of frequencies with fixed values {\it
chosen in advance}, which are incommensurable between themselves
as well as with each one of the linear frequencies $\Omega_{q}$ of the
remaining modes $q \notin {\cal D}_0$. We then determine formal
solutions $Q_q(t)$, $q=1,\ldots,N-1$ containing only trigonometric
terms of the form $\cos(n\cdot(\omega t +\phi))$, where
$n\equiv(n_1,n_2,\ldots,n_s)$ is an $s$--dimensional integer vector
and $\omega\equiv(\omega_{q_1},\ldots,\omega_{q_s})$,
$\phi\equiv(\phi_{q_1},\ldots,\phi_{q_s})$ are the frequency and
phase vectors respectively.

According to the PL method, these solutions are written as series
in powers of a small parameter $\mu= \alpha/\sqrt{2N}$, or
$\mu=\beta/2 N$, namely:
\begin{equation}\label{qser}
Q_q(t)=\sum_{k=0}^{\infty} \mu^k Q_q^{(k)}(t)
,~~~q=1,\ldots,N-1~~.
\end{equation}
The series terms $Q_q^{(k)}$ are computed step by step. The zero
order terms are set as
\begin{equation}\label{qq0}
Q_q^{(0)}(t)=\left\{
\begin{array}{rl}
A_q\cos (\omega_{q} t+\phi_q), & \mbox{if}~q\in{\cal D}_0\\
0,                                    & \mbox{otherwise}~~.
\end{array}
\right.
\end{equation}
We emphasize that the amplitudes $A_{q_i}$, $i=1,\ldots,s$ are
unknown quantities to be specified at the end of the process. In
the computer--algebraic program, $A_{q_i}$ are symbols carried all
along the construction of the PL series, while the frequencies
$\omega_{q_i}$ are substituted at the beginning by their selected
numerical values. However, according to the PL method, the frequencies 
$\omega_{q_i}$ must also be expressed in the form of a series in 
powers of the amplitudes $A_{q_i}$, namely
\begin{equation}\label{omeser}
\omega_{q_i}=\Omega_{q_i}
+ \sum_{k=1}^{\infty} \mu^k
\omega_{q_i}^{(k)}(A_{q_1},\ldots ,A_{q_s} )~~.
\end{equation}
The functions $\omega_{q_i}^{(k)}(A_{q_1},\ldots ,A_{q_s})$ are
polynomials of the amplitudes $A_{q_1},\ldots ,A_{q_s}$, of order
$k+1$ in the $\alpha$--case, or $2k+1$ in the $\beta$--case.

Substituting Eqs.(\ref{qser}) and (\ref{omeser}) in the
equations of motion (\ref{eqmo}), we find the following
equations to be solved at order $k$: for the FPU--$\alpha$
model we have
\begin{eqnarray}\label{moda}
\ddot {Q}_q^{(k)}+\omega _q^2Q_{q}^{(k)}&=&
\sum _{n_{1}=1}^{k}\sum _{n_{2}=0} ^{n_{1}} \omega _{q}^{(n_{2})}
\omega _{q}^{(n_{1} - n_{2})}Q_{q}^{(k-n_{1})} \nonumber\\
&-&\Omega_{q} \sum_{l,m=1}^{N-1}\Omega _l\Omega _m
B_{qlm}\sum_{\mathop{n_{1,2}=0}\limits_
{n_1+n_2=k-1}}^{k-1}Q_l^{(n_1)}Q_m^{(n_2)},
~~~\mbox{if $q\in{\cal D}_0$}\nonumber\\
\ddot {Q}_q^{(k)}+\Omega_q^2Q_{q}^{(k)}&=&
-\Omega_{q} \sum_{l,m=1}^{N-1}\Omega _l\Omega _m
B_{qlm}\sum_{\mathop{n_{1,2}=0}\limits_
{n_1+n_2=k-1}}^{k-1}Q_l^{(n_1)}Q_m^{(n_2)},
~~~\mbox{if $q\notin{\cal D}_0$}\nonumber\\
\end{eqnarray}
while for the FPU--$\beta$ model we have
\begin{eqnarray}\label{modb}
\ddot {Q}_q^{(k)}+\omega_q^2Q_q^{(k)}&=&
\sum _{n_{1}=1}^{k}\sum_{n_{2}=0} ^{n_{1}} \omega _{q}^{(n_{2})}
\omega _{q}^{(n_{1} - n_{2})}Q_q^{(k-n_{1})}\nonumber \\
&-&\Omega _q \sum_{l,m,n=1}^{N-1}\Omega _l\Omega _m \Omega
_nC_{qlmn}\sum_{\mathop{n_{1,2,3}=0}\limits_
{n_1+n_2+n_3=k-1}}^{k-1}Q_l^{(n_1)}Q_m^{(n_2)}Q_n^{(n_3)},
~~~\mbox{if $q\in{\cal D}_0$}\nonumber\\
\ddot {Q}_q^{(k)}+\Omega_q^2Q_q^{(k)}&=&
-\Omega _q \sum_{l,m,n=1}^{N-1}\Omega _l\Omega _m \Omega
_nC_{qlmn}\sum_{\mathop{n_{1,2,3}=0}\limits_
{n_1+n_2+n_3=k-1}}^{k-1}Q_l^{(n_1)}Q_m^{(n_2)}Q_n^{(n_3)},
~~~\mbox{if $q\notin{\cal D}_0$.}\nonumber\\
\end{eqnarray}

By integrating either Eqs.(\ref{moda}) or (\ref{modb}), secular
terms of the form $t\sin(\omega_q t)$ appear in the right hand 
side (when $q \in {\cal D}_0$), at even orders in the
FPU--$\alpha$ and at all orders in the FPU--$\beta$. The
requirement to eliminate all secular terms leads to an algebraic
expression for the frequency correction terms
$\omega_{q_i}^{(k)}$, $i=1,\ldots,s$. After eliminating the
secular terms, direct integration of Eqs.(\ref{moda}) or
(\ref{modb}) yields the solution of the series terms
$Q_q^{(k)}(t)$. 

Each term in both series (\ref{qser}) and (\ref{omeser}) depends,  
now, on the yet unspecified amplitudes $A_{q_i}$. However, since 
the numerical values of the frequencies $\omega_{q_i}$ are specified 
in advance, the series (\ref{omeser}) can be solved for the 
amplitudes $A_{q_i}$. In practice, we solve the set of equations 
resulting from finite truncations of the series (\ref{omeser}). 
Thus, the amplitudes $A_{q_i}$ are also specified with finite 
accuracy. After determining the values of the $A_{q_i}$, 
substitution into the series (\ref{qser}) yields also numerical 
coefficients for all series terms $Q_q^{(k)}(t)$. Thus, we 
specify approximately a $q$--torus solution given by finite 
truncations of the functions $Q_q(t)$ for all $q=1,\ldots,N-1$.

\subsection{Convergence and precision tests -- Torus dimension}

The existence of a solution of Eqs.(\ref{omeser}) for a fixed choice 
of frequency values $\omega_{q_i}$, along with the convergence of the 
series (\ref{qser}) for that particular solution, constitutes a proof 
that a $q$--torus with the so chosen frequencies exists. 

In practice, we can hardly provide such a proof by rigorous means. 
Instead, as mentioned already we work with finite truncations of 
the PL series. In \cite{chrietal2010} we computed only low order 
truncations, i.e. for orders not higher than three. In order 
to probe the behavior of our series we developed a 
computer--algebraic program able to perform high order 
computations of the PL series. In the limiting case of 
one-dimensional $q$--tori, i.e., $q$--breathers, we were 
able to reach expansion orders as high as 200. However, 
more interesting is the case of two-- (or more) dimensional 
tori, where quite small divisors are present. Despite this 
fact, it is known from theoretical works \cite{eli1997}, 
\cite{gal1994} that the PL series exhibit {\it near-cancellations} 
between terms of large size generated in the series by the 
recursive appearance of small divisors. Such near-cancellations 
have been shown to lead to the convergence of the PL series for 
diophantine frequencies in simple Hamiltonian models.

\begin{figure}
\centering
\includegraphics[scale=0.20 ]{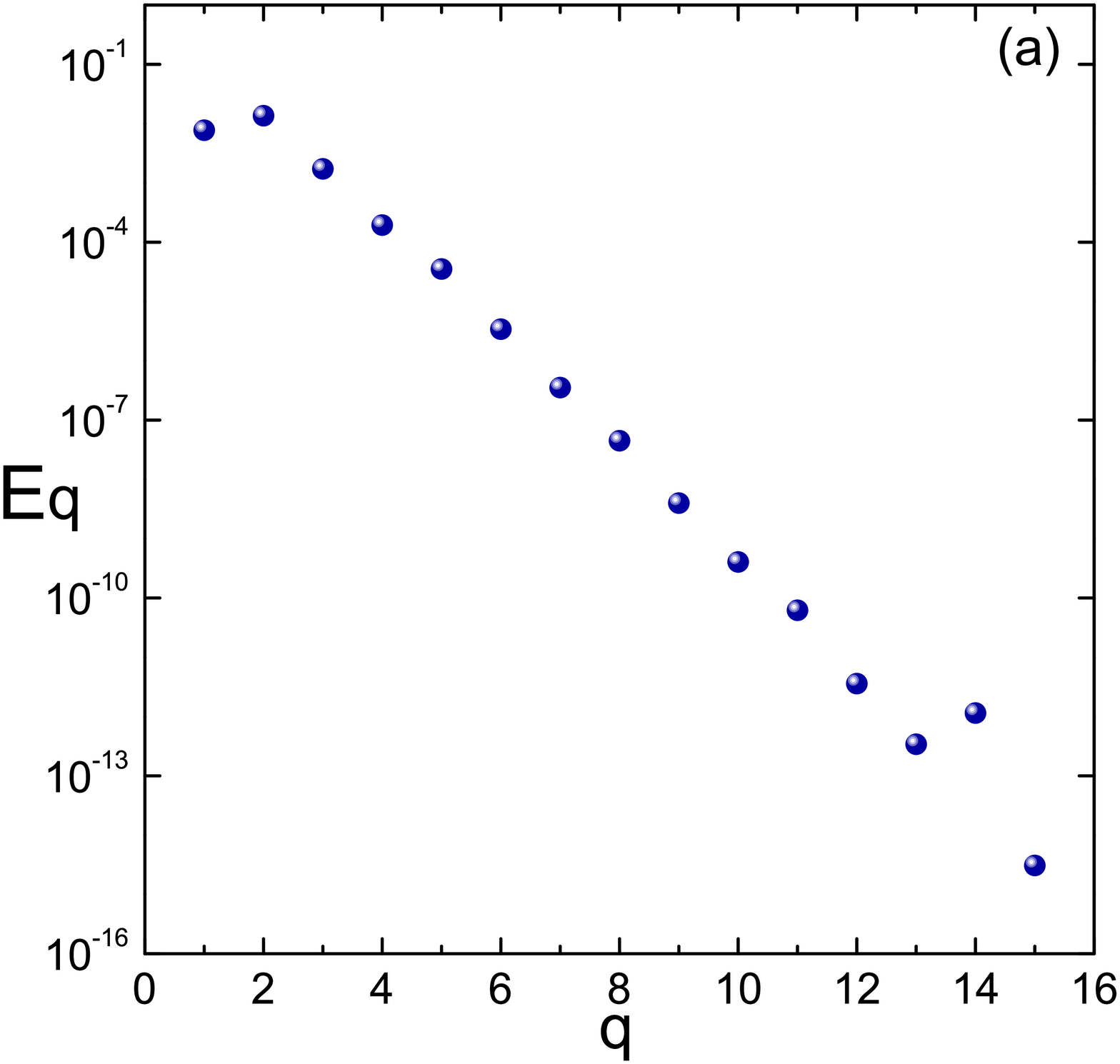}
\includegraphics[scale=0.20 ]{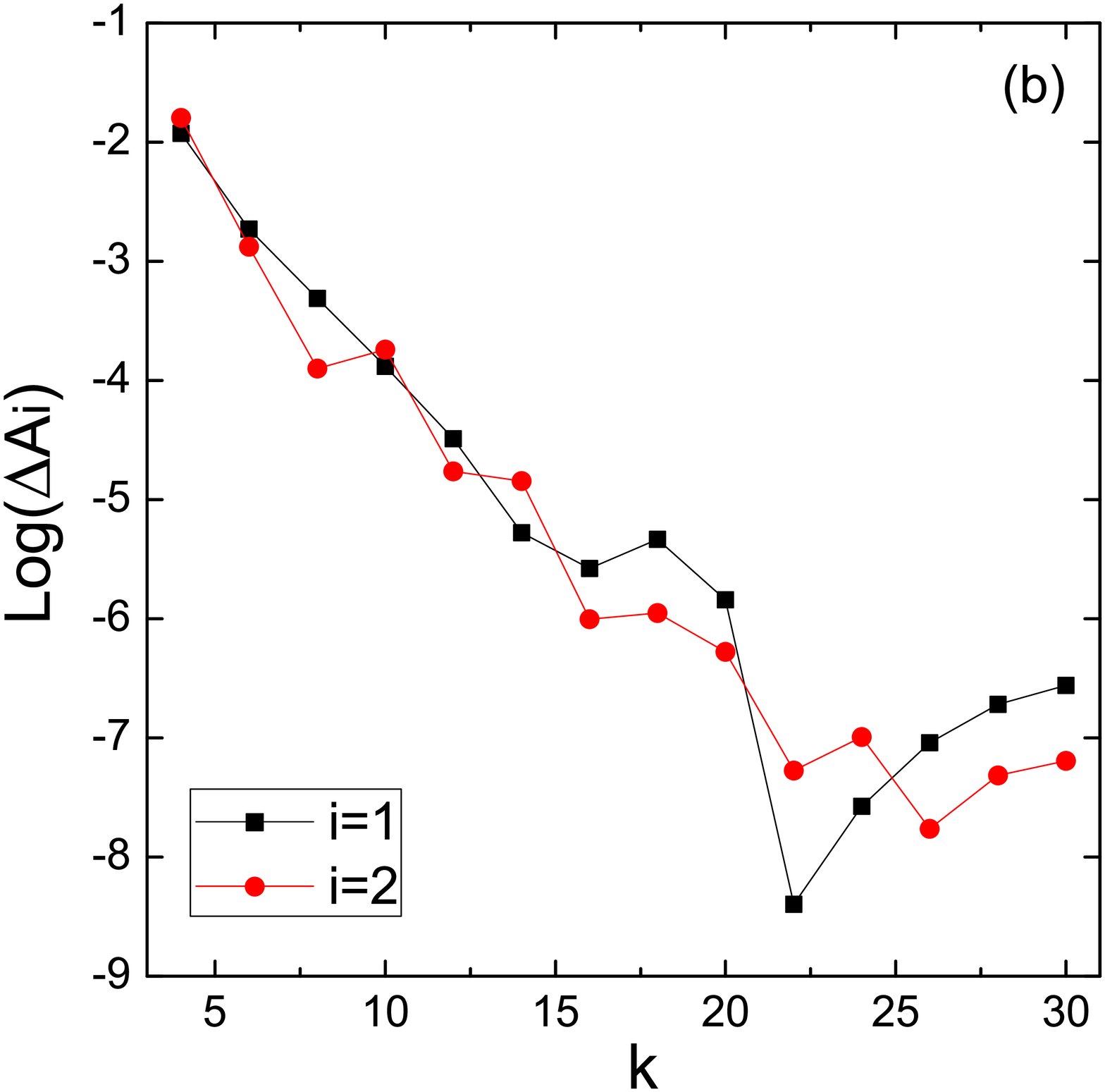} 
\includegraphics[scale=0.20 ]{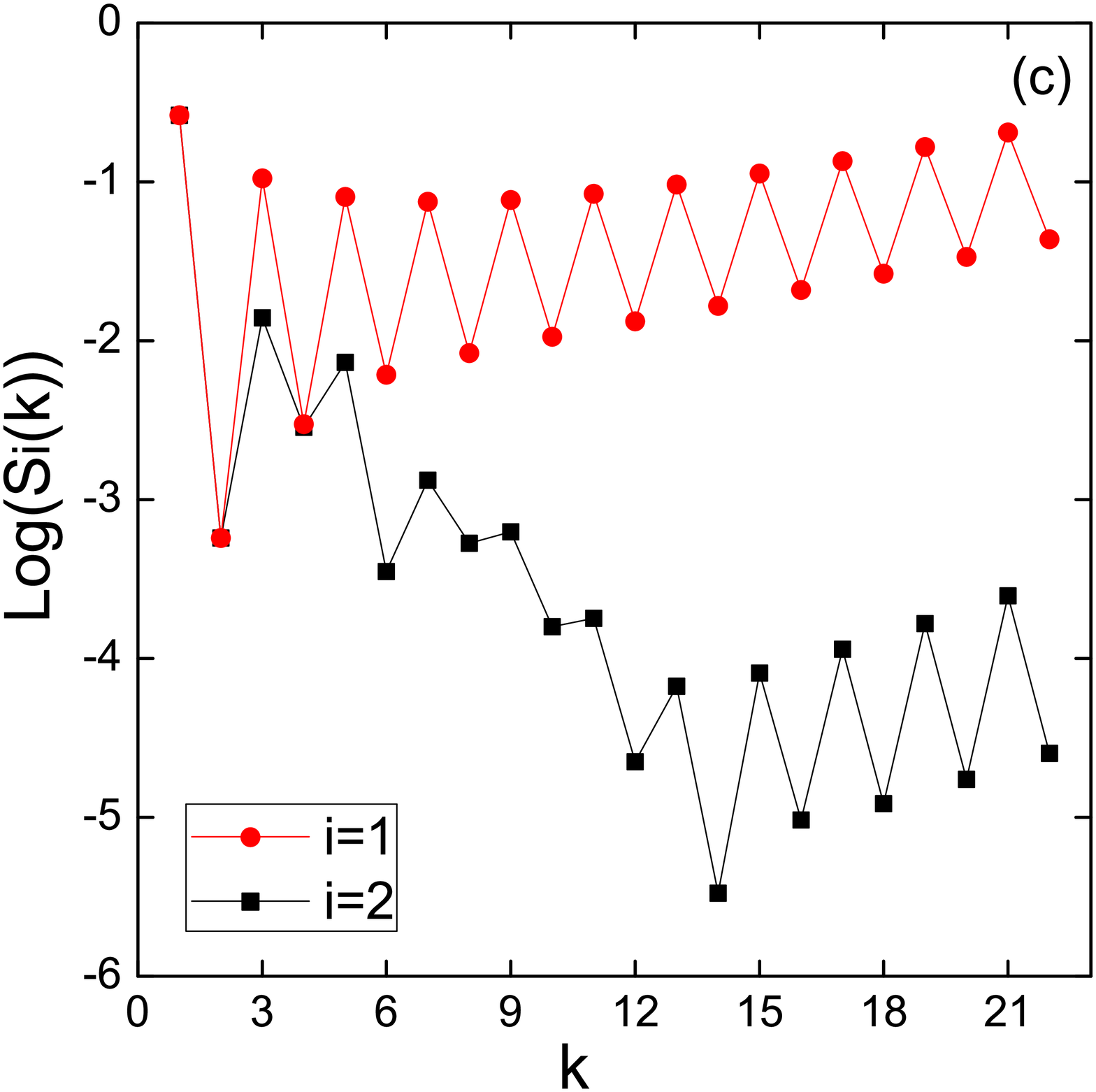} \\
\includegraphics[scale=0.20 ]{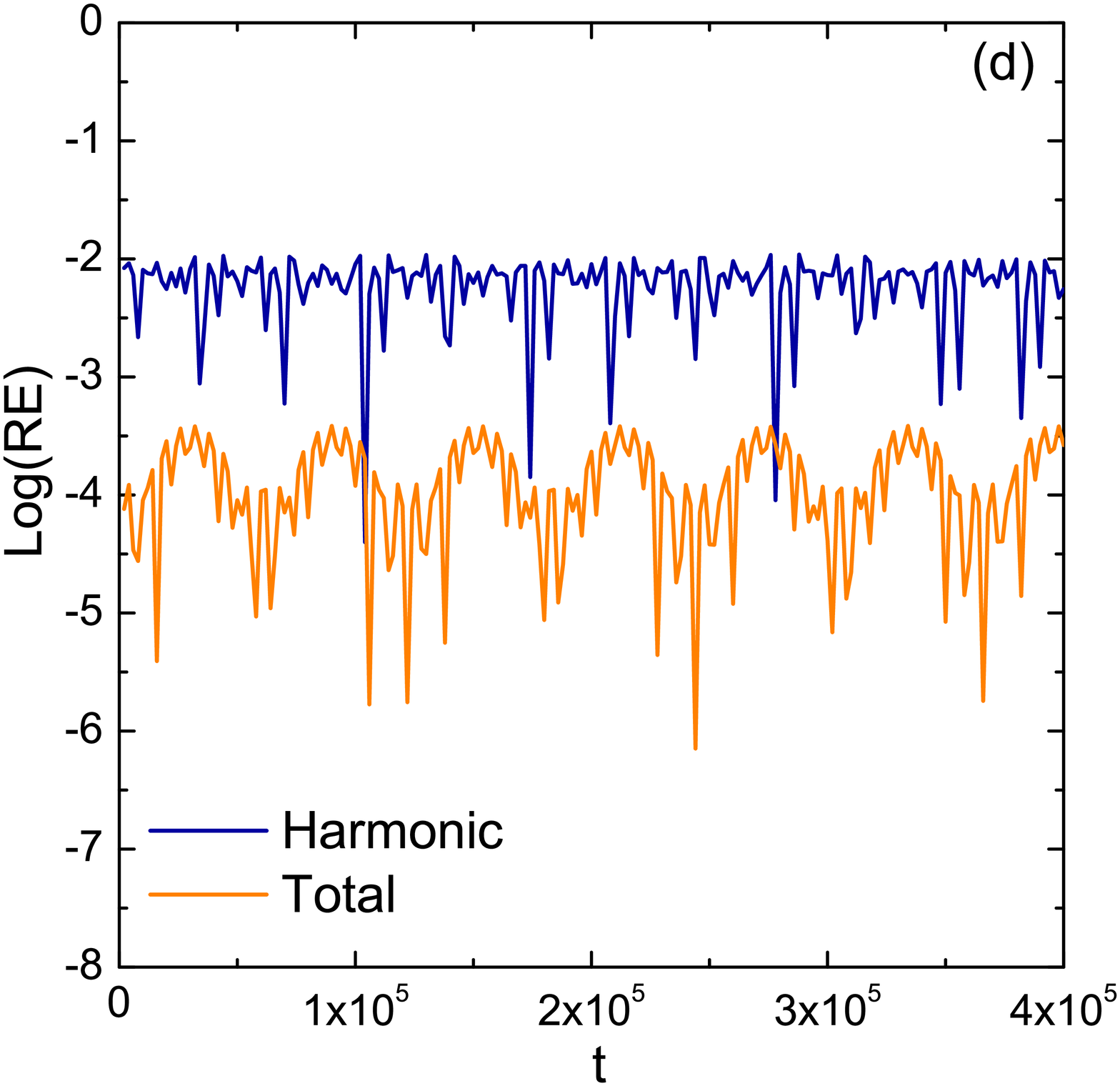}
\includegraphics[scale=0.20 ]{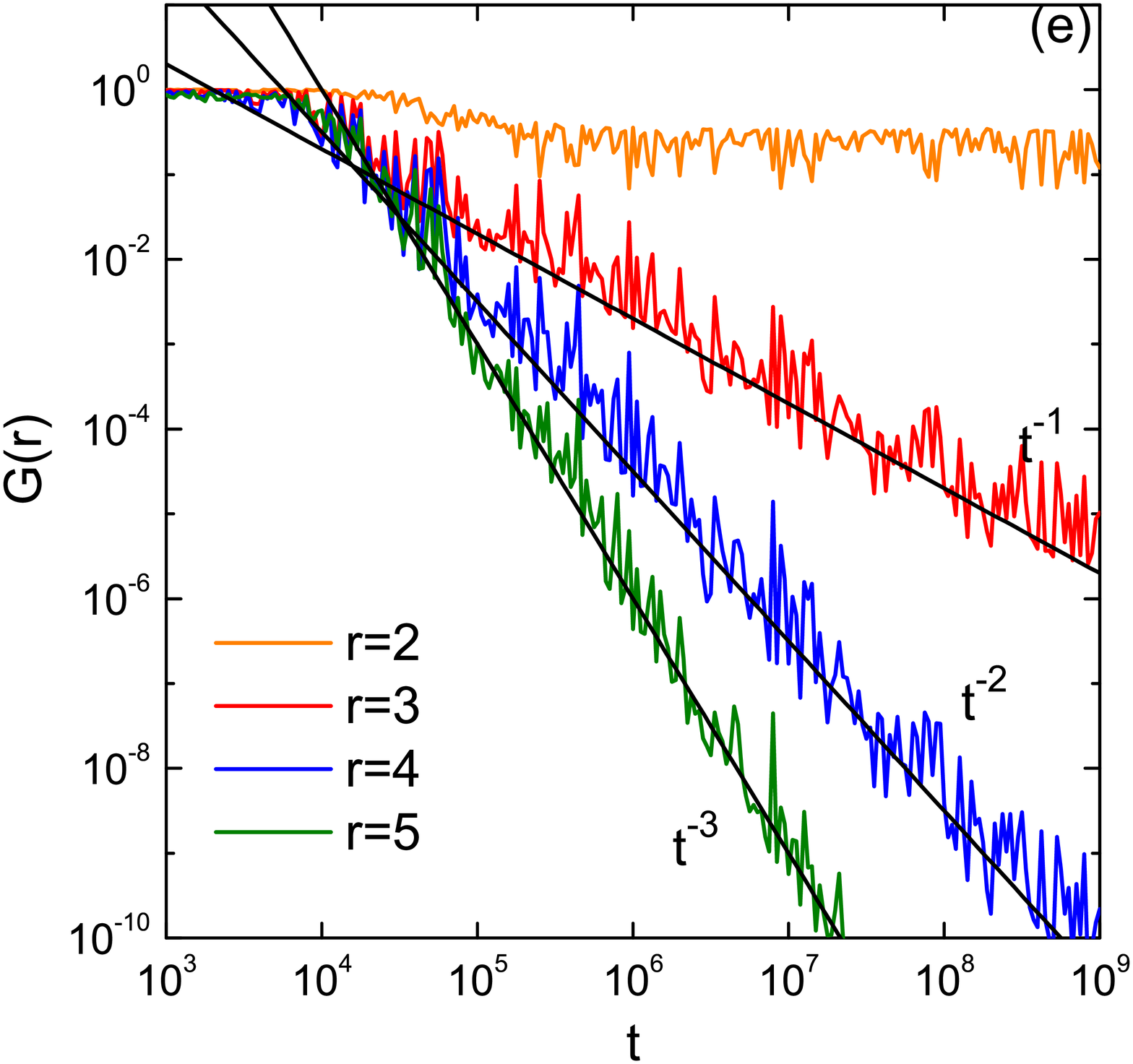}
\caption{ (a) The exponentially localized energy spectrum 
of a 2--torus. (b) The logarithm of the absolute errors 
$\Delta A_i (k) =|A_i(k+2)-A_i(k)|$ for $i=1,2$ versus the order $k$.
(c) The logarithm of the sums $S_1$ and $S_2$ defined in Eq.(\ref{serabs2}) 
and Eq.(\ref{serabs3}) respectively versus $k$. (d) The relative errors 
in time of the harmonic and the total energy calculated by the PL method.
(e) The GALI indices for $r=2,\ldots,5$. \label{canc}  }
\end{figure}

In our particular model, we observe numerically the appearance of 
near-cancellations in our computed PL series. One example is shown 
in Fig.\ref{canc}, referring to $N=16$, $a=0.33$, where, at the 
zeroth order of the PL series, we `excite', i.e. consider a 
non-zero amplitude at the zeroth order for the modes $q_1=1$ and 
$q_2=2$. We construct a series representing a 2D $q$--torus with 
frequencies $\omega_1=0.19626$, $\omega_2=0.39046$. These frequencies 
are commensurable beyond the fifth decimal digit, but they can be 
considered as practically incommensurable regarding the divisors 
they generate up to our maximum reached expansion order $k=30$. 

The resulting solution corresponds to a total energy $E=0.02345$. 
Computing the average harmonic energy for each 
mode (up to a time $T$), given by 
$\overline{E}_q=(1/T)\int_0^T (P_q(t)^2+\Omega_q^2 Q_q(t)^2)dt$, 
where $Q_q(t)$, $P_q(t)=\dot{Q}_q(t)$ are the theoretical 
functions determined by the truncated PL series, we arrive  
at the profile shown in Fig.\ref{canc}(a). This profile exhibits 
exponential localization, since the energy of the $q$--th mode 
decays exponentially with $q$. This phenomenon will be studied 
with more detailed examples in sections \ref{4toros} to 
\ref{numqtor} below. 

In order, now, to probe the behavior of our series with increasing 
truncation order, we perform a number of numerical tests leading to  
Figs.\ref{canc}(b),(c),(d), and (e). 

Figure \ref{canc}(b) shows, on a logarithmic scale, the absolute 
errors in the determination of the amplitudes $A_1$,$A_2$ 
using numerical solutions of finite truncations 
of Eqs.(\ref{omeser}) at various (increasing) orders. 
The plotted quantities are $\Delta A_1(k)= |A_1(k+2)-A_1(k)|$ and 
$\Delta A_2(k)=|A_2(k+2)-A_2(k)|$, 
where $A_1(k),A_2(k)$ and $A_1(k+2),A_2(k+2)$ are approximations 
to the amplitudes as determined by solving numerically 
Eqs.(\ref{omeser}) truncated at the $k-th$ and $k+2$nd 
orders respectively. We note that the error in the determination 
of the amplitudes decreases as the truncation order increases, 
up to the order $k=24$, while afterwards the error slightly 
increases. The increase appears after the error reaches a 
minimum level $\Delta A\sim 10^{-8}$--$10^{-7}$, while the 
frequencies themselves become commensurable at this level of 
precision. Thus, in subsequent calculations we use the values 
of the amplitudes found at the order $k=24$, up to which the 
behavior of the series appears as convergent. The computed 
values of the amplitudes are $A_1=0.5848897$, $A_2=0.4115552$.  

Figure \ref{canc}(c) shows now the main effect regarding the 
appearance of near-cancellations in our series. After completing 
the calculations up to the $k$--th order, we re-cast all computed 
series terms $Q_q^{(k)}(t)$ under the form:
\begin{equation}\label{serabs1}
Q_q^{(k)}(t)= \sum_{m_1,m_2}
\left(\sum_{i,j} h^{(k)}_{m_1,m_2,i,j} A_1^i A_2^j\right)
\cos[m_1(\omega_1t +\phi_1)+m_2(\omega_2t +\phi_2)]~~,
\end{equation}
where the sums are over integers $m_1,m_2,i,j$ whose range 
depends on $k$. Then, focusing for example on $q=1$, the upper 
curve in Fig.\ref{canc}(c) shows the value of the sum:
\begin{equation}\label{serabs2}
S_1(k)=\sum_{m_1,m_2}\sum_{i,j} 
|h^{(k)}_{m_1,m_2,i,j} A_1^i A_2^j|
\end{equation}
while the lower curve shows the value of the sum:
\begin{equation}\label{serabs3}
S_2(k)=\sum_{m_1,m_2}\Bigg|\left(\sum_{i,j} 
h^{(k)}_{m_1,m_2,i,j} A_1^i A_2^j\right)\Bigg|~~.
\end{equation}
In words, the upper sum represents the absolute sum of all 
individual Fourier coefficients appearing in the series at 
the order $k$, while, in the lower sum, all coefficients 
corresponding to the same harmonics 
$\cos[m_1(\omega_1t +\phi_1)+m_2(\omega_2t +\phi_2)]$
are grouped first together and summed algebraically (as 
actually happens in the real series after substitution of 
the numerical values of the amplitudes). We now observe that 
the latter summation leads to a near--cancellation of terms 
of increasing size, leaving a residual which decreases 
as $k$ increases. As a result, the cancellation takes 
place up to four orders of magnitude compared to the size 
of each independent term of the PL series at the order $k=14$. 
The overall size of the terms of $Q_1^{(k)}$ at $k=14$ is 
about $10^{-5}$. However, as shown in Fig.\ref{canc}(c), beyond 
the order 14 numerically we cannot observe a cancellation better 
than one part in $10^4$. We attribute this fact to the 
finite precision by which the amplitudes $A_1$, $A_2$ have 
been specified. Nevertheless, despite a slight increase after 
$k=14$, the upper and lower curves in Fig.\ref{canc}(c) appear 
to move one parallel to the other on a logarithmic scale, 
indicating that the cancellations take place at all computed 
orders beyond $k=14$. 

As an independent test of the precision of our series 
computations, Fig.\ref{canc}(d) shows the time evolution of the 
relative harmonic and total energy, $RE=(E(t)-E(0))/E(0)$,
found by substituting the functions $Q_q(t)$, $P_q(t)$, as computed by 
our truncated PL series, into the harmonic part or the total 
Hamiltonian (\ref{fpuham2}). For an exact solution, the harmonic 
energy undergoes some time variations due to the presence of 
the cubic terms in the Hamiltonian, whose size in our case is 
of order $\sim 10^{-2}$. On the other hand, the value of the 
full Hamiltonian energy should be a preserved quantity. Using 
our truncated series, we observe some fluctuations in the total 
energy which indicate an error of the level of $10^{-4}$. 
On the other hand, the total harmonic energy undergoes 
fluctuations at the expected level $10^{-2}$. 

As a final test, we implement (as in \cite{chrietal2010}) the method 
of the Generalized Alignment Index (GALI) \cite{skoetal2007}, which 
determines the dimension of a {\it stable} low-dimensional torus by 
the temporal behavior of a set of indices (denoted G$_{2}$, G$_{3}$, 
\ldots) computed via an integration of the variational equations of 
motion along with the original ones. We recall that each index $G_r$ 
represents the $r$--volume of $r$ unitary deviation vectors. In the 
case of chaotic orbits, all indices $G_r$ decay exponentially in time. 
However, in the case of regular quasi-periodic orbits, the indices 
$G_r$ obtain an asymptotically constant value in time for $r$ smaller 
or equal to the dimension of the torus on which the orbit lies, while 
they decay by power laws for $r$ larger than the torus dimension 
(see  \cite{chri06,skoetal2007,skoetal2008} for more details). 

Figure \ref{canc}(e) shows the behavior of the indices $G_r$ for $r=2, 
3,4,5$, for an orbit integrated numerically with initial conditions 
as specified by our truncated series solution at the time $t=0$. We 
observe that even after a quite long integration time ($10^9$), the 
index $G_2$ appears so be stabilized to a nearly constant value, while 
all other indices decay in time by the power laws $t^{-1}$, 
$t^{-2}$, and $t^{-3}$ respectively. This behavior implies that 
our initial conditions lie indeed on a 2D-torus embedded in the 
30--dimensional phase space of the considered FPU system. 

\section{A further example. Comparison of $q$--torus 
and `FPU trajectory'}
\label{4toros}
In order to construct the two-dimensional $q$--torus PL series solution  
of the previous section, a specific choice of values for the frequencies 
$\omega_1$ and $\omega_2$ was made. In general, such choices lead 
to formal series of the form (\ref{omeser}) for which there is no 
a priori guarantee that i) real-valued solutions for the amplitudes 
$A_{q_i}$ exist, and ii) if they exist, that they lead to convergent 
series (\ref{qser}). In order to circumvent this difficulty, we have 
developed a step-by-step algorithm by which we specify appropriate 
sets of numerical values $\omega_{q_i}$ for constructing $q$--torus 
solutions. For fixed $N$ and choice of $q_i$, this algorithm allows 
to move in frequency space, by specifying sets of values $\omega_{q_i}$ 
of progressively higher difference from the unperturbed frequencies 
$\Omega_{q_i}$. Using the so-determined values of the frequencies 
$\omega_{q_i}$ we nearly always find real-valued solutions $A_{q_i}$ 
of the equations (\ref{omeser}) (in truncated form). This, in turn, 
allows to determine series of the form (\ref{qser}) exhibiting 
an apparently convergent behavior according to all numerical criteria 
set in section \ref{plseries}. In this way we construct approximate 
$q$--torus solutions corresponding to progressively higher values of 
the energy.
 
The above step-by-step algorithm for the determination of frequency values is 
discussed in \ref{AppA}. Implementing this algorithm in the FPU--$\alpha$ 
model with $N=32$, and an initial excitation of the 4 lowest frequency 
modes $q_1=1$, $q_2=2$, $q_3=3$, $q_4=4$, we determine sets of values 
$\omega_{1}$, $\omega_{2}$, $\omega_{3}$, $\omega_{4}$ yielding $q$--torus 
solutions corresponding to progressively higher energy. For each set 
we solve numerically the truncated Eqs.(\ref{omeser}) in order to
specify the corresponding values of the amplitudes $A_{1}$, 
$A_{2}$, $A_{3}$, $A_{4}$. 

\begin{figure}
\centering
\includegraphics[scale=0.18 ]{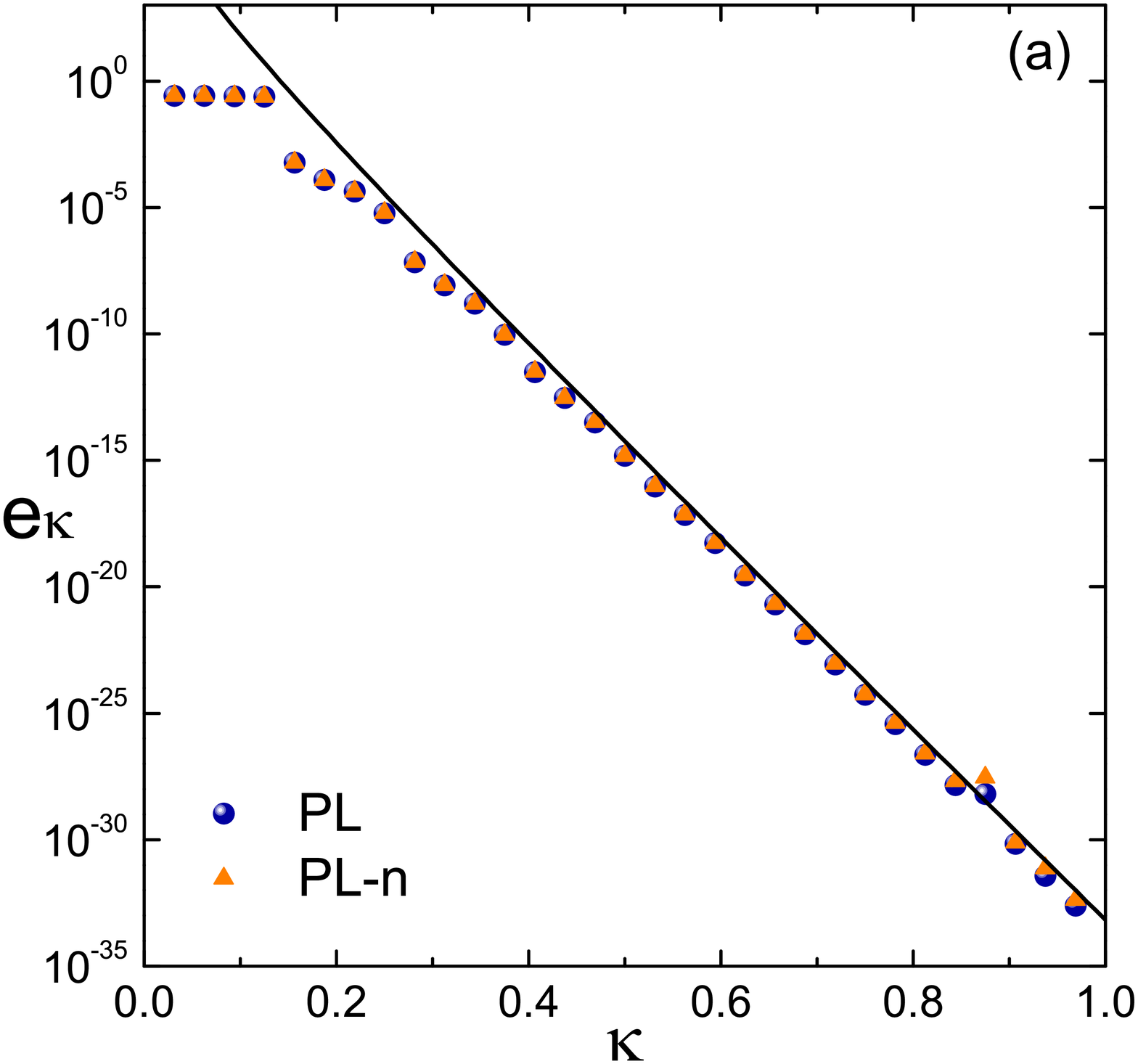}
\includegraphics[scale=0.18 ]{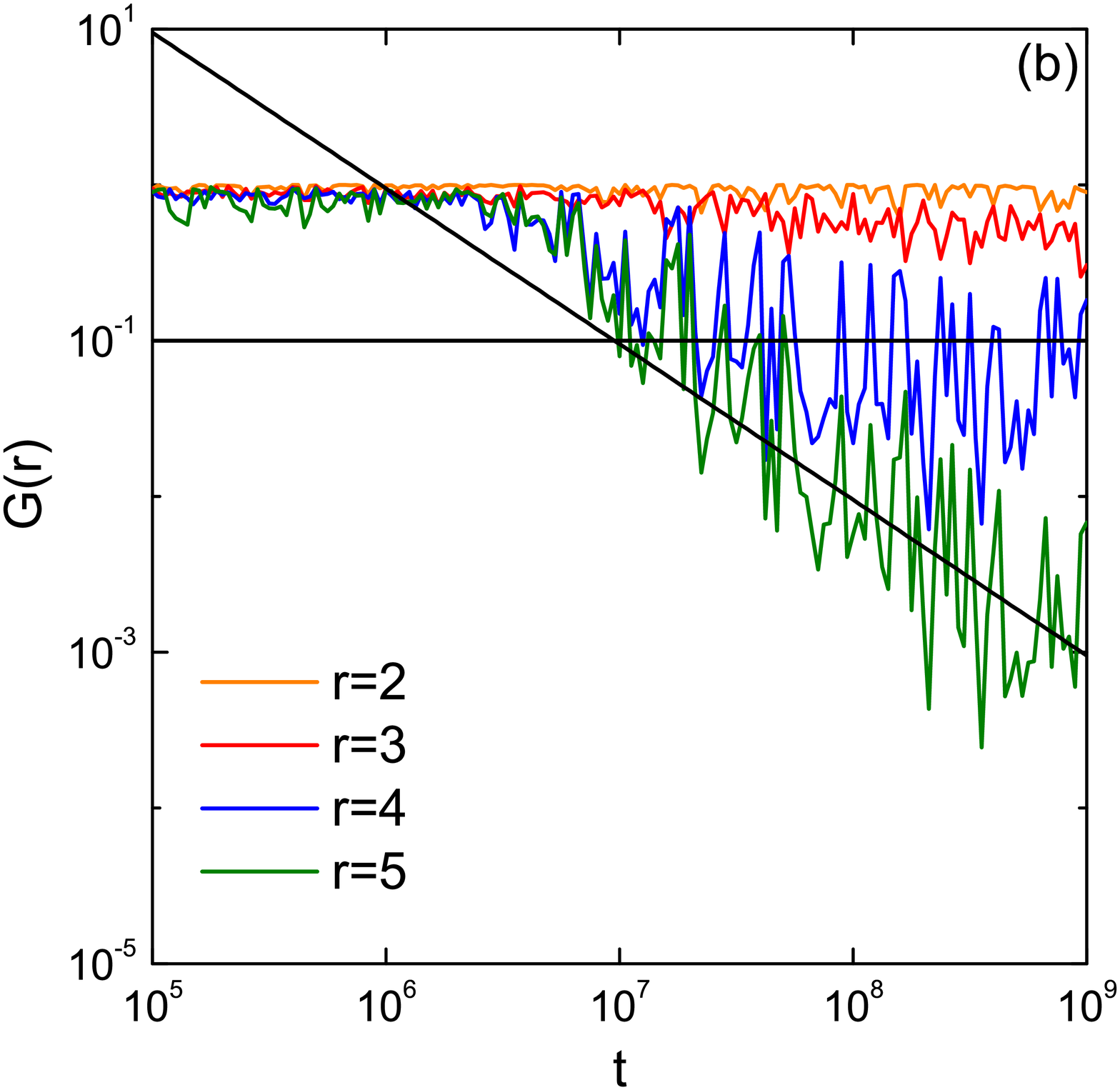}
\includegraphics[scale=0.18 ]{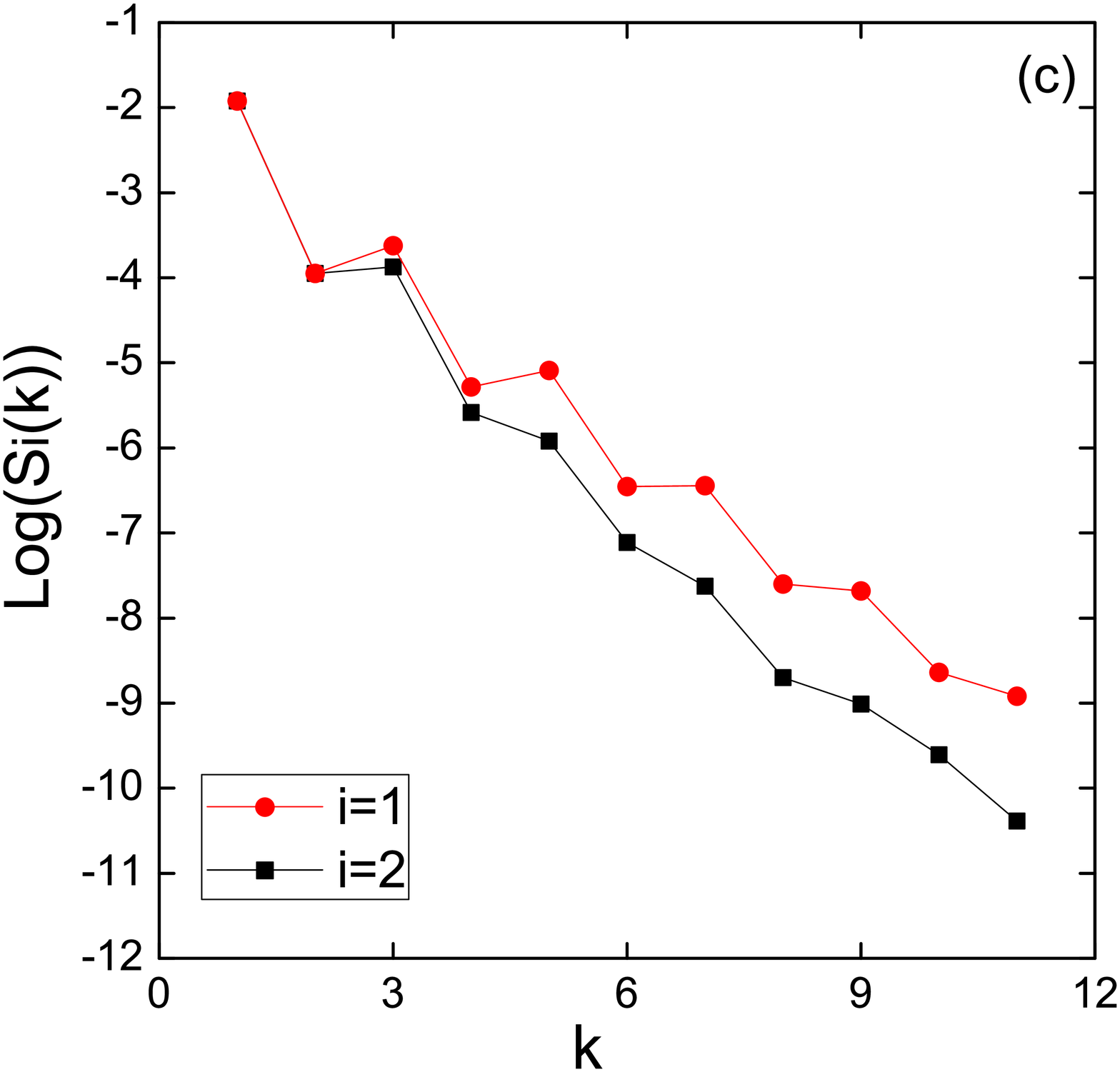}
\caption{ (a) Normalized averaged exponential spectra $e_{\kappa }$
versus $\kappa =q/N $ for the system with ${\cal D}_0 = \{1,2,3,4\}$,
$\alpha =0.33$, $E=0.000182466$, $N=32$, for a total time
$T=10^6$. The blue spheres correspond to a $q$--torus construction
with the truncated PL series, denoted $Q_q^{PL,11}(t)$. The orange
triangles correspond to a numerical solution, denoted $Q_q^{PLn,11}(t)$,
obtained by numerically integrating the FPU equations of motion
with initial conditions $Q_q^{PL,11}(0)$, $P_q^{PL,11}(0)$. 
The continuous line is $\log (E_q/E)= 3.49545 - 36.6613\kappa  
-2 \log (\kappa )$. (b) GALI indices (see text) for $r=2,3,4,5$ 
calculated for the orbit of the orange spectrum in (a). 
(c) The logarithm of the sums $S_1$ and $S_2$ defined in Eq.(\ref{serabs2}) 
and Eq.(\ref{serabs3}) respectively versus $k$. 
\label{fig:qexample}  }
\end{figure}

Figure \ref{fig:qexample} refers to a $q$--torus solution of this 
form computed for the numerical values for the frequencies and amplitudes 
given in the first group of the table in \ref{AppF}. The corresponding 
series (\ref{qser}) are truncated at the order $k_0=11$. We refer  
to such {\it analytical} (i.e. truncated series) solutions by the 
notation $Q_q^{PL,k_0}(t)$. However, in a number of tests we also 
use orbits obtained by {\it numerical}\footnote{We used all over 
the paper the Yoshida symplectic splitting algorithm of forth order.}
integration of analytical initial conditions, i.e. the initial 
conditions $Q_q^{PL,k_0}(0)$. We denote such orbits by 
$Q_q^{PLn,k_0}(t)$. 
 
In Fig.\ref{fig:qexample}(a) the blue spheres refer to the analytical 
orbit $Q_q^{PL,11}(t)$, while the orange spheres refer to the numerical 
orbit $Q_q^{PLn,11}(t)$. The figure shows the averaged in time and 
normalized harmonic energies $e_{\kappa } = E_{\kappa}/E$ for the 
above solutions as a function of the rescaled wavenumber $\kappa=q/N$. 
We observe that $Q_q^{PL,k_0}(0)$ and  $Q_q^{PLn,k_0}(t)$ have no 
distinguishable difference in their profiles $e_{\kappa}$. This implies 
that the analytical or numerical determination of the orbit using the 
same initial conditions leads to an invariant profile $e_{\kappa}$, 
as expected for a solution lying on a $q$--torus\footnote{Actually, not only the two profiles match, but also 
the solutions themselves $Q_q^{PL,11}(t)$ and $Q_q^{PLn,11}(t)$ 
are almost identical.}.

The main feature to observe in Fig.\ref{fig:qexample} is, again, 
exponential localization. Namely, we observe a strong localization 
of the energy in the first four modes, followed by an exponential 
decay of $e_{\kappa}$ versus the re--scaled wavenumber $\kappa$. We 
emphasize that this is the profile corresponding to a $q$--torus  
solution computed by analytical means, i.e., by truncated PL 
series. As shown in subsection \ref{acaseqtor} below, for such 
solutions we can predict theoretically the exponential slope 
of the localization profile. This prediction is shown in 
Fig.\ref{fig:qexample}(a) by a continuous line with negative 
slope. 

Besides comparing the solutions $Q_q^{PL,11}(t)$ and 
$Q_q^{PLn,11}(t)$, here, as in section \ref{plseries}, 
we perform two more tests: 

i) We compute the GALI indices. Fig.\ref{fig:qexample}(b) shows 
the evolution of the GALI indices $G_r$ for $r=2,3,4,5$. 
The indices $G_2$ and $G_3$ converge very rapidly to a constant 
value, while $G_4$ oscillates also around a constant average 
value after an initial decay lasting for a rather long time, 
$t=3\cdot 10^7$. On the other hand, at times greater than the 
time of stabilization of $G_4$, the index $G_5$ clearly decays 
as $1/t$, up to at least the time $10^9$. Thus we conclude that 
the numerical orbit $Q_q^{PLn,11}(t)$ lies on a torus of 
dimension $s=4$. In fact, since we only have a finite precision 
in the initial conditions, a more precise statement is that the 
orbit follows a $4$--torus dynamics for times up to a time 
$10^9$, i.e., no large scale chaotic diffusion phenomena arise 
for the orbit within this timescale.

ii) We probe the convergence of the PL series using the sums 
$S_1(k)$ and $S_2(k)$ (Eqs.(\ref{serabs2}) and (\ref{serabs3})
respectively, 
Fig.\ref{fig:qexample}(c)). In the present case both sums 
$S_1$ and $S_2$ decay up to the maximum considered truncation 
order $k_0=11$. However, we observe again the phenomenon of 
cancellations, which results in a difference of about two 
orders of magnitude between $S_1$ and $S_2$ at $k_0=11$.

\subsection{Comparison with a FPU--trajectory.
Stages of dynamics \label{fputra}}

A question of central relevance concerns the behavior of nearby 
trajectories to a $q$--torus solution. A case of particular 
interest regards the the so--called `FPU--trajectories' 
\cite{flaetal2005}. These are trajectories arising from initial 
conditions in which we excite initially only a small subset of modes. 
In the example of the $q$--torus solution of Fig.\ref{fig:qexample}, 
a corresponding nearby FPU--trajectory arises by setting initially 
$Q_q(0)=A_q$, for $q=1,2,3,4$, and $Q_q(0)=0$ for $4<q<32$. We stress 
that such an initial condition cannot be confused with the one 
of the $q$--torus solution itself, given by $Q_q(0)=Q_q^{PL,11}(0)\neq 0$ 
for {\it all} $q$ with $1\leq q<32$. In fact, in the latter 
case {\it all} modes have some energy initially, while in the 
case of the FPU--trajectory only the four first modes have 
energy initially. 
 
Nevertheless, in the case of FPU-trajectories we observe 
an energy flow from the initially excited modes to the remaining 
modes in $q$--space. As a result, the energy localization 
profiles after a long time present quite similar features 
with those of the nearby $q$--tori solutions. 

In \cite{dresden} a detailed study was made of the dynamics of 
FPU--trajectories arising from the initial excitation of just 
one mode. Such FPU--trajectories are nearby to one-dimensional 
$q$--tori, i.e., $q$--breathers. It was found that their 
evolution can be separated into two main so-called 
{\it stages of dynamics}, whose distinction becomes more evident 
by observing the time evolution of the energy acquired by 
each one of the normal modes. During the {\it first stage}, 
energy is transfered from the initially excited (low frequency) 
mode to the rest of modes, the transfer taking place via a 
resonant mechanism (see \cite{dresden}). The first 
stage stops after a relatively short time, beyond which 
the energy spectrum of the system appears to be practically 
frozen to an exponentially decaying profile. In \cite{dresden} 
it was emphasized that this process is integrable-like, since 
the same process occurs in the so-called Toda lattice model, 
which is integrable. However, in the FPU case the first stage 
of dynamics is followed by a {\it second stage} of dynamics, 
during which a weakly chaotic (and slow) drift of energy to the 
high frequency modes takes place. The time scale of the second 
stage of dynamics, which leads the system towards energy 
equipartition, was the main subject of study in \cite{benetal2011,dresden}.

\begin{figure}
\centering
\includegraphics[scale=0.20 ]{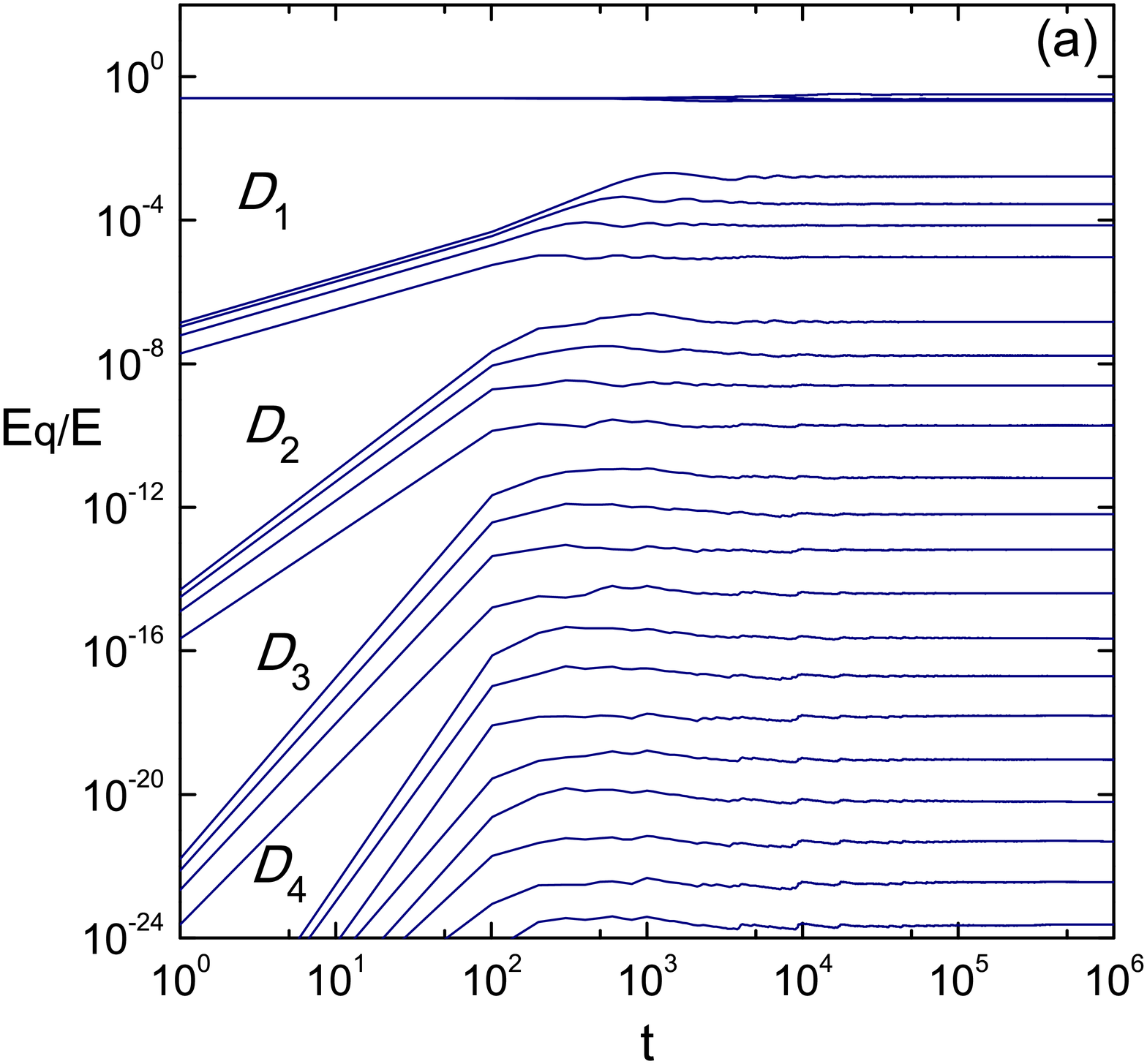}
\includegraphics[scale=0.20 ]{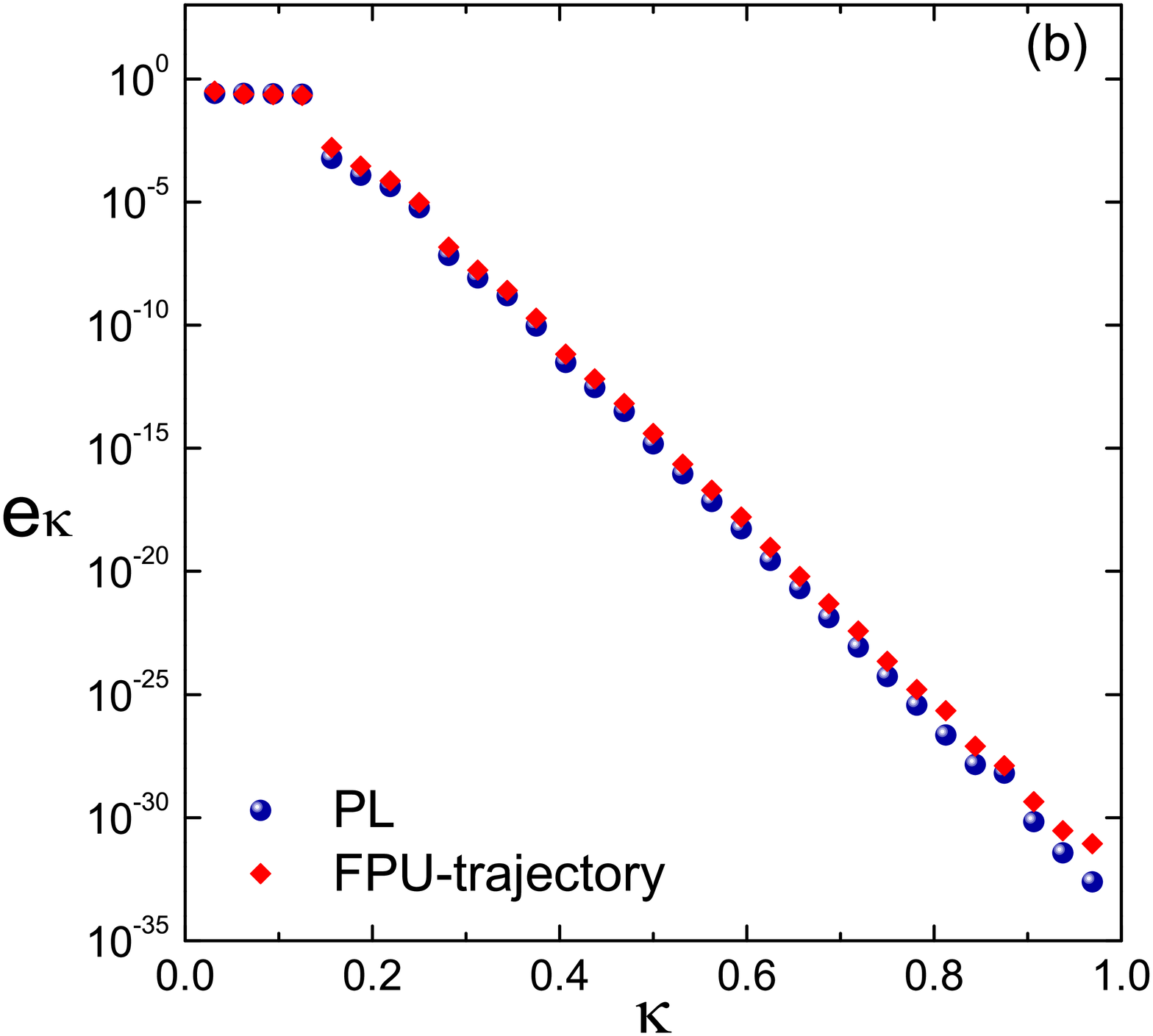}
\caption{ (a) Time evolution of the normalized averaged energies
$\overline{E}_q(t)/E$, in double logarithmic scale,
for the FPU--trajectory arising from the excitation of the first
$s=4$ modes with initial conditions $Q_q(0)=A_q$ if $q=1,2,3,4$ and 
$Q_q(0)=0$ else, and system's parameters as in Fig.\ref{fig:qexample} 
for the $q$--torus solution. (b) Comparison of the normalized
averaged energy spectra $e_{\kappa}$ versus $\kappa$ of the
$q$--torus solution (blue) and the corresponding FPU--trajectory
(orange) over a total integration time $T=10^6$.
\label{fig:dif}}
\end{figure}

We will now show that the distinction of two stages of dynamics 
applies to FPU--trajectories not only close to $q$--breathers, 
but also close to $q$--tori of dimension larger than one. 
An example, referring to a FPU--trajectory 
near the 4--torus of Fig.\ref{fig:qexample}, is shown in 
Fig.\ref{fig:dif}. Fig.\ref{fig:dif}(a) shows the evolution
of the normalized time averaged energies $\overline{E}_q(T)/E$ 
for all the modes $q=1,\ldots,N-1$, in the case of the 
FPU--trajectory, during the `first stage' of dynamics. The energies 
of all modes besides $q=1,2,3,4$ are initially equal to zero. 
Thus, the initial distance in phase space between the FPU--trajectory 
and the $q$--torus trajectory is of order $O(\mu)$, where in this 
case $\mu\approx 4\times 10^{-2}$. One can observe that the modes 
$q=1,2,3,4$, which shared exclusively the total energy at $T=0$, 
continue to share most of the energy at all subsequent times $T$. 
As a result, $\overline{E}_q(T)/E$ remains practically constant 
for these 4 modes, exhibiting only small variations (of order 
$10^{-4}$) which correspond to the energy gradually transferred 
to the remaining modes. However, the energy transfer to the 
remaining modes modes is also evident. This takes place in a 
rather short time interval, of order $T=10^3$ for the modes
$q=5,6,7,8$, and $T=10^2$ for the remaining ones.

A key feature, now, of the energy transfer process during the 
first stage of dynamics is the formation of well distinguished 
{\it groups of modes}. Besides the group ${\cal D}_0=\{1,2,3,4\}$, 
which shared the energy initially, by the time evolutions of the 
energies in Fig.\ref{fig:dif} we clearly distinguish the groups   
${\cal D}_1=\{5,6,7,8 \}$, ${\cal D}_2 =\{ 9,10,11,12 \}$ etc, 
or, in general, ${\cal D}_k=\{ks+1,\ldots, (k+1)s\}$.
It is observed that, during this stage, for all modes  $q$ in the same
group ${\cal D}_k$, $k>0$ the quantity $\overline{E}_q/E$ grows in
time by almost the same power--law, i.e. the energy spectra behave as
 $\overline{E}_q(t)/E\sim  t^{c_q(k)}$, where the exponent is also 
almost constant within a group: $c_q(k)={3\over 2} k + \epsilon_q $, 
with $|\epsilon_q|<<1$ for all $q\in{\cal D}_k$. 

As shown in section \ref{sequence}, the above groups ${\cal D}_k$ 
correspond exactly to the so-called `sequence of mode excitations' 
associated with a $q$--torus construction via PL series with 
initial excitation ${\cal D}_0$. That is, {\it the sequence of 
groups formed in the energy transfer process for a FPU-trajectory,  
during  the `first stage of dynamics', coincides with the formal 
`sequence of mode excitations' appearing in the PL series 
construction of nearby $q$--tori with the same initial excitation 
${\cal D}_0$}. 

It must be emphasized that, 
in the case of a FPU-trajectory, the separation of the modes 
into groups is a {\it dynamical} phenomenon related to the 
process of energy transfer. On the contrary, in the case of 
$q$--tori, the groups only concern a {\it formal} aspect of 
perturbation theory, as analyzed in section \ref{sequence} 
below. The fact that the final groups defined in either case 
are the same, indicates some deeper connection between the 
$q$--torus solutions and the FPU--trajectories, whose evolution 
during the first stage of dynamics is integrable-like. Nevertheless, 
most FPU--trajectories enter eventually into the stage of approach 
to energy equipartition, whereby their time evolution is 
weakly chaotic, while the solutions lying on $q$--tori 
maintain their regular character at all times $t$.

The averaged normalized energy spectrum for the FPU--trajectory 
at the end of the first stage of dynamics appears to be stabilized 
to a form persisting for quite long times. This stabilized spectrum
exhibits also an exponential profile, very similar to the $q$--torus 
profile shown in Fig.\ref{fig:qexample}(a). The two profiles are 
superposed in Fig.\ref{fig:dif}(b). We observe that the groups 
${\cal D}_k$ are well distinguished in the spectrum of the 
FPU--trajectory, and they coincide with the ones of the $q$--torus 
solution. To the theoretical interpretation of the latter groups 
we now turn our attention. 

\section{Sequence of mode excitations}
\label{sequence}
It is well known that for some special choices of initial
conditions, the resulting FPU--trajectories (in both the $\alpha$
and $\beta$ models) take place on lower dimensional invariant
sub--manifolds of the FPU phase space. This is due to the
existence of {\it discrete symmetries}, which give rise to
explicit low--dimensional FPU solutions. An extensive study on
such symmetries is made in \cite{chechin2005,poggi1997,rink2003} 
(in the latter such solutions are called `bushes of normal modes'). 
A particular case are {\it periodic} trajectories, arising from 
exciting only one of the modes $q_0=N/3$, $N/2$ or $2N/3$ in the 
$\beta$ model \cite{Pasta73} (see also \cite{chechin2005,poggi1997}).
Solutions like the above lead, by definition, to energy localization, 
since the energy remains always distributed among a small subset 
of modes.

On the other hand, as pointed out in the examples of the previous 
sections, energy localization occurs also for sets of initial 
conditions not obeying any obvious or simple symmetry. As in 
\cite{chrietal2010}, we now study this phenomenon using the concept 
of {\it propagation} of some initial {\it excitation} in the PL series 
for low--dimensional tori. We can briefly state our main result as follows: 
through the study of propagation, we can define a hierarchy of groups 
of modes participating in a $q$--torus solution, such that all the
modes in one group share a similar (in order of magnitude) amount
of energy, while distinct groups share quite different amounts of
energy. The hierarchy of these groups allows us to predict the
whole localization profile via a leading order analysis of the
associated PL series. However, as shown in the previous section, 
it also allows us to characterize the {\it paths of energy transfer} 
in $q$--space for FPU trajectories neighboring some $q$--torus 
solution, from an initial excitation up to the moment when a 
metastable profile is established for the FPU--trajectories 
(like in the example of Fig.\ref{fig:dif}(a)).

We start by the following formal definitions:\\
\\
{\bf Definition 1:} A mode $q$ is said to be {\it excited at the
$n$--th order of the PL scheme}, iff in the series (\ref{qser})
it is $Q_q^{(k)}(t)=0$ for all $k<n$, and $Q_q^{(n)}(t)\neq 0$.
In addition, $Q_q^{(n)}(t)$ is called {\it leading order term}
of the series (\ref{qser}). \\
\\
{\bf Definition 2:} Let ${\cal D}_0$ be a set of modes excited at
the zero order of the PL scheme according to Eqs.(\ref{qq0}).
The sequence of sets ${\cal D}_k$, $k=1,2,\ldots$, where ${\cal
D}_k$ consists of modes excited at the $k$--th order of the PL
scheme, is called {\it sequence of mode excitations}. \\
\\
{\bf Definition 3:} Let $k_0$ be a positive integer. We call {\it
tail modes} with respect to $k_0$ the modes belonging to the set
$\cup_{k\geq k_0} {\cal D}_k$. \footnote{In \cite{dresden},
studying solutions corresponding to an initial excitation of the
first normal mode, tail modes were called those belonging to the
last third of the spectrum. This is equivalent to
state that $k_0=[2N/3]$.} \\
\\
{\bf Definition 4:} We call {\it FPU--trajectory} with initial
excitation in the set of modes ${\cal D}_0\equiv\{ q_1,\ldots,q_s\}$
the trajectory resulting from the equations of motion (\ref{eqmo})
for the set of initial conditions $Q_{q}(0)=A_{q}\cos\phi_{q}$,
$P_{q}(0)=-A_{q}\Omega_q\sin\phi_{q}$, for some set of amplitudes
$A_q$ and phases $\phi_q$, if $q\in{\cal D}_0$, and $Q_q(0)=P_q(0)=0$
if $q\notin{\cal D}_0$.
\footnote{As rendered clear with specific examples throughout the 
paper, in the case of $q$--tori one has in general $Q_q(0)\neq 0$ 
or $P_q(0)\neq 0$ for {\it all} $q=1,\ldots, N-1$. In fact, for the 
modes belonging to the $k$--th set ${\cal D}_k$ (see Definition 2) we 
have in general $Q_q(t)\simeq O(\mu^k)$ for all times $t$, including 
$t=0$. This implies that in $q$--tori all modes have some energy already 
at $t=0$. On the contrary, according to the Definition 4, in 
FPU--trajectories only the modes belonging to ${\cal D}_0$ share 
the total energy at $t=0$. In that sense, the use of the term `excitation' 
for FPU--trajectories is literal, i.e. ${\cal D}_0$ refers to the
modes excited at $t=0$. The origin of the term `FPU--trajectories'
is that these are trajectories with initial conditions
of the same type as those considered in the original FPU paper.}\\

In order to determine the sequence ${\cal D}_k$ produced by a
particular initial excitation ${\cal D}_0$, we first define the
set
\begin{eqnarray}\label{set}
 \Sigma ^r= \{ -1,+1 \}^{r}~~~.
\end{eqnarray}
The elements of $\Sigma ^r$ are $r$--dimensional vectors of the
form $\sigma^{(r)} = (\sigma_1, \ldots, \sigma_r)$, where
$\sigma_i =1$ or $-1$, $i=1, \ldots, r$. Furthermore, for an
$r$--vector $x\equiv(x_1,\ldots,x_r)$ we define $\sigma^{(r)}
\cdot x$ as the Euclidean product $\sigma^{(r)} \cdot x = \sigma_1
x_1 + \ldots + \sigma_r x_r $. We can now prove the following \\
\\
{\bf Proposition:} {\it Let ${\cal D}_0=\{q_1,q_2,\ldots,q_s\}$,
$1\leq q_1 < \ldots < q_s\leq N -1$ be the set of seed modes of an
initial excitation yielding a formal PL solution associated with
trajectories on an $s$--dimensional $q$--torus. Let $M_k$ be the set
\begin{eqnarray}\label{app}
M_k = \left\{ \Bigg| 2 \left[ \frac{\mid  \sigma ^{(r)}
q^{(r)} \mid +N-1}{2N} \right] N - \mid  \sigma ^{(r)} q^{(r)}
\mid \Bigg|
: \sigma^{(r)} \in \Sigma ^r,~~q^{(r)} \in  {\cal D}_0^r
\right\}
 \end{eqnarray}
where $r=r(k)$ with $r(k)=k+1$ for the FPU--$\alpha$ model and
$r(k)=2k+1$ for the FPU--$\beta$ model. Then, the sequence of mode 
excitations ${\cal D}_k$ corresponding to the initial choice 
${\cal D}_0$ is defined by the recursive relations
\begin{equation}\label{dk}
{\cal D}_k =M_k \setminus  \bigcup_{0 \leq j \leq k-1}
{\cal D}_{j},~~k=1,2,\ldots
\end{equation}
}

An explicit proof of the above proposition is given in
\cite{chrietal2010} for the FPU--$\beta$, and in \ref{AppB} 
for the FPU--$\alpha$.  We note that $\left[ \cdot \right]$
is the integer part of a number and $M_k$ represents
the set of all modes for which the right hand side of Eqs.
(\ref{moda}) and (\ref{modb}) is non--zero, i.e. the modes
yielding some non--zero contribution to the energy spectrum
up to the $k$--th order of the PL series (some modes excited
at previous orders up to $k$ might also belong to $M_k$).

Some examples clarify the use of Eq.(\ref{dk}):\\

i) {\it q--breathers}: If we choose ${\cal D}_0 = \{q_0\}$,
the so--induced PL solution corresponds to a one--dimensional
torus, i.e. a periodic orbit. In this case we find:
\begin{equation}\label{qbre}
{\cal D}_k=\{q_k\}~\mbox{with}~
q_k= \Bigg| 2  \left[  \frac{r q_0+N-1}{2N} \right] 
N - r q_0 \Bigg|,
\end{equation}
where $r(k)=k+1$ in the FPU--$\alpha$, or $r(k)=2k+1$ in the
FPU--$\beta$. This rule coincides with the one given in
\cite{flachtiz} for an arbitrary seed mode $q_0$. From Eq.
(\ref{qbre}) we readily find that, if $q_0$ is even, only even
modes become excited at subsequent orders in both the $\alpha$ and
$\beta$ models. On the other hand, if $q_0$ is odd, in the
$\alpha$ model both odd and even modes become excited, while in
the $\beta$ model only odd modes become excited. The localization
properties of solutions corresponding to $q$--breather excitations
will be discussed in detail in Section \ref{trajbreath} below.\\

ii) {\it Example with two seed modes:} Suppose ${\cal D}_0 =
\{q_1,q_2\} = \{3,5\}$ for $N$ large in FPU--$\alpha$. At first
order ($k=1$) it is $r(1)=2$. In order to determine $M_1$, we
consider all possible combinations of the symbols $\sigma^{(2)}
\in \Sigma ^2$ and $q^{(2)} \in {\cal D}_0^2$. These are
$\{(1,1),(1,-1),(-1,1),(-1,-1)\}$ and
$\{(3,3),(3,5),(5,3),(5,5)\}$ respectively. Then from
Eq.(\ref{app}) we find $M_1 = \{ 2,3,5,6,8,10 \}$. Since ${\cal
D}_0 = \{ 3, 5 \}$, from Eq.(\ref{dk}) we have ${\cal D}_1 = M_1
\setminus  {\cal D}_0 = \{ 2,6,8,10 \}$. Repeating the above 
procedure for $k=2$, we find $M_2= \{ 1, 3, 5, 7, 9, 11, 13, 15 \}$
and ${\cal D}_2 = M_2 \setminus  {\cal D}_0\cup {\cal D}_1 = \{ 1,
7, 9, 11, 13, 15 \}$. In the same way we proceed to subsequent
orders $k=3,4,\ldots$ In the FPU--$\beta$ one follows the same
steps, but for $r(k)=2k+1$.\\

iii) {\it Excitation representing a low--frequency packet of seed
modes}: Let us consider as an initial excitation the packet of
modes ${\cal D}_0 = \{ 1,2,3,4 \}$ for $N$ large. Following the same procedure
as above, in the FPU--$\alpha$ we find the sequence of excitations
${\cal D}_1 = \{ 5,6,7,8 \}$, ${\cal D}_2 = \{ 9,10,11,12 \}$,
etc. We notice that at each order modes are excited {\it in
groups}. In the same way, the $\beta$ model yields ${\cal D}_1
= \{5,\ldots,12 \}$, ${\cal D}_2 = \{ 13,\ldots,20\}$,  etc.

These propagation rules can be generalized for $s$--dimensional
$q$--tori corresponding to low--frequency packets of modes. The
seed mode excitation ${\cal D}_0 = \{1,2,\ldots,s\}$ generates  ${\cal
D}_k=\{ks+1,\ldots, (k+1)s\}$ in the FPU--$\alpha$, and ${\cal
D}_k=\{(2k-1)s+1,\ldots, (2k+1)s\}$ in the FPU--$\beta$, with
$k\geq 1$. Furthermore, we observe that ${\cal D}_k^{\beta } =
{\cal D}_{2k-1}^{\alpha } \cup {\cal D}_{2k}^{\alpha }$. As shown
in the next subsection, these rules imply that the resulting
$q$--tori solutions exhibit {\it exponential energy localization
profiles}.\\

iv) {\it Excitation representing a high--frequency packet of seed
modes}: As an example, let us consider ${\cal D}_0
=\{28,29,30,31\}$ in the $N=32$ dimensional chain. In the
FPU--$\alpha$ we find ${\cal D}_1 =\{1,\ldots,8\}$, then ${\cal
D}_2 =\{ 20,\ldots,27\}$, etc. The general rule, by setting the last
$s$ modes ${\cal D}_0 = \{N-s,\ldots,N-1\}$ as seed modes, is:
${\cal D}_k=\{(k-1)s+1,\ldots, (k+1)s\}$, if $k=2n+1$ and ${\cal
D}_k=\{N-(k+1)s,\ldots, N-(k-1)s-1\}$, if $k=2n$. By the same way, 
in the FPU--$\beta$  we find ${\cal D}_k=\{N-(k+1)s,\ldots,
N-(k-1)s-1\}$, $\forall k$. Comparing the two models, we see that
${\cal D}_k^{\beta } = {\cal D}_{2k}^{\alpha }$.\\

v) {\it Discrete symmetry solutions:} Suppose ${\cal D}_0 =
\{N/2\}$, or ${\cal D}_0 = \{2N/3\}$. We then find ${\cal D}_k=
{\cal D}_0$ for all $k=1,2,\ldots$ in both models, while in
FPU--$\beta$ the condition ${\cal D}_k= {\cal D}_0$ holds also for
${\cal D}_0 = \{N/3\}$. The $q$--breather solutions for these
cases correspond to the  `nonlinear normal modes' of the
FPU system  \cite{Pasta73,chechin2005,poggi1997},
that coincide with the FPU--trajectories resulting from the same
seed mode.\\

We note finally, that for $q$--breathers we have a general
relation connecting the sequences of excitations ${\cal D}_k$ in
the $\alpha$ and in the $\beta$ model, starting from the same seed
mode. Namely, from Eq.(\ref{qbre}) we find that ${\cal
D}_k^{\beta} = {\cal D}_{2k}^{\alpha}$. In words, the mode excited
at the $k$--th order in the FPU--$\beta $ is the same as the mode
excited at the $2k$--th order in the FPU--$\alpha$. This relation
holds also for excitations of small packets around $N/4$, $N/2$
and $3N/4$, but it does not hold in the case (iii) (low-frequency
packets of modes). In fact, for an arbitrary excitation we have $M^{\alpha }_{2k}
=M^{\beta }_{k} $, but we only have ${\cal D}_k^{\beta } = {\cal
D}_{2k}^{\alpha}$ provided that $M^{\alpha }_{2k-1} \cap
M^{\alpha }_{2k} = \emptyset$.

\section{Numerical examples. Localization profiles}
\label{numqtor}

In this section we provide various numerical tests on the energy profiles 
and dynamics of the $q$--tori and their neighboring FPU--trajectories. 
We are interested to examine several generic localizations profiles, rising
by the excitation of arbitrary modes, consecutive or isolated, that 
form different localization patterns in $q$--space. In particular, we
derive the precise sequence of modes that become excited in subsequent orders
by $s$ consecutive modes, chosen to be in the i) beginning, 
ii) one fourth, iii) middle and iv) three fourths of the spectrum,
while in case i) we predict the localization law of the energy profile.
Finally, few examples on $q$--breathers, as particular cases of 
one dimensional $q$--tori, are given.

The frequencies and amplitudes used throughout all examples below are 
listed in \ref{AppF}. 

\subsection{$q$--Tori low frequency packet solutions and
exponential energy localization}
\label{lowfreqmod}
The FPU--trajectory examined in Section \ref{4toros} is an example 
of a class of solutions
of particular interest in the literature (see \cite{benetal2011,
beretal2004,beretal2005,deletal1999}),
namely solutions corresponding to the initial excitation of
a {\it packet} of low--frequency modes. In fact, it is numerically 
found that, for values of the specific energy beyond
some threshold, low--frequency packets of modes are formed
naturally, even if  initially we excite only {\it one}, e.g.
the $q=1$ mode. Some questions of central interest in the
literature concern the dependence of: i) the width of natural
packets and ii) the exponential slope of the energy spectrum
of the remaining modes, on system's parameters $E$, $N$ etc 
(see \cite{lichetal2008} for a
review). 

In the sequel we examine the form of localization profiles
for $q$--torus solutions associated with an initial excitation
of a low--frequency packet of modes. We obtain theoretical
results based on a leading order term analysis of the PL
series for $q$--tori (see \ref{leading}). Furthermore, 
we compare these results with ones found numerically for 
FPU--trajectories with a similar initial excitation.
\subsubsection{FPU--$\alpha$ model}
\label{acaseqtor}

Our main result for the FPU--$\alpha$ can be stated as follows:
in Section \ref{sequence} it was mentioned that, for $q$--tori, an initial
excitation ${\cal D}_0=\{ 1,2,\ldots ,s\}$ in the PL series
leads to the sequence of mode excitations 
${\cal D}_k=\{ks+1,\ldots,(k+1)s\}$. Starting, now, 
from the median mode in the group ${\cal D}_k$, i.e. the
mode $q_{mid}=ks+[{s}/{2}]$, we can obtain estimates of the size of the
leading term $Q_{q_{mid}}^{(k)}$ given by Eq.(\ref{Qqkr})  
and derive estimates on the magnitude of the harmonic
energy $E_{q_{mid}}^{(k)}$, which is hereafter
denoted by $E^{(k)}$. Then, we find:
\begin{equation}\label{eksa}
E^{(k)} \simeq  \frac{(k+1/2)^2 \varepsilon}{M}
\left(\frac{\alpha^2 N^4 \varepsilon}{\pi^4 s^4} \right)^k
\end{equation}
where  $M=s/N$ is the fraction of initially excited modes with
respect to the total number of modes.
The derivation of  Eq. (\ref{eksa}) is given in \ref{AppE}.

\begin{figure}
\centering
\includegraphics[scale=0.20 ]{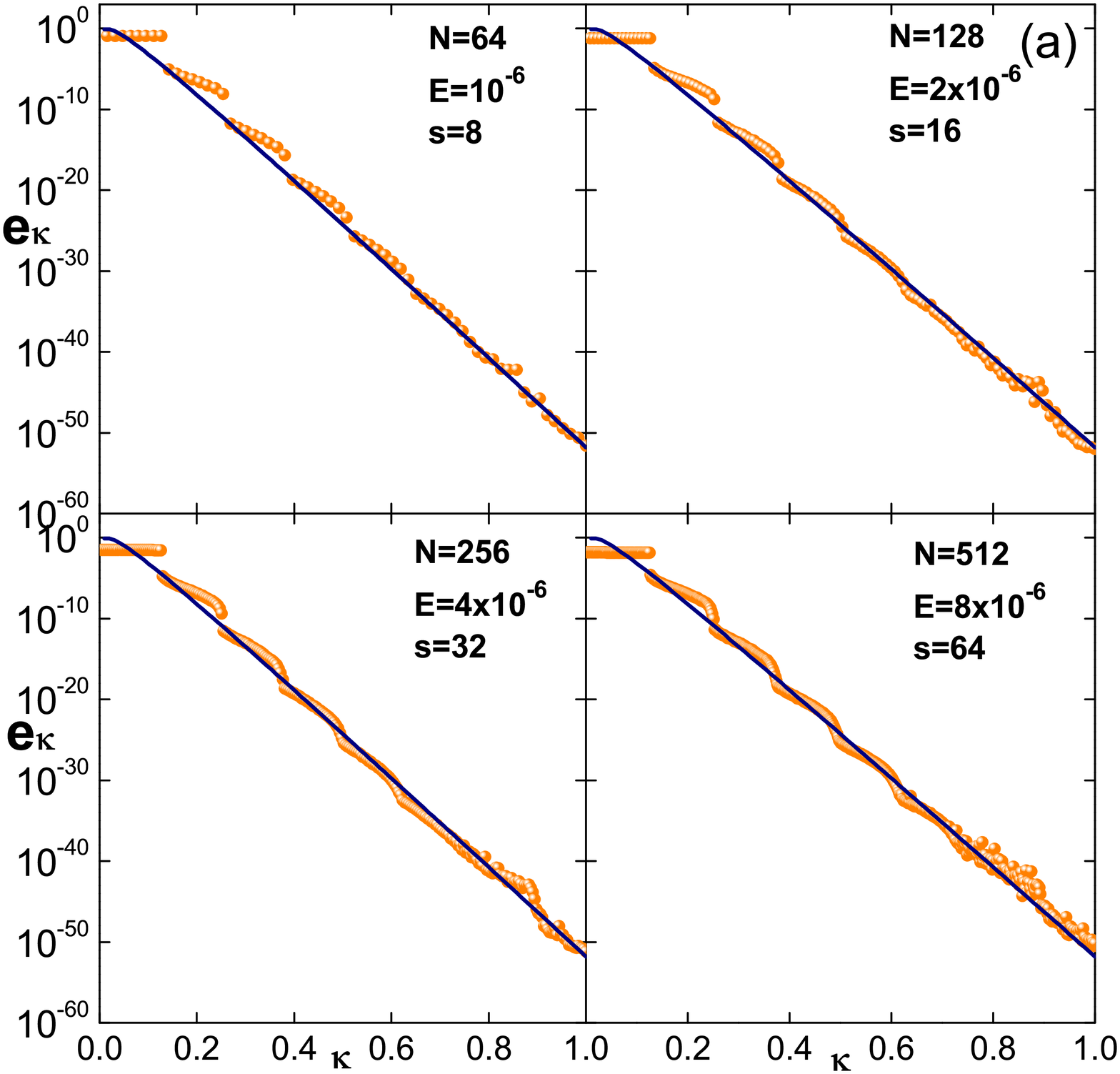}
\includegraphics[scale=0.20 ]{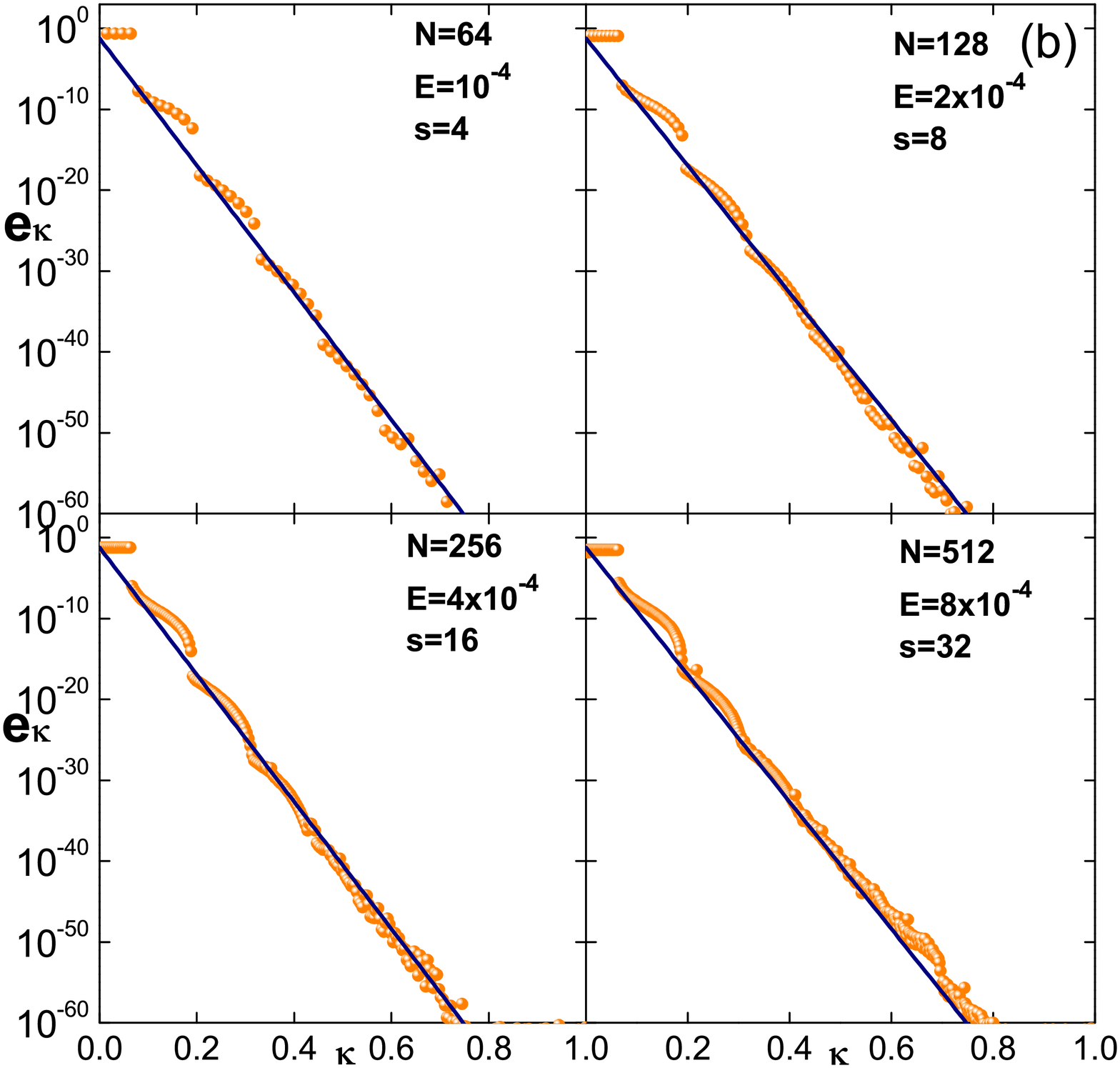}
\caption{\label{fig:atori} Normalized averaged energy spectra
$e_{\kappa}$ versus $\kappa$ for FPU--trajectories (orange spheres)
in both the FPU--$\alpha$ and FPU--$\beta$ models, keeping the specific
energy $\varepsilon$
and the ratio of initially excited modes $M$ fixed. The parameters
are: (a) FPU--$\alpha$ model with $\alpha=0.33$, $M=1/8$,
$\varepsilon=1.5625 \times 10^{-8}$. (b) FPU--$\beta$ model with
$\beta=0.3$, $M=1/16$ and $\varepsilon=1.5625 \times 10^{-6}$. In (a),
the blue line in all panels corresponds to the theoretical prediction
of Eq.(\ref{eksa2}) based on the leading order term analysis for
$q$--tori. This is given by $\log e_{\kappa }=-56.27\kappa +4.46+2\log
\kappa $. In (b) the blue line corresponds to a similar prediction for
the FPU--$\beta$, given by Eq.(\ref{eksp2}). We find $\log e_{\kappa }
=-78.64\kappa -1.204$.}
\end{figure}
Normalizing Eq.(\ref{eksa}) we obtain an equivalent expression 
for $e_{\kappa } = E^{(k)} / E$ 
\begin{equation}\label{eksa2}
\log e_{\kappa } \simeq   \frac{\log \lambda}{M} \kappa +2 \log \kappa
- \log ( \lambda^{1/2} M^3 N)
\end{equation}
where  $\kappa =q/N$ is  re--scaled wavenumber and
 $\lambda = \alpha^2 \varepsilon/\pi^4 M^4$. The main
prediction is that if $\alpha, \varepsilon$, and the fraction
$M=s/N$ are kept fixed, {\it the normalized energy
profiles of $q$--tori remain unaltered as $N$ increases.}

This prediction becomes hardly possible to test by a direct
construction of the $q$--tori solutions via PL series, because
as $N$ increases, we quite soon encounter the limits of computer
memory required for storing the coefficients produced by the
computer--algebraic program. However, taking into account the
evidence presented in subsection \ref{fputra}, that
FPU--trajectories with the same initial excitations as
$q$--tori exhibit similar localization profiles, we can
test numerically the extent up to which the invariance
of the averaged normalized energy spectrum holds, at least 
for FPU--trajectories.

Such a test is made in Fig.\ref{fig:atori}.
In (a) we give the normalized averaged energy spectrum $e_{\kappa }$
as a function of the re--scaled wavenumber $\kappa =q/N$ for an
FPU--trajectory of $\alpha=0.33$,
$M=1/8$, and $\varepsilon = 1.5625 \times 10^{-8}$. We
computed these trajectories by progressively increasing
$N$, namely $N=64$, $128$, $256$ and $512$. The energy spectra of 
Fig.\ref{fig:atori} are all evaluated at 
$T=10^6$. The solid line corresponds to the fitting law
of Eq. (\ref{eksa2}), which, for the adopted parameters,
takes the form indicated in the figure caption. The main
remark is that {\it the same} line fits all re--scaled
normalized spectra, for different $N$ (while the fraction
of initially excited modes $M=1/8$ as well as the specific
energy $\varepsilon = 1.5625 \times 10^{-8}$ are kept constant).

A relevant question of central interest regards the upper limit 
in the specific energy for which the normalized spectra for the
FPU--trajectories continue to exhibit exponential localization.
Eq.(\ref{eksa}) allows us to obtain an upper limit, by requiring
that $\lambda=\alpha^2\varepsilon/\pi^4 M^4<1$. However,
as we approach the upper limit
$\varepsilon=\pi^4M^4/\alpha^2$, the analysis based on
only the leading order terms of the PL series ceases to
be valid, since important contributions to the energy
spectrum are made also by the higher order terms in each
mode's series expansion of Eq.(\ref{qser}). 

The condition $\lambda<1$ implies $s>\alpha^{1/2}\varepsilon^{1/4}N$. 
Thus Eq.(\ref{eksa}) applies when the initially excited packet 
has a width larger than the width of the so--called natural packets 
\cite{benetal2011,beretal2004,beretal2005}. 
In the case of natural packets, it is found that by the excitation of a 
number of low frequency modes satisfying $s<\alpha^{1/2}\varepsilon^{1/4}N$,
the so resulting energy spectrum exhibits a plateau of width
$\alpha^{1/2}\varepsilon^{1/4}N$ (larger than the initially excited modes).
Furthermore, there is evidence that the slope $\tilde{\sigma}  $ of the 
exponential energy localization profile 
$e_{\kappa } \sim \exp(- \tilde{\sigma}  \cdot  \kappa )$ depends linearly on
$\alpha^{-1/2}\varepsilon^{-1/4}$ \cite{ponbam2005}. This is in contrast to
the slope  which refers to $q$--tori solutions of Eq.(\ref{eksa}), 
that depends logarithmically on 
$[\alpha^{-1/2}\varepsilon^{-1/4}]^{-4}$
and points out that {\it different choices in the fraction of the 
initially excited low--frequency modes result 
in different exponential laws}. 

\subsubsection{FPU--$\beta$ model}
\label{bcaseqtor}
For the sake of completeness we report the results of our previous 
work \cite{chrietal2010}, concerning
exponential energy localization in the FPU--$\beta$ model. Assuming
the initial excitation to be ${\cal D}_0=\{1,2,\ldots,s\}$, the
sequence of mode excitations in the PL series at the orders $k=1,2,
\ldots$ is ${\cal D}_k=\{(2k-1)s+1,\ldots,(2k+1)s\}$.
Estimating the size of the median mode $q_{mid,k} =2ks$ in each
group ${\cal D}_k$ via Eq.(\ref{Qqkr}), we are lead to an estimate
for the energy spectra of $q$--tori with the above excitation,
namely
\begin{equation}\label{eksp}
E^{(k)} \simeq  {\varepsilon \over M}
\left({ \beta^2\varepsilon^2\over \pi^4M^4}\right)^k~~ 
\end{equation}
where $M=s/N$.
Using re-scaled variables as in the $\alpha $ case, Eq.(\ref{eksp})
takes the form
\begin{equation}\label{eksp2}
\log e_{\kappa }\simeq\frac{\log\lambda}{M}\kappa -\log s
\end{equation}
where $\lambda={\beta\varepsilon/\pi^2M^2}$. We find a similar
result as in the $\alpha$ case, namely Eq.(\ref{eksp2}) implies
that by keeping both the specific energy $\varepsilon$ and the
fraction of excited modes $M$ fixed, while $N$ increases, the
normalized exponential profile remains invariant.
Again, in order that the analysis be valid, one must have
$s>N(\beta\varepsilon)^{1/2}$, implying that the initial
excitation should be in a regime quite different from that of natural
packets. In fact, in the present case as well, Eq.(\ref{eksp2}) 
describes correctly the localization profile provided that 
$\lambda<<1$ (see also \cite{chrietal2010}).

As a numerical test of the above predictions we use again
numerical computations based on FPU--trajectories rather than
exact $q$--tori solutions. The four panels of Fig.\ref{fig:atori}(b)
show the normalized averaged spectra $e_{\kappa }$ of FPU--trajectories,
along with the predictions of Eq.(\ref{eksp2}), for the fixed
values $M=s/N=1/16$, and $\varepsilon=1.5625 \cdot 10^{-6}$.
We observe again 
that the spectrum remains practically invariant with increasing $N$.

\subsection{Localization patterns for arbitrary initial excitations}
\label{patternqtor}

So far, we focused on $q$--tori, and their neighboring FPU--trajectories
corresponding to initial excitations in the low--frequency part of
the spectrum. However, it is possible to see that energy
localization appears also in cases where the initial excitation
has quite different features than in the case of packets of
low--frequency modes. In particular, we will examine now $q$--tori
solutions in the FPU--$\alpha$ system produced by an initial
excitation ${\cal D}_0$ consisting of a small set of $s$ modes
{\it arbitrarily distributed} in $q$--space. We give several
such examples, in which we vary $s$, $N$, $E$, as well as
${\cal D}_0$. As in the example of Fig.\ref{fig:qexample}(a),
in all present cases we compare the averaged normalized energy
spectra $e_{\kappa }^{PL}$ obtained with the PL series, with
the ones $e_{\kappa}^{PLn}$ obtained by numerical integration
of the equations of motion for the initial conditions
$Q_q ^{PL}(0)$, $P_q^{PL}(0)$, $q=1,\ldots,N-1$. \\

\begin{figure}
\centering
\includegraphics[scale=0.20 ]{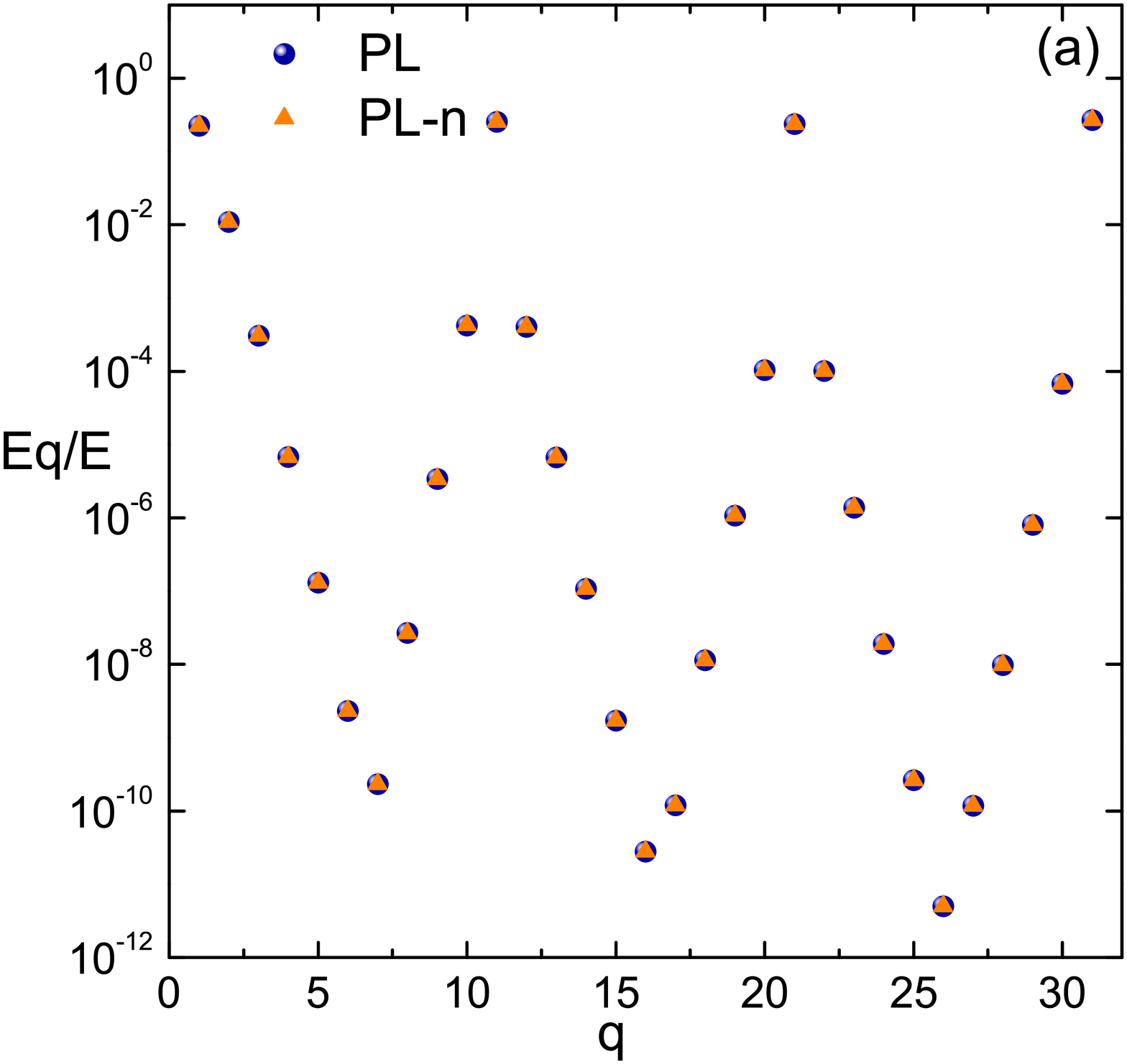}
\includegraphics[scale=0.20 ]{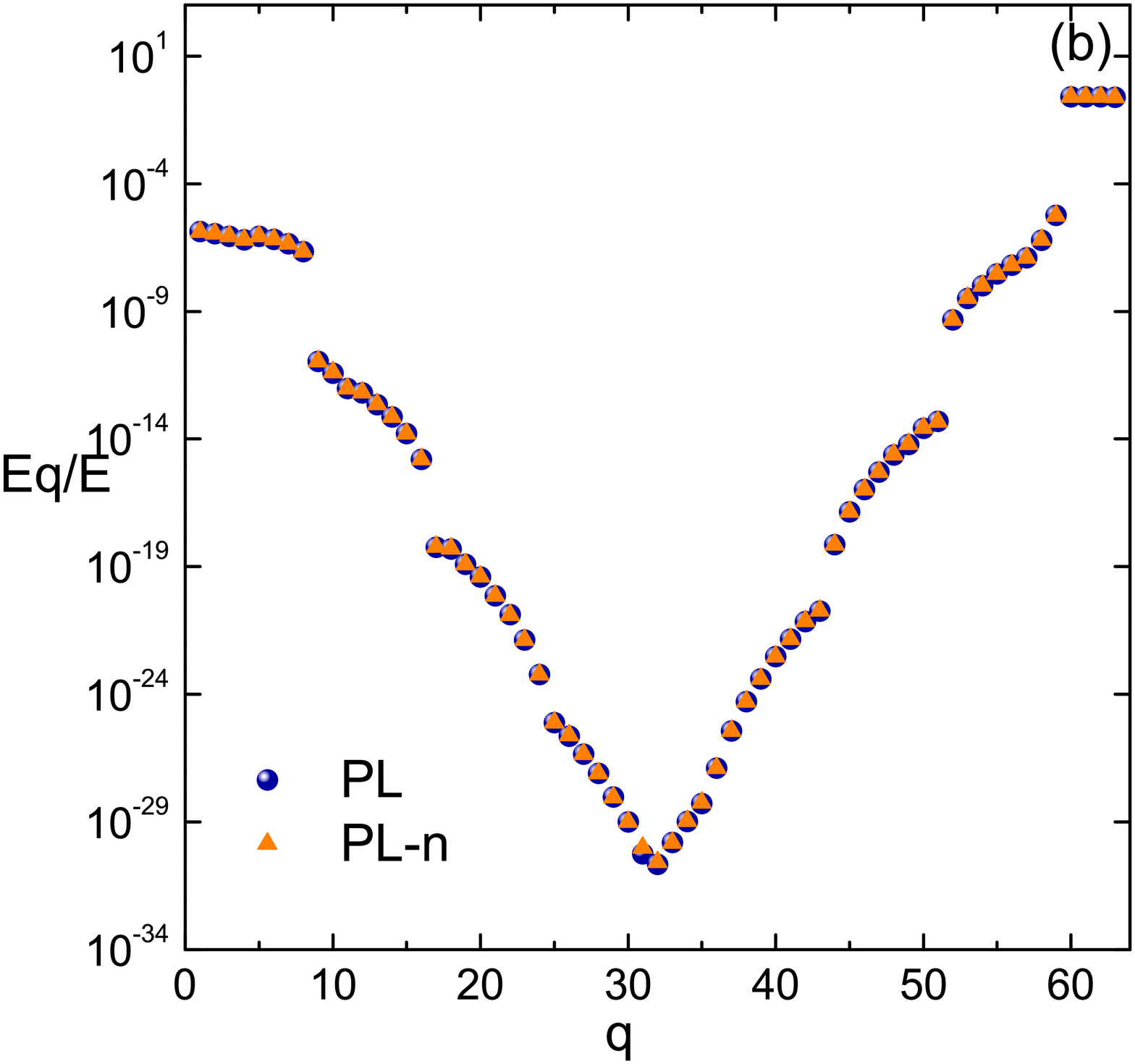} \\
\includegraphics[scale=0.20 ]{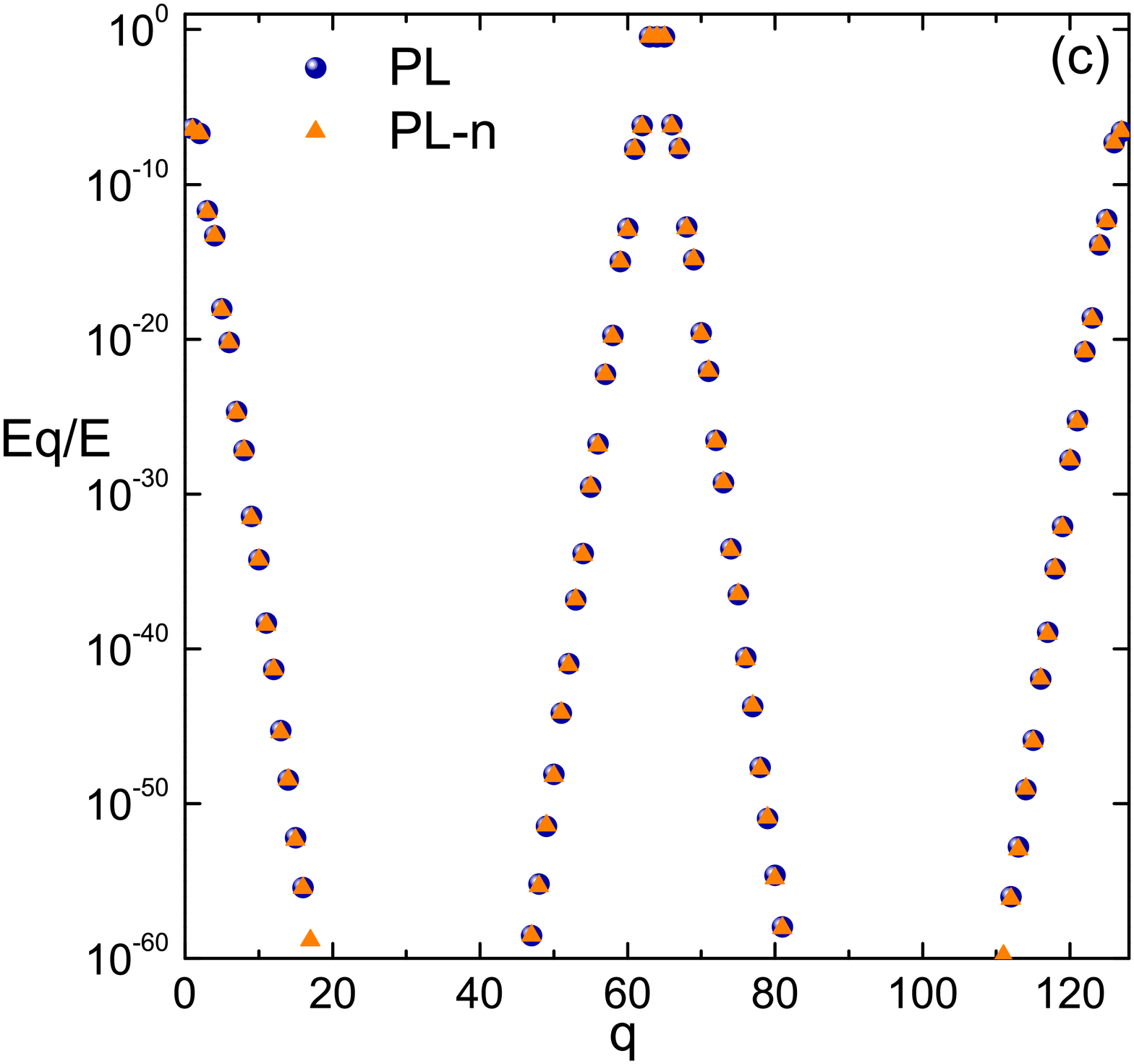}
\includegraphics[scale=0.20 ]{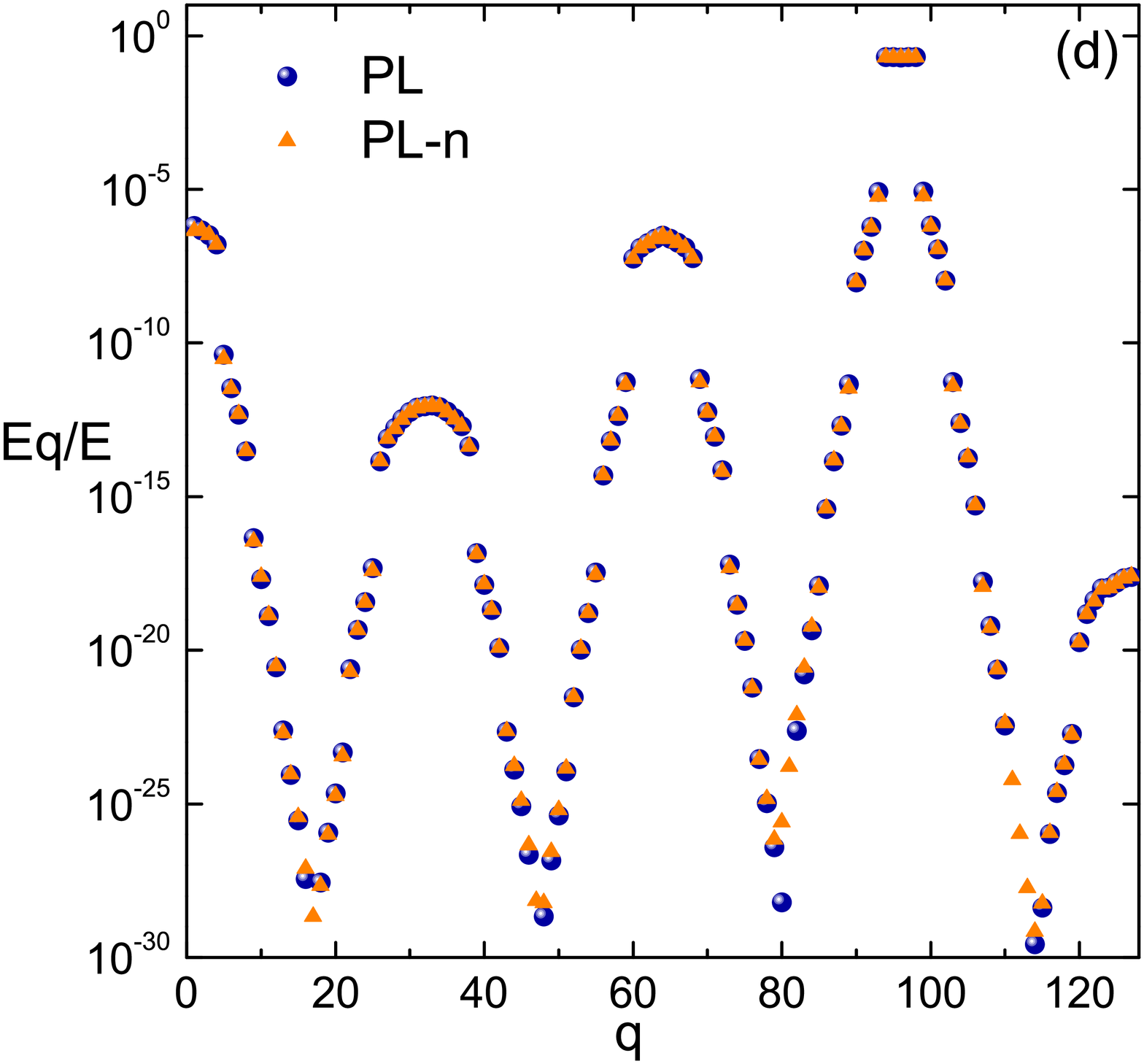}
\caption{Normalized averaged energy spectra $E_q$ versus $q$, deduced
by the PL series $q$--torus solution (blue spheres), or a numerical
integration (PLn) with initial conditions on a $q$--torus (orange triangles,
see text). (a) $N=32$, $\alpha =0.33$, ${\cal D}_0 = \{1,11,21,31\}$, 
(b) $N=64$, $\alpha=1$ with ${\cal D}_0 = \{60,61,62,63\}$. (c) $N=128$,
$\alpha=1$ with ${\cal D}_0 = \{63,64,65 \}$. (d) $N=128$, $\alpha=1$
and ${\cal D}_0 = \{94,95,96,97,98\}$. The chosen frequency values,
and resulting amplitudes and energies in each case are given in
\ref{AppF}. \label{fig:nnn}  }
\end{figure}

{\it Example 1: evenly distributed initial excitation.}
In the FPU--$\alpha$ with $N=32$, we construct a
$q$--torus PL series starting from the 0--th order excitation
${\cal D}_0 = \{1,11,21,31\}$, when the frequency values for 
$\omega_1 $, $\omega_{11} $, $\omega_{21}
$, $\omega_{31} $ are chosen
as in the second group of \ref{AppF}. The  truncation order here is
$k_0=11$. Solving numerically Eqs.(\ref{omeser}) we specify
the values of the amplitudes $A_1$, $A_{11}
$, $A_{21} $ and $A_{31}$.
The total energy is $E=0.001563$.

Fig.\ref{fig:nnn}(a) shows the averaged normalized energy
spectrum for the above $q$--torus solution. Energy localization
is manifestly present, since we observe the formation of four
peaks of the energy spectrum around the seed modes 1,11,21 and
31. The localization pattern is readily understood by computing
the sequence of mode excitations ${\cal D}_k$ deduced by the
proposition of Section \ref{sequence}. Namely, we find
${\cal D}_1 =\{2,10,12,20,22,30\}$,
${\cal D}_2 = \{3,9,13,19,23,29 \}$,
${\cal D}_3 = \{ 4,8,14,18,24,28 \}$,
${\cal D}_4 = \{5,7,15,17,25,27 \}$ etc.
We observe that consecutive modes, adjacent (on either side)
to the initially excited ones, are excited at subsequent 
orders of perturbation theory. Thus, starting for example
from the mode $q=11$, the modes $q=10,12$ are excited at
first order, $q=9,13$ at second order, etc. This explains
the formation of the peaks in the spectrum. In fact, the
pairs $\{10,12\}$, $\{9,13 \}$, etc. share quite similar energies,
corresponding to excitation amplitudes $O(\mu)$, $O(\mu^2)$,
$\ldots$ As a result, the local form of the energy spectrum
on either side of one peak is exponential.

Finally, as evident in Fig.\ref{fig:nnn}(a), we find a very
precise agreement between the normalized spectrum corresponding
to the analytical solution $Q_{q }^{PL}(t)$, and the
one $Q_{q}^{PLn}(t)$ obtained by numerical integration of
the initial conditions on the $q$--torus. This fact indicates
that at the truncation order $k_0=11$ the solution has converged
to a good accuracy.\\

{\it Example 2: initial excitation in the high--frequency
part of the spectrum.}  We consider a $q$--torus solution
found by PL series in the case $N=64$, $\alpha=1$, when
${\cal D}_0 = \{60,61,62,63\}$, while the choice in the
frequencies and the resulting amplitudes are
shown in the third group of \ref{AppF}. The truncation
order of the series is $k_0=8$ and the energy is $E = 0.000883$. 

Fig.\ref{fig:nnn}(b) shows the resulting averaged normalized
energy spectrum for the above $q$--torus solution. This displays
several features similar to the case of a low--frequency excitation.
Namely, we observe the formation of groups of consecutive modes sharing a
similar amount of energy. The sequence
of mode excitations in this case 
turns out to be ${\cal D}_1=\{1,\ldots, 8\}$, 
${\cal D}_2=\{52,\ldots, 59\}$, ${\cal D}_3=\{9,\ldots, 16\}$,
${\cal D}_4=\{44,\ldots, 51\}$, etc. 

Finally, we note again the exponential fall of the energy along
two separate branches of the spectrum, namely a low--frequency
and a high--frequency branch. \\

{\it Example 3: excitation in the middle part of the spectrum.}
We consider the middle modes initial excitation ${\cal D}_0 =
\{63,64,65 \}$ in the FPU--$\alpha$ with $\alpha=1$, $N=128$,
and with frequencies and amplitudes displayed
 in the fourth group of \ref{AppF}. The PL series are truncated
at $k_0=16$ and the energy of the system 
is $E=7.63469\times 10^{-5}$.

In the averaged normalized energy spectrum $E_{q }^{PL}/E$
(Fig.\ref{fig:nnn}(c)) we observe that three energy peaks are formed:
by the lowest, the highest and the middle modes.
The spectrum of our numerical solution $E_{q }^{PLn}/E$
follows $E_{q }^{PL}/E$ until values of the order $10^{-60}$.
We find that the sequence of mode excitations here is
${\cal D}_1 = \{ 1,2, 126,127 \}$, ${\cal D}_2 = \{ 61,62,66,67\}$,
${\cal D}_3 = \{ 3,4,124,125 \}$, ${\cal D}_4 = \{ 59, 60, 68,69 \}$, etc.\\

{\it Example 4: excitation in the 3/4 part of the spectrum.}
We consider, as before, $N=128$, $\alpha =1$, and an initial
excitation ${\cal D}_0=\{94,95,96,$ $97,98\}$. The chosen
frequency values and the so--resulting amplitudes are shown 
in the fifth group of \ref{AppF}.
The truncation order is $k_0=9$ and the energy is $E=0.000649478$.

At subsequent orders, we now find the sequence of mode excitations
${\cal D}_1=\{1,2,3,4 \}\cup\{60,\ldots,68\}$, ${\cal D}_2=\{ 26,\ldots
,38\}\cup\{90,\ldots ,93\}\cup\{99,\ldots,102\}$,
${\cal D}_3=\{5,\ldots,8 \}\cup\{56,\ldots ,72\}\cup
\{120,\ldots,127\}$, ${\cal D}_4=\{22,\ldots ,25 \} \cup
\{39,\ldots,42\}$ $\cup\{86,\ldots ,89\}\cup [103,\ldots ,106
\}$, etc. This leads to the localization pattern shown in
Fig.\ref{fig:nnn}(d). \\

\begin{figure}
\centering
\includegraphics[scale=0.15 ]{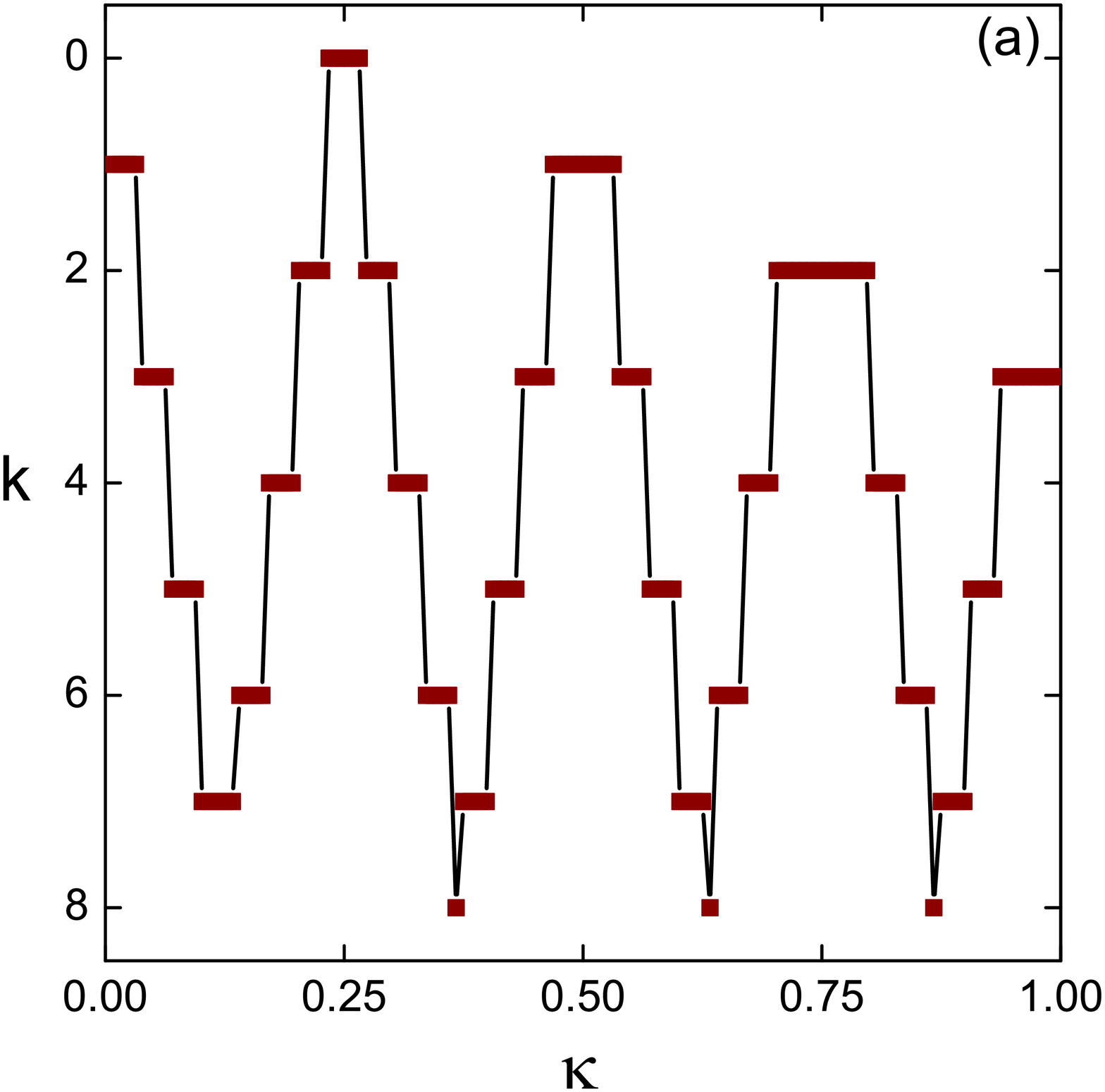}
\includegraphics[scale=0.15 ]{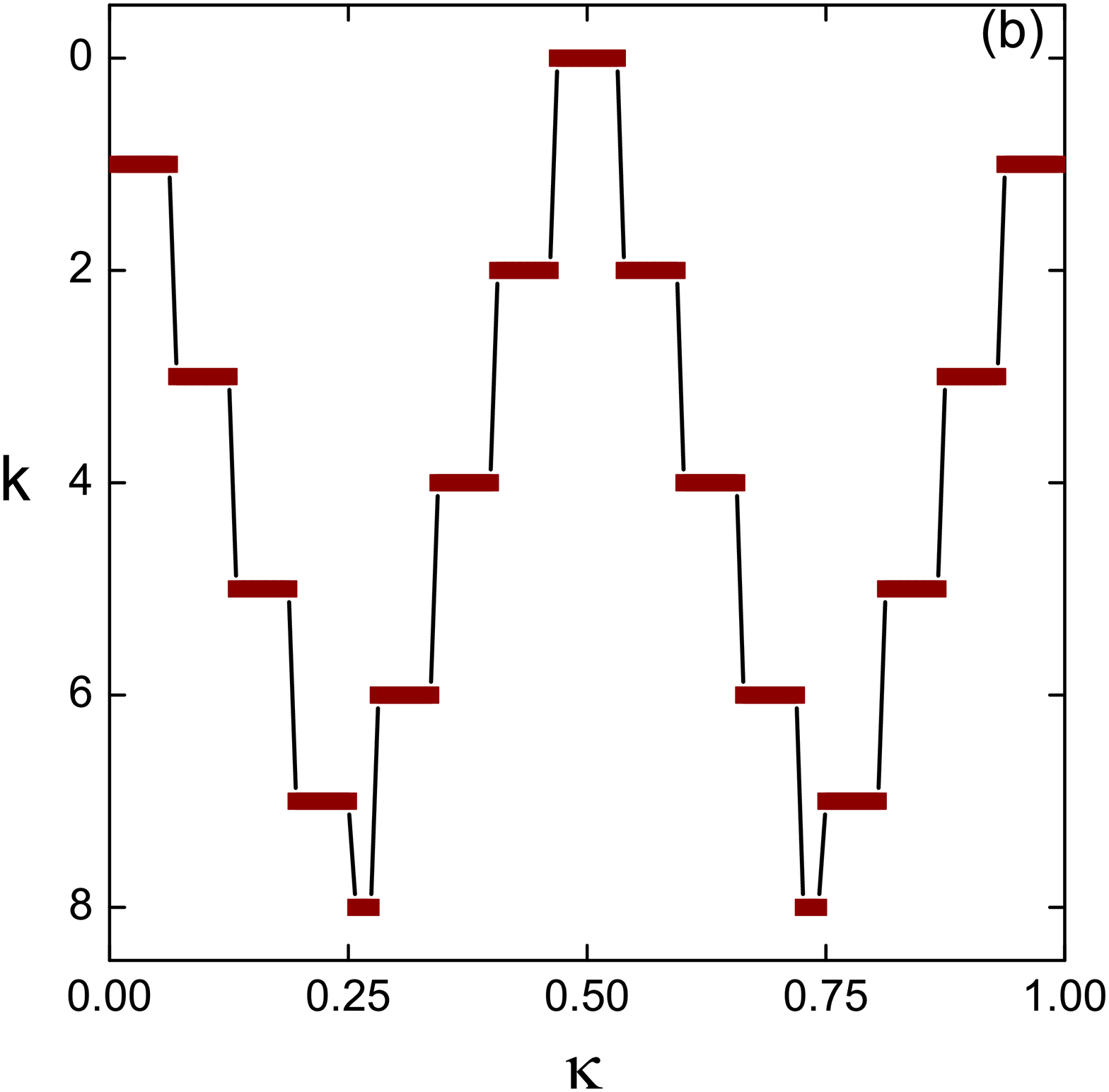}
\includegraphics[scale=0.15 ]{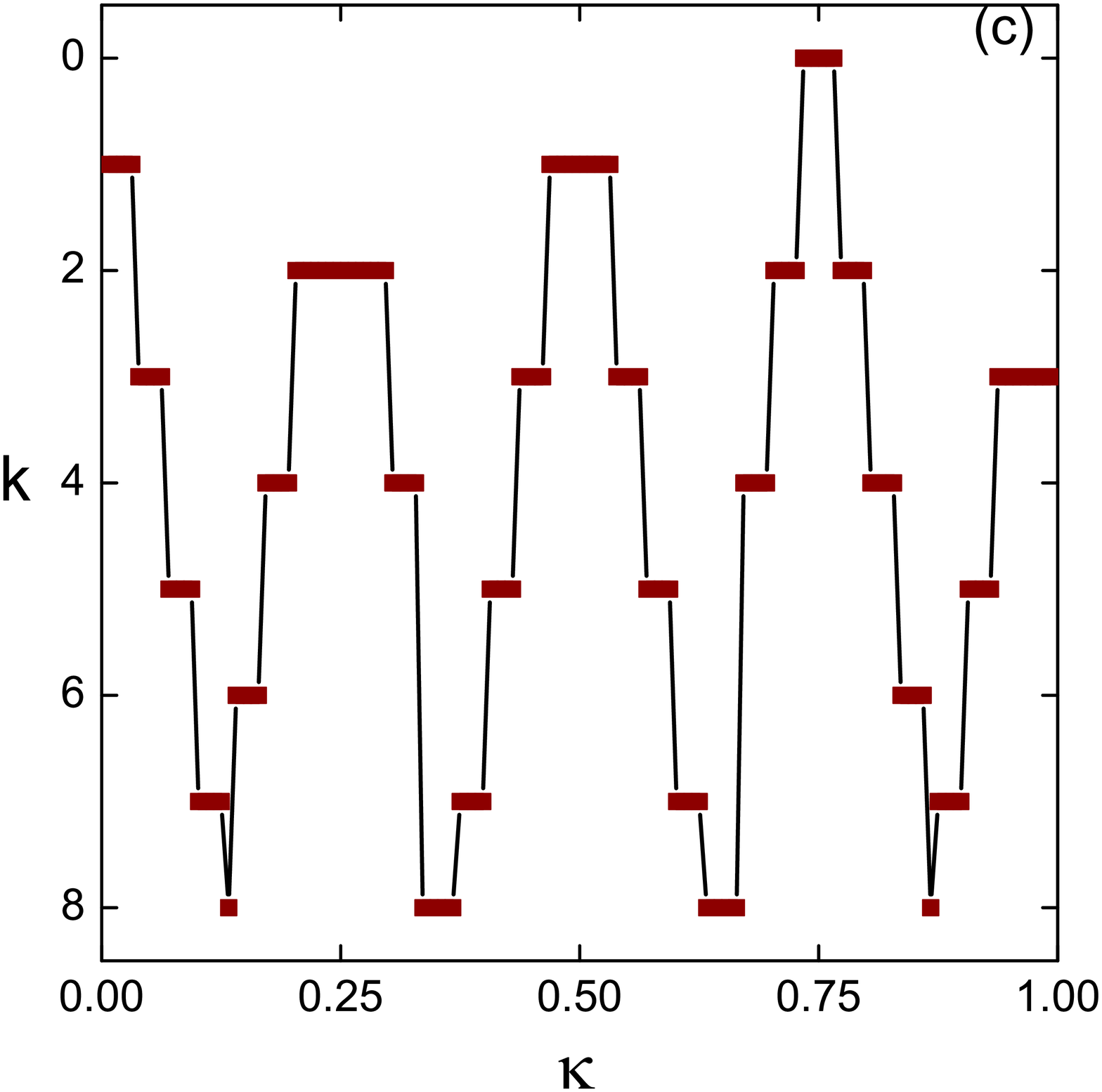}
\caption{Propagation of modes for an initial excitation ${\cal D}_0 =
[\kappa _0 - \epsilon , \kappa _0 + \epsilon]$ when (a) $\kappa _0 =1/4$,
(b) $\kappa _0=1/2$ and (c) $\kappa _0 = 3/4$. In all panels, the ordinate
yields the order $k$ of perturbation theory at which the (re--scaled) mode $\kappa =q/N$ 
 is first excited. Since in the exponential energy
localization regime we always have $\log e_\kappa \sim \kappa $, the patterns
shown in the present panels are similar to the energy spectra
for the same excitations plotted in semi--logarithmic scale.
\label{fig:promod}  }
\end{figure}

\subsection{Patterns from generalized packet excitation}
\label{general}

We now generalize results on the localization patterns formed by 
initial excitations of packets of arbitrary width, around
the locations in $q$--space corresponding the one fourth,
half and three thirds of the spectrum. We suppose
that each packet is of the form ${\cal D}_0 = [\kappa_0 -
\epsilon , \kappa_0 + \epsilon]$, having a width equal to
$s/N=2\epsilon $, $\epsilon<<1$ in the normalized
$q$--space  $\kappa =q/N\in[0,1]$.  In order to  specify the sequence of
mode excitations  ${\cal D}_k$ at subsequent orders, we first specify
the sets $M_k$ defined in Eq.(\ref{app}), whereby  the sets ${\cal D}_k$
are immediately derived by the relation ${\cal D}_k =M_k \setminus
\cup_{0 \leq j \leq k-1}M_j$. Examining in detail the case of excitations
around the re--scaled wavenumbers $\kappa_0=1/4$, $1/2$, or
$3/4$ we have:

1) For $\kappa_0 = 1/4$ we find
$M_k = [0, \epsilon (k+1)] \cup [1/2 - \epsilon (k+1),
1/2 + \epsilon (k+1) ] \cup [1-\epsilon (k+1),1] $, if $k=2n+1$ and $k>1$,
or $M_k =  [1/4 - \epsilon (k+1), 1/4 + \epsilon (k+1)]
\cup [3/4 - \epsilon (k+1), 3/4 + \epsilon (k+1) ]$, if $k=2n$.
Only the case of $k=1$ differs, for which it turns out that $M_1=
[0, 2\epsilon] \cup [1/2 - 2\epsilon,  1/2 + 2 \epsilon ]$, i.e.
the last modes are not yet excited.
The resulting localization pattern displays three peaks around
$\kappa=0$, $1/2$, and $1$, produced at odd orders, and two
peaks around $1/4$ and $3/4$, produced at even orders. The total
pattern is shown in Fig.\ref{fig:promod}(a). In this figure, the
ordinate in all panels indicates the order $k$ of the PL
series at which the corresponding mode, of wavenumber $\kappa=q/N$, is
first excited. In fact, according to the leading order
term analysis of the PL series discussed above we have
$\log e_\kappa\sim\kappa$. Thus, the patterns shown in all
panels of Fig.\ref{fig:promod} are similar to the averaged
normalized energy spectra for the corresponding excitations
when plotted in semi--logarithmic scale.

2) For $\kappa_0=1/2$ we find
$M_k=[0, \epsilon (k+1)] \cup [1-\epsilon
(k+1),1]$,  if $k=2n+1$, or
$M_k=[1/2-\epsilon (k+1),1/2+\epsilon (k+1)]$,
if $k=2n$.
Thus, we have two localization peaks around $\kappa=0$ and
$\kappa=1$ created by contributions at odd orders of the
PL series, and one more peak around $\kappa=1/2$ for
contributions at even orders. The overall localization
pattern is shown in Fig.\ref{fig:promod}(b).

3) For $\kappa_0=3/4$ we find similar results as in case (1),
i.e.
$M_1$ is $[0, 2\epsilon] \cup [1/2 - 2\epsilon,  1/2 + 2 \epsilon ]$
and then
$M_k = [0, \epsilon (k+1)] \cup [1/2 - \epsilon (k+1),
1/2 + \epsilon (k+1) ] \cup [1-\epsilon (k+1),1] $,  if
$k=2n+1$ and $k>1$, and
$M_k =  [1/4 - \epsilon (k+1), 1/4 + \epsilon (k+1)]
\cup [3/4 - \epsilon (k+1), 3/4 + \epsilon (k+1) ] $,  if $k=2n$.
In fact, the only difference with respect to case (1) concerns
the initial excitation at the order $k=0$.
The resulting localization pattern is shown in
Fig.\ref{fig:promod}(c).

\subsection{$q$--breathers and FPU--trajectories}
\label{trajbreath}

The existence, stability, and energy localization properties
of $q$--breathers were studied extensively in
\cite{flaetal2005}--\cite{flachtiz}, \cite{flaetal2006b}, 
\cite{kanetal2007}. It was found that $q$--breathers 
have quite similar energy localization profiles as their nearby 
FPU--trajectories. Furthermore, the $q$--breathers are periodic 
orbits whose existence, for an arbitrarily high energy,  
is guaranteed by the Lyapunov's theorem. Therefore, their 
existence extends well beyond the domain of convergence of their 
associated PL series. 

However, the PL series can still be quite useful in studying 
analytically some properties of $q$--breathers. In the sequel we examine 
$q$--breathers as a particular case of one--dimensional $q$--tori. 
A computational advantage is that, at any fixed order $k$, the number 
of terms in the resulting series is substantially smaller for $q$--breathers
than for $q$--tori of any other dimension $s>1$. This fact allows us to 
construct the series up to a very high order (in the case $N=32$ we were 
able to compute examples of PL series for $q$--breathers up to the 
truncation order $k_0=250$). Even so, the convergence of the resulting 
series is very slow, and in practice we obtain little gain in precision
after a truncation order near $k_0=50$. In most of our trial examples, 
the precision achieved for the computation of initial conditions on a 
$q$--breather using PL series is 4 to 6 significant digits, while in 
some cases we reach 10 significant digits. However, using 
these numbers as {\it initial guess}, we are able to determine many 
more via a root--finding technique. Let us note that a root-finding 
determination is possible only in the case of $q$--breathers, which 
are periodic orbits, while we cannot use such technique in the case 
of $q$--tori of dimension higher than one. 

\subsubsection{The breather $q=1$}
\label{q1breather}

\begin{figure}
\centering
\includegraphics[scale=0.20 ]{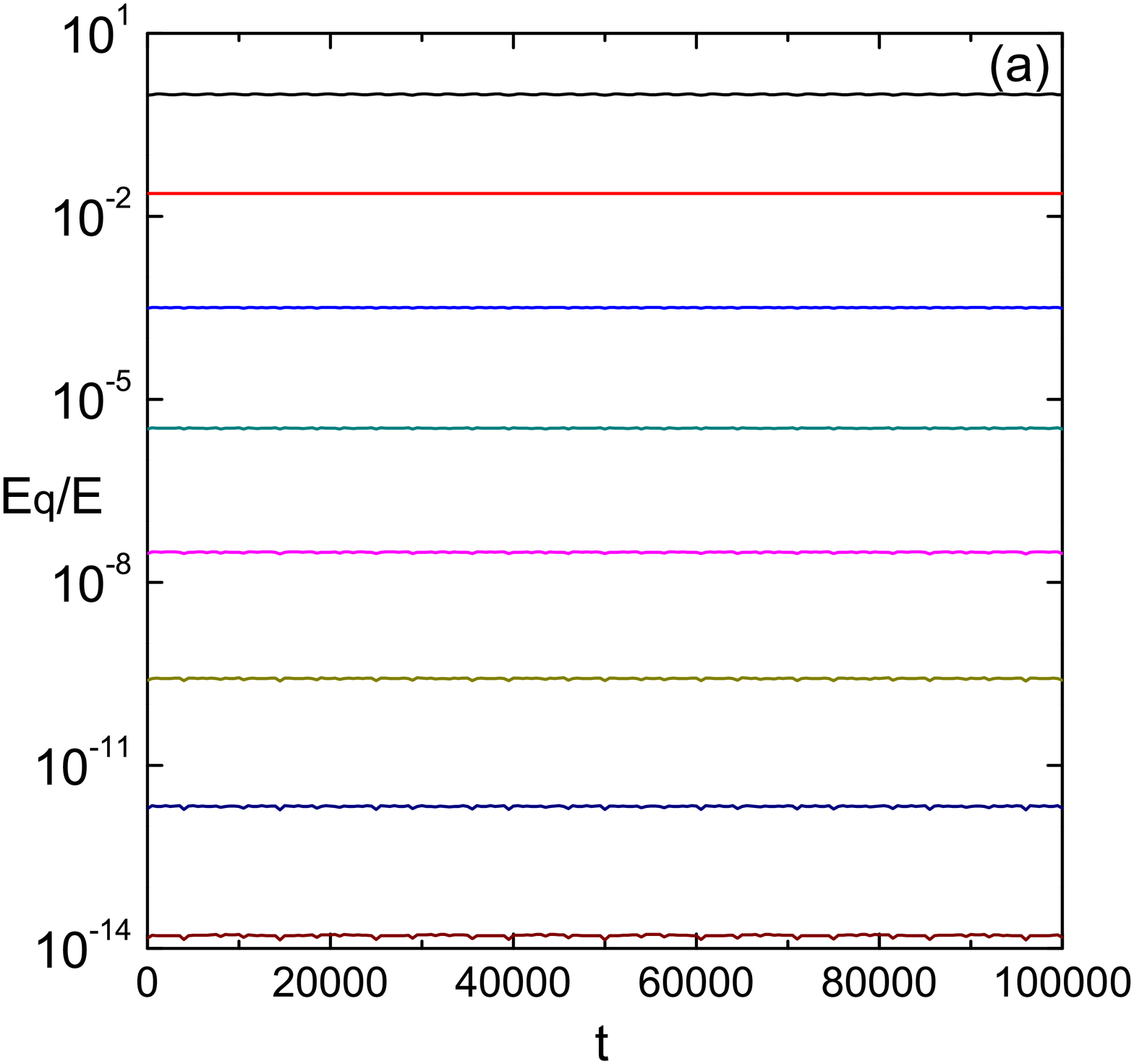}
\includegraphics[scale=0.20 ]{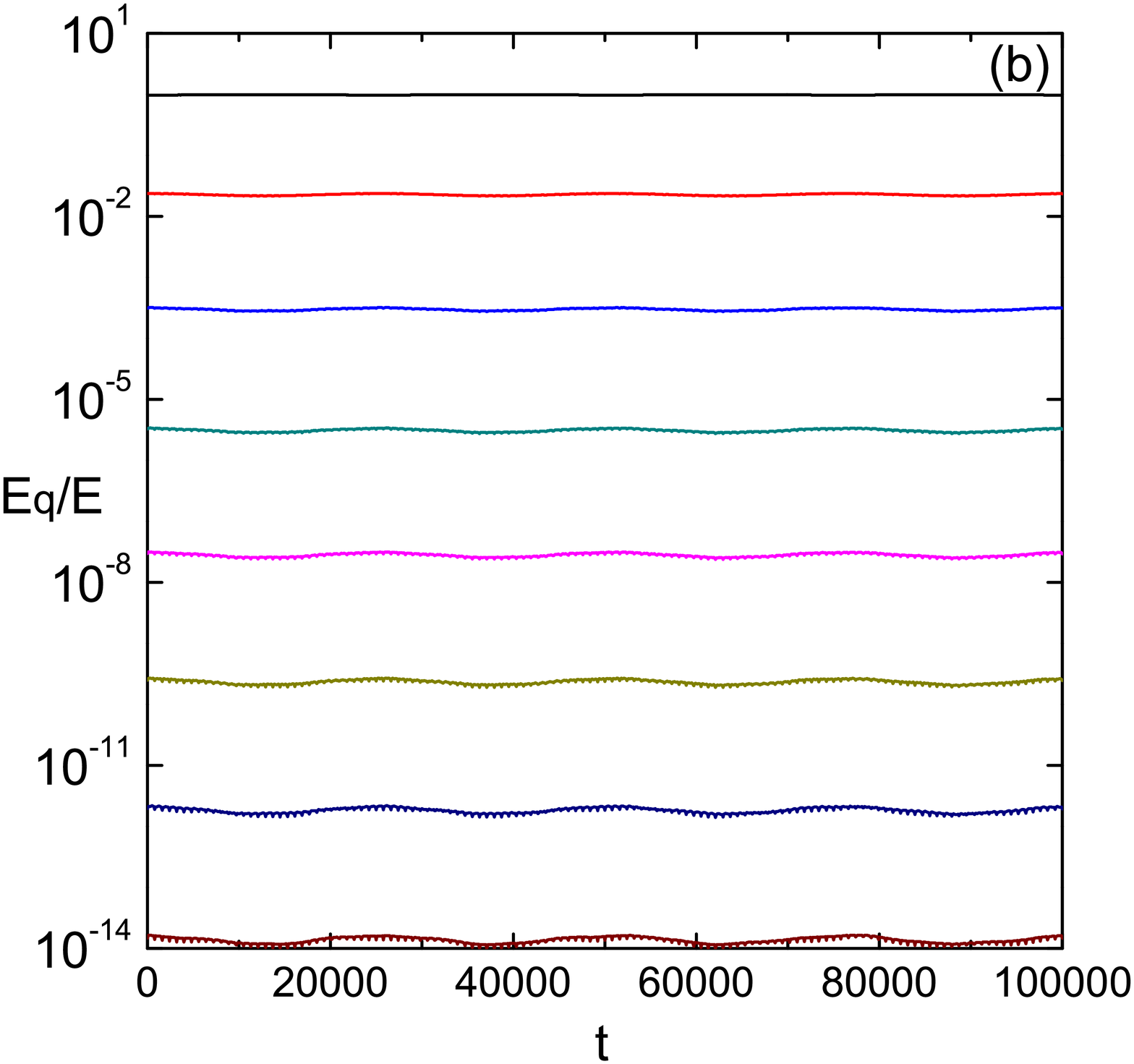}\\
\includegraphics[scale=0.20 ]{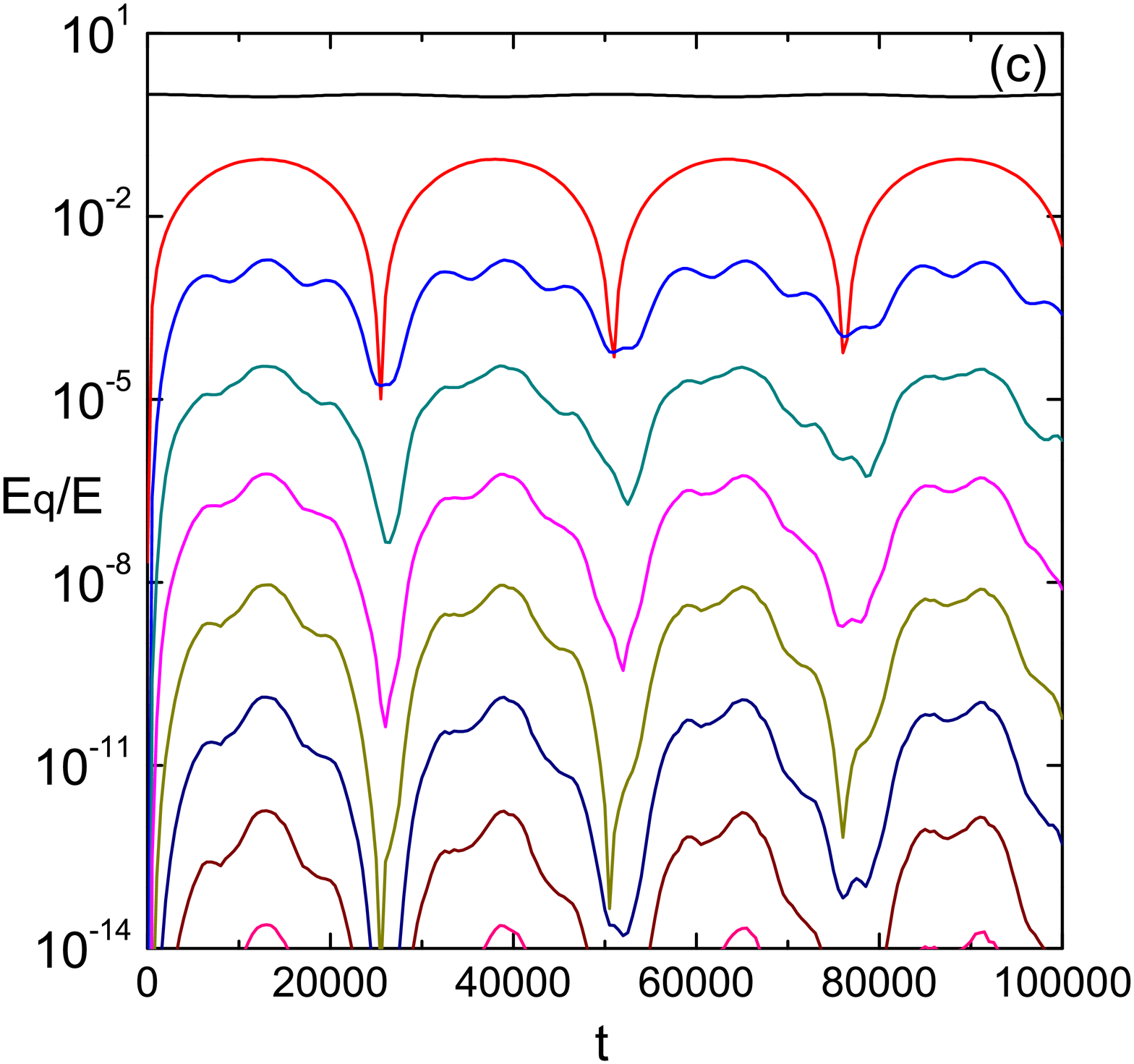}
\includegraphics[scale=0.20 ]{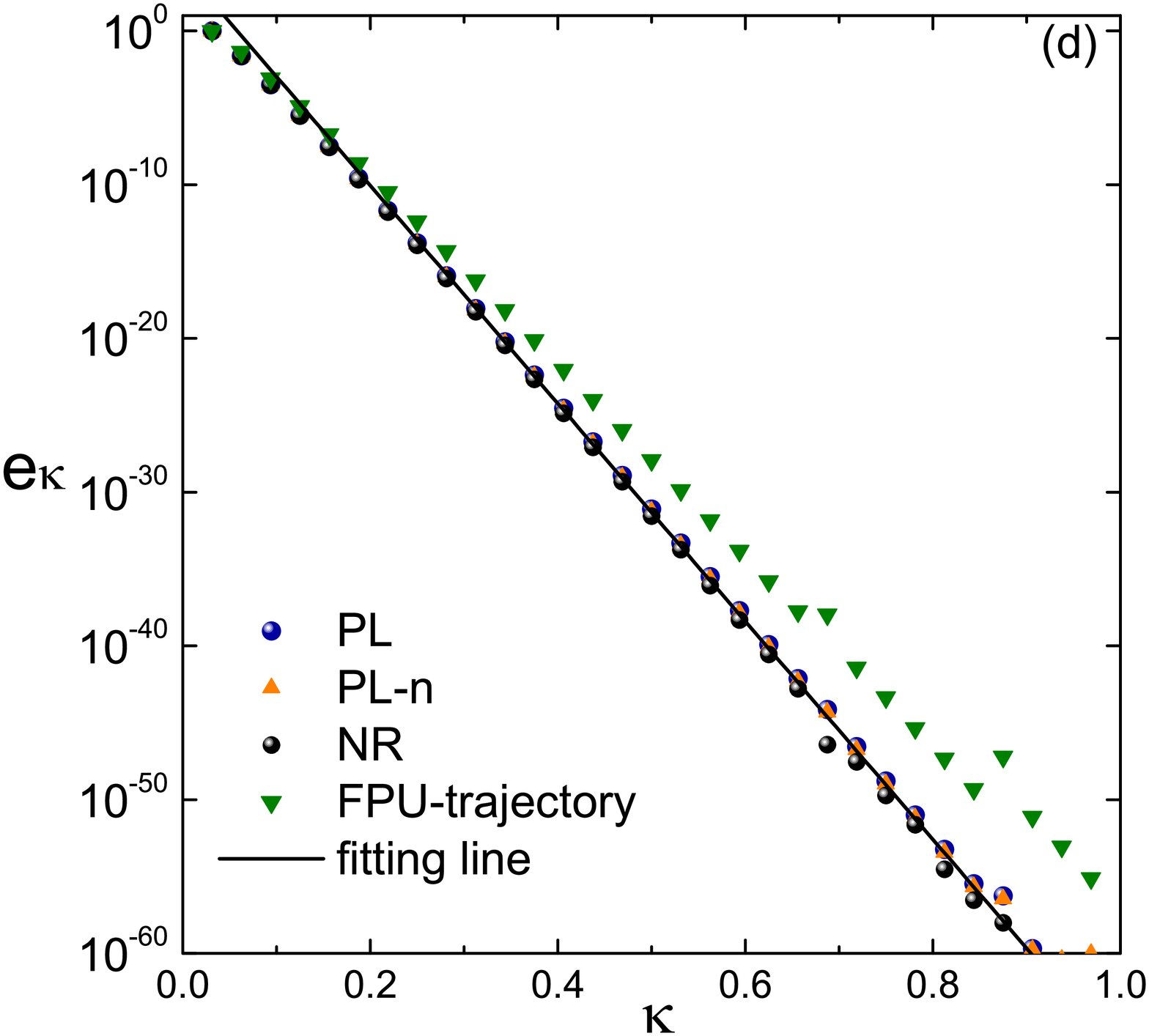}
\caption{  FPU--$\alpha $ system with $\alpha =0.33$, $N=32$ and
$E=0.00016635$. In Panels (a), (b) and (c) is shown the evolution
of the normalized instantaneous spectra $E_q(t)/E$ of modes
$q=1,\ldots ,8$ for the PL solution, the numerical integration of
FPU--dynamics for the same initial conditions with (a) and for the
FPU--trajectory with seed mode $q =1$, respectively (the energy $E_q$
has the highest value for $q =1$ and it progressively decreases for
$q=2,3,\dots$). In panel (d) is the exponential
profile of the normalized averaged energy spectra
$e_{\kappa }$ versus $\kappa $ of panel (a) with blue spheres,
of (b) with orange triangles, of
(c) with green triangles, as well as the $q$--breather found by
Newton--Raphson (NR) with black spheres. The continuous line is
the fitting law (\ref{brelaw}). \label{fig:qbreath2}  }
\end{figure}

As a first example, we consider the construction of a $q$--breather  
with `seed mode' $q_0=1$. The sequence of mode excitations here is $q_k=k+1$. 
Fig.\ref{fig:qbreath2} refers to a calculation for $N=32$,
$\alpha =0.33$ and $E=0.00016635$, for the truncation order
 $k_0=51$. The choice of $\omega_1$, as well as the 
resulting amplitude $A_1$, are given in \ref{AppF}. 
As for the overall precision in the analytic determination of
the periodic orbit using PL series at the truncation order 
$k_0=51$, we find that this trajectory returns to its initial
conditions after the time $T=2\pi/\omega_1$, with $\omega_1$
specified by the PL method, up to 10 significant digits, or 
more using the values $Q_q^{PL,51}(0),P_q^{PL,51}(0)$ as initial 
guess values for a numerical (Newton--Raphson) determination of the 
periodic orbit.

Fig.\ref{fig:qbreath2}(a),(b) and (c) show the evolution
of the normalized harmonic energies $E_q(t)/E$, for $q=1,\ldots,8$
in three different computations. Namely, in (a) we compute
$E_q (t)$ by the analytical solution $Q_q^{PL,51}(t)$ as found by the
truncated PL series. In (b), we integrate numerically the initial conditions
$Q_q^{PL,51}(0)$, $P_q^{PL,51}(0)$. Finally, in (c) we consider a FPU--trajectory
rising by the simple initial condition (corresponding to $q_0=1$): 
$x_n(0)=A_1\sin(\pi n/N)$, $y_n(0)=0$, $n=1,\ldots,N-1$.

The main remark, by a direct comparison of the three panels, is an
important difference in the temporal behavior of the $q$--breather
(in Fig.\ref{fig:qbreath2} (a) and (b)) from that of the corresponding
 FPU--trajectory  (Fig.\ref{fig:qbreath2} (c)). Namely, the
energies $E_q (t)$, $q=1,\ldots,8$ remain practically constant
in the case of the $q$--breather, while they behave as
quasi--periodic functions in the case of the FPU--trajectory.
In fact, the energies $E_q(t)$ for the latter oscillate around
mean values following closely the energy values of the
$q$--breather solution, but with an amplitude causing variations
of  more than one orders of magnitude.

Fig.\ref{fig:qbreath2}(d) shows a comparison of the averaged
normalized energy spectra in all four computations, namely
(i) $Q_q^{PL,51}(t)$, (ii) $Q_q^{PLn,51}(t)$, (iii) the
FPU--trajectory, and (iv) the periodic orbit with initial
conditions as determined by the Newton--Raphson. 
The solid line in Fig.\ref{fig:qbreath2}(d)
corresponds to the exponential law 
\begin{equation}\label{brelaw}
 E_q = \gamma ^{ q-1} q^2 E_1,~~ where ~~\gamma =
 \alpha^2 N^4 \varepsilon / \pi ^4 ,
\end{equation}
suggested as a fitting law in \cite{flaetal2006}. 

\subsubsection{The breather $q=25$}
\label{q25breather}

\begin{figure}
\centering
\includegraphics[scale=0.20 ]{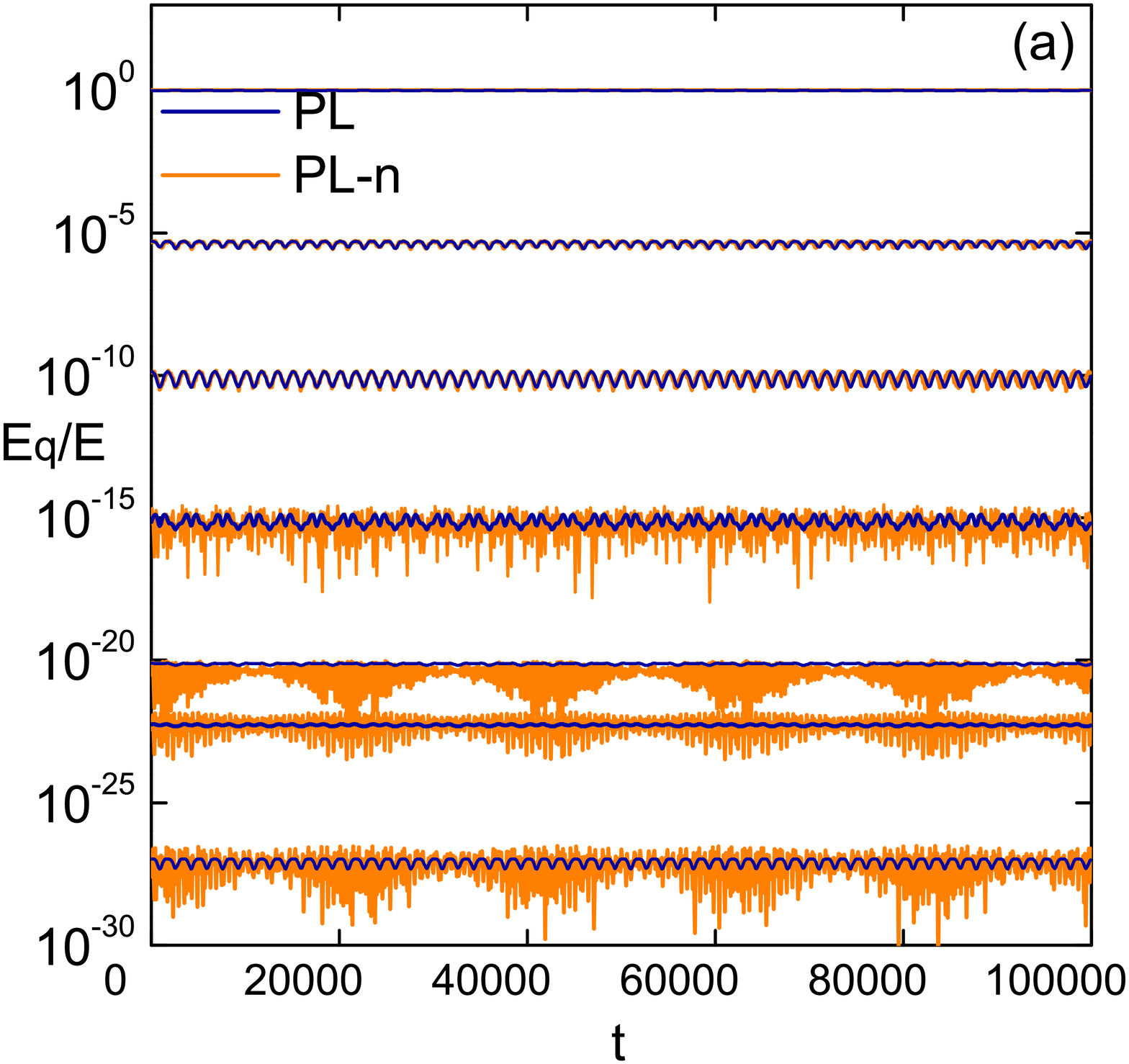}
\includegraphics[scale=0.20 ]{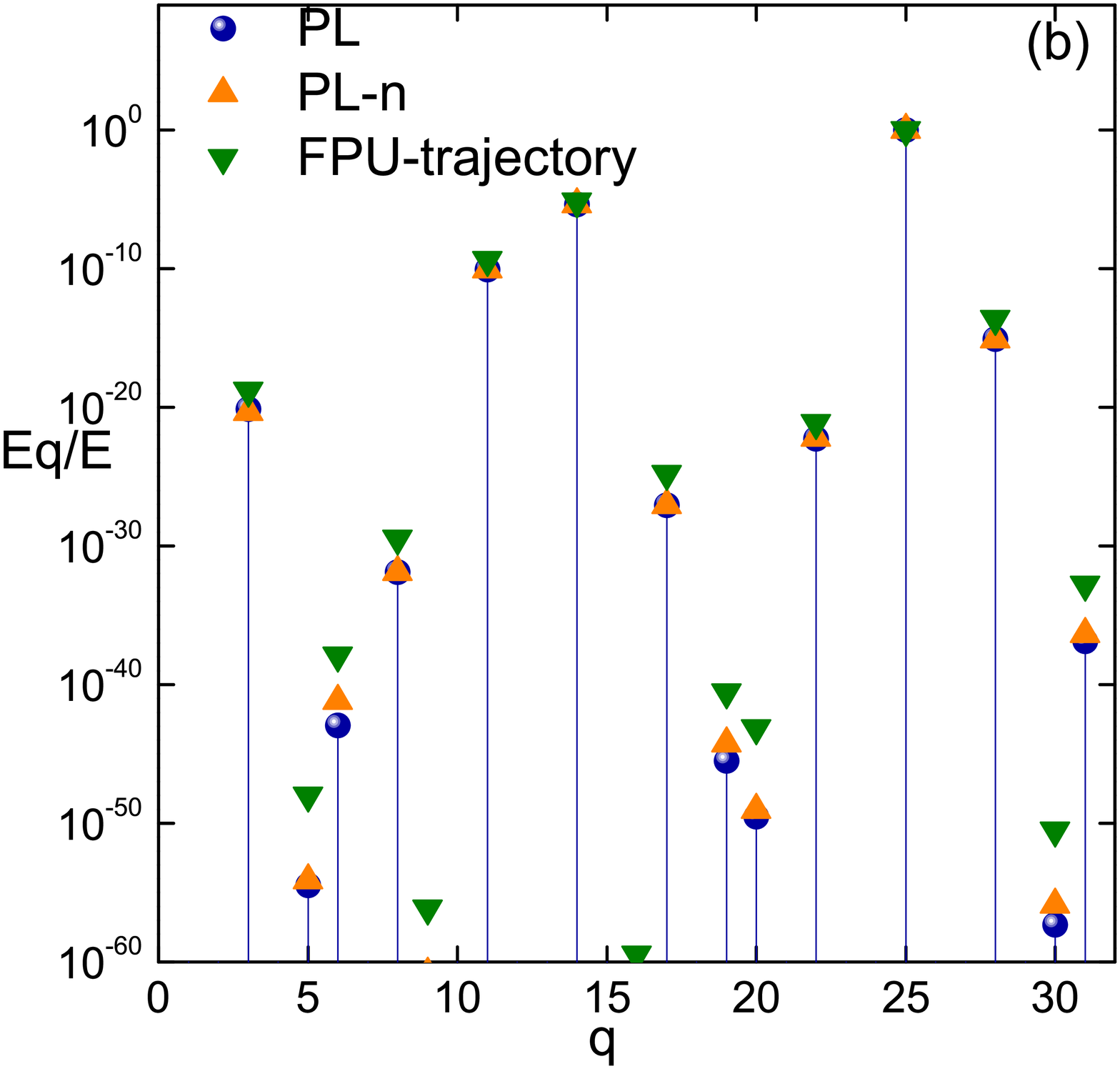}\\
\caption{FPU--$\alpha $ system with $\alpha =0.33$, $N=32$,
$E=0.00465079$ with seed mode $q_0=25$. In panel (a) it is shown the
evolution of the normalized instantaneous spectra $E_q(t)/E$ of
modes  $25,14,11,28,3,22,17$ for the PL solution of the
$q$--breather (blue) and the numerical integration (PLn) 
of the initial condition $Q_q^{PL}(0)$, $P_q^{PL}(0)$ (orange).
In panel (b) is the exponential profile of the normalized
averaged energy spectra $\overline{ E} _{q }/E$ versus $q$ of PL
with blue spheres, of PLn with orange triangles and of the FPU--trajectory
with green triangles. \label{fig:qbreath1}  }
\end{figure}

Fig.\ref{fig:qbreath1} refers, now, to a different $q$--breather
example, in which we choose to initially excite a mode at an arbitrary
position in $q$--space, namely $q_0=25$, for the system with
$\alpha =0.33$, $N=32$ and $E=0.00465079$. Again $\omega_{25}$
and $A_{25}$ are found in the table of \ref{AppF}, while the truncation 
order here is $k_0=60$. The sequence of mode excitations derived from 
Eq.(\ref{qbre}) up to the $10$--th order of perturbation theory is 
$q_1=14$, $q_2=11$, $q_3=28$, $q_4=3$, $q_5=22$, $q_6=17$, $q_7=8$, 
$q_8=31$, $q_9=6$, and $q_{10}=19$. 

Fig.\ref{fig:qbreath1}(a) shows the temporal evolution of the normalized
energies $E_q(t)/E$ of the first seven modes in the sequence of
excitations, namely $q_0, \ldots ,q_6$, for the solutions $Q_q^{PL,60}(t)$
(blue) and $Q_q^{PLn,60}(t)$ (orange). Differences observed between
the energy spectra which are below $10^{-15}$ show that the numerical 
integration cannot preserve precisely the analytical construction
and therefore the energy spectra make some small oscillations.

Fig.\ref{fig:qbreath1}(b) now shows the normalized averaged spectra
for the solutions (i) $Q_q^{PL,60}(t)$, (ii) $Q_q^{PLn,60}(t)$,
and (iii) a FPU--trajectory rising by the seed mode excitation $q_0=25$. 
We clearly see again that the FPU--trajectory's spectrum deviates from 
the $q$--breather's one at modes corresponding to a higher order in the 
excitation sequence. Still, however, the hierarchy of modes 
in the energy spectrum is preserved.

\begin{figure}
\centering
\includegraphics[scale=0.2 ]{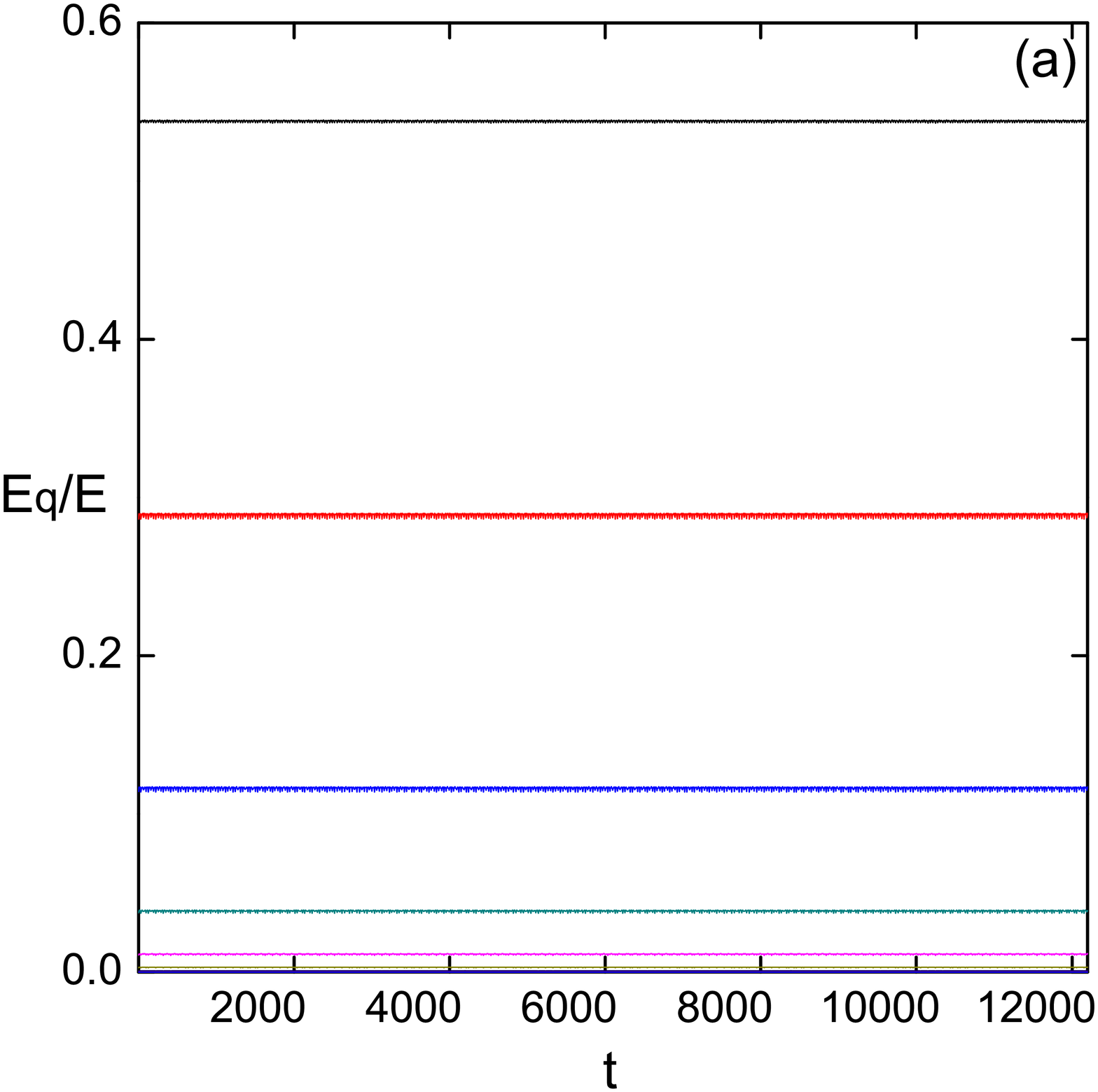}
\includegraphics[scale=0.2 ]{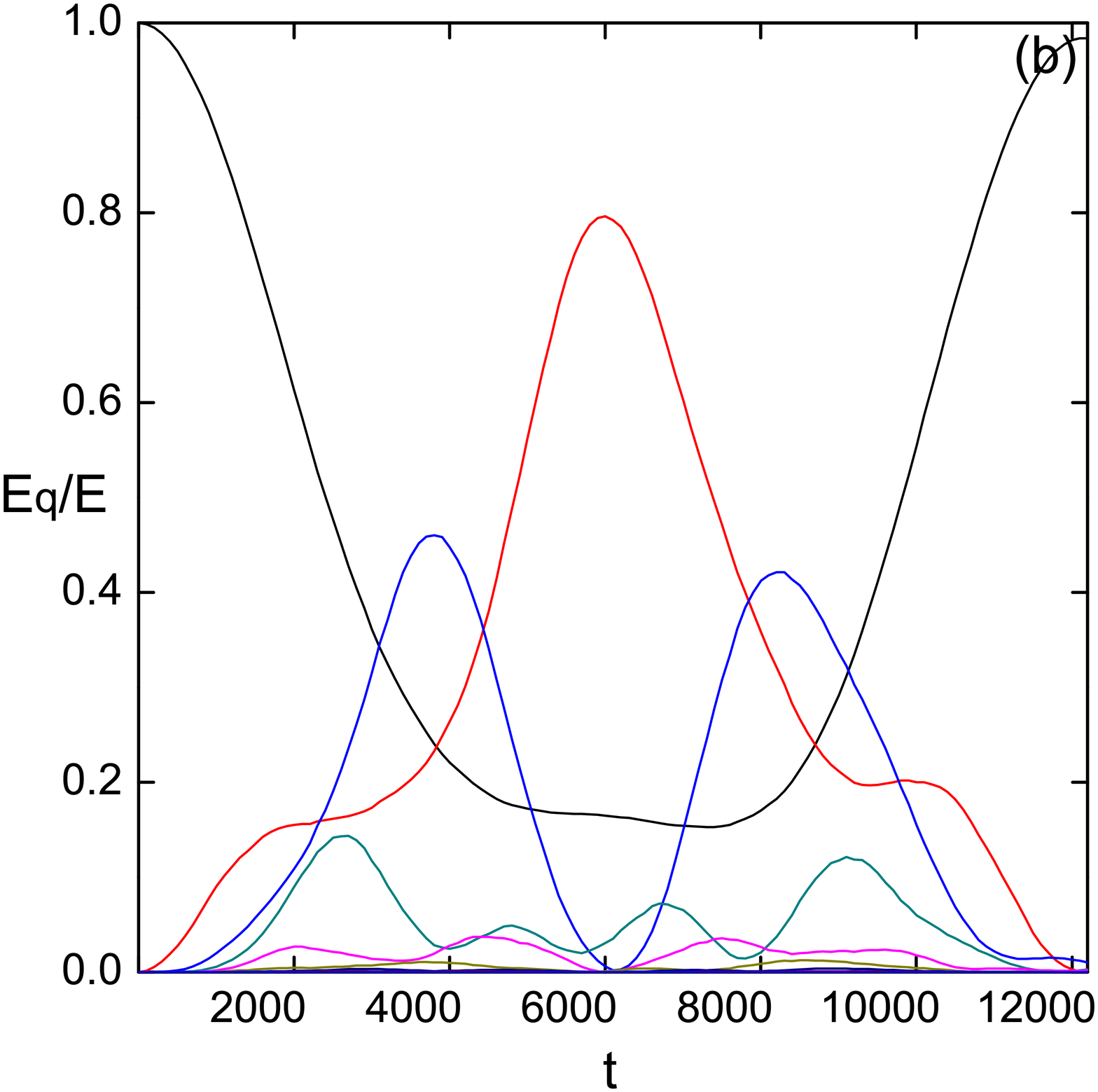}\\
\includegraphics[scale=0.2 ]{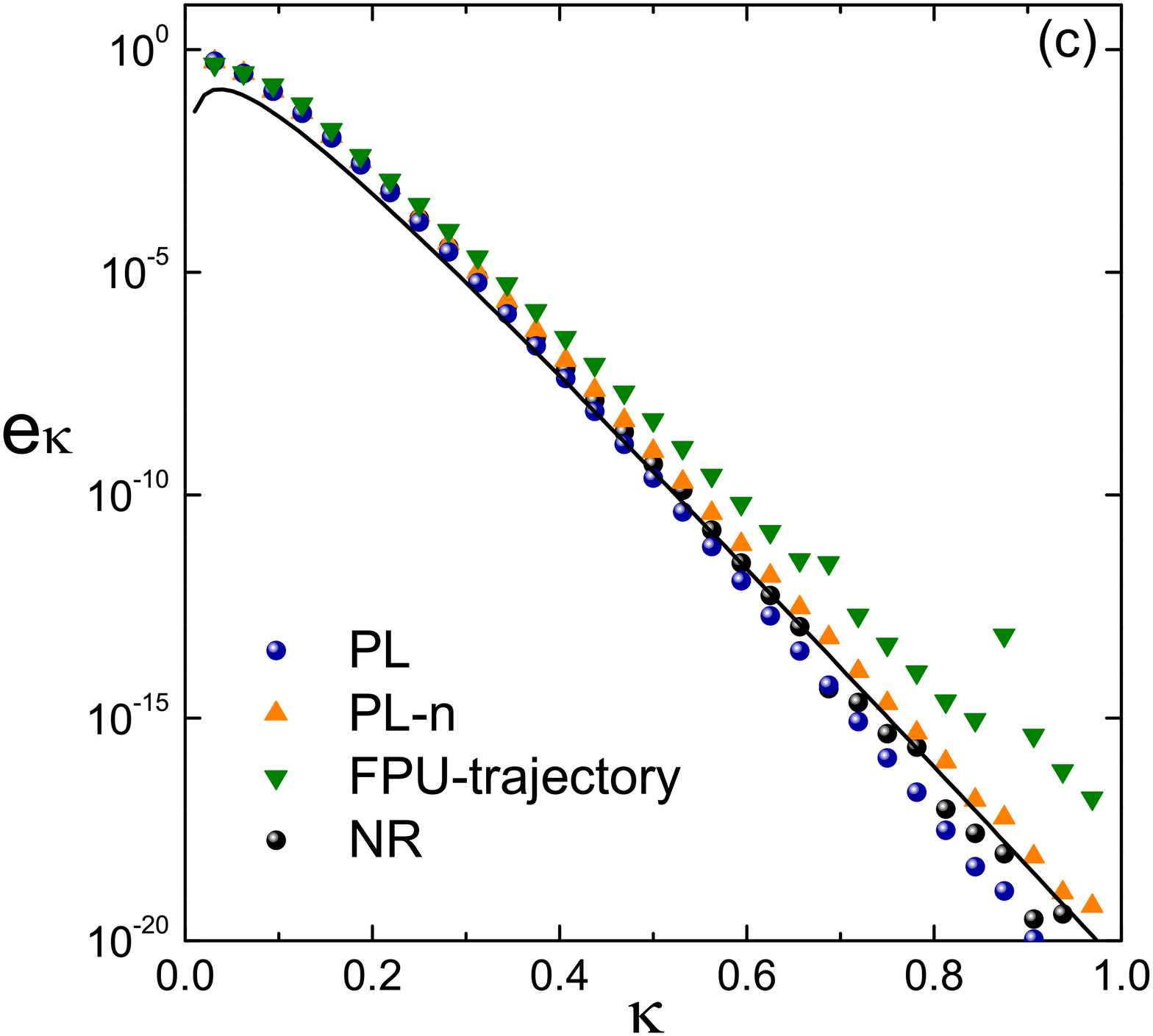}
\includegraphics[scale=0.2 ]{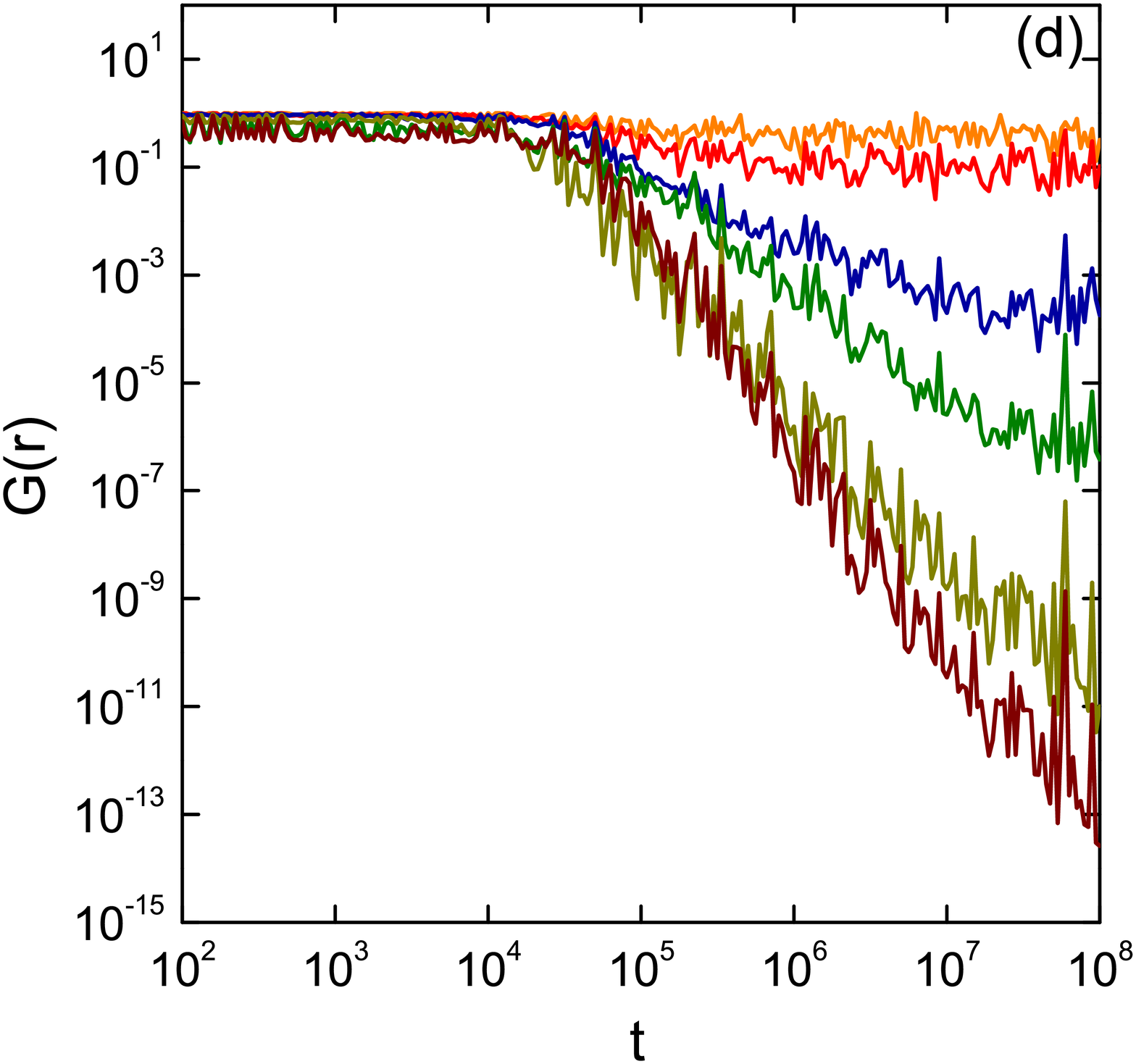}
\caption{FPU--$\alpha $ with $N=32$, $\alpha=0.33$, $E=0.014915$
and seed mode $q_0=1$. Panels (a) and (b) are for the evolution of
the normalized instantaneous spectra $E_q(t)/E$ for the
$q$--breather and the FPU--trajectory, respectively. In panel (c)
is the exponential profile of the normalized averaged energy
spectra $e_{\kappa }$ of PL with blue spheres, 
the numerical integration (PLn) 
of the initial condition $Q_q^{PL}(0)$, $P_q^{PL}(0)$ with
orange triangles, of FPU--trajectory with green triangles and of
NR with black spheres. The continuous line is
the fitting law (\ref{brelaw}).
 In (d) is the evolution of the GALI indices
$G_2,G_3,\ldots, G_7$. \label{fig:qbreath}  }
\end{figure}

\subsubsection{The original FPU--trajectory: how many frequencies?}
We finally discuss a FPU--trajectory similar to the classical 
experiment of Fermi Pasta and Ulam in \cite{feretal1955}, 
leading to the observation of the celebrated {\it FPU--recurrences} 
(Fig.\ref{fig:qbreath}(b)). Practically, we return to the
example of Subsection \ref{q1breather}, but for a higher energy. 
The PL construction of the corresponding $q$--breather to this 
FPU--trajectory was made for $E=0.014915$ and truncation order 
$k_0=242$ ($\omega_1$ and $A_1$ are given in \ref{AppF}).

Comparing their energy spectra in Fig.\ref{fig:qbreath}(c), 
we see a quite good agreement, especially to the low--frequency 
modes. While these two objects are lying close in phase space,
we would like to emphasize that they exhibit a different 
dynamical behavior. It can be observed, from the temporal
evolution of their energy spectra, that the case of the 
FPU--trajectories shown in Fig.\ref{fig:qbreath}(b), 
contrary to the almost constant $q$--breather's spectra of Fig.\ref{fig:qbreath}(a),
clearly show a recursive behavior, leading to a nearly complete
return to their initial values at the time $t=12000$. 


By implementing the GALI method, we find that the
number of incommensurable frequencies that govern this FPU--trajectory
is five, a fact implying that it lies on a 5--dimensional torus.
In particular, Fig.\ref{fig:qbreath}(d) shows that the indices $G_2$ and
$G_3$ are constant from the beginning, while $G_4$ and $G_5$ reach an asymptotic constant value at
nearly $t=5\times 10^{7}$, giving evidence for a 5--dimensional torus 
with 3 dominant actions and 2 compactified in smaller scales.

We note, in this latter respect, that according to \cite{beretal2005},
an excitation as in Fig.\ref{fig:qbreath} should lead to the
formation of a `natural packet' of modes exhibiting a sort
of internal equipartition. However, the prediction for the
packet width yields $\alpha ^{1/2}\varepsilon ^{1/4} N \simeq 2.7 \simeq
3$, which is smaller than the dimension of the torus
as suggested by the GALI indicators. On the other hand, 
a very recent study by Genta et al. \cite{gentaetal} shows
that the cutoff of the packet's width is proportional to
 $a^{2/5} \varepsilon ^{1/5} N  \simeq 4.43$ (between 4 and 5).



We thus conclude that the dynamics of the FPU trajectory of Fig.\ref{fig:qbreath}(b) shares
a number of common features with a $q$--breather with a similar initial excitation,
but also a number of common features with low--dimensional objects of dimension higher than one.
In fact, the numerical indications are that the dimension of the associated object is around 5.
However, in the present case the energy is quite high and our attempt to construct a torus
solution analytically does not appear to lead to a convergent PL series.
We thus leave open for future study the question of truly separating between solutions
of different dimension that approximate the dynamics of FPU trajectories in this regime.

\section{Conclusions} \label{concl}

In the present paper, we study the dynamical features and
localization properties of low--dimensional invariant 
objects of the FPU phase space called {\it $q$--tori}. 
Our main findings can be summarized as follows:

1) We use the method of Poincar\'{e} -- Lindstedt (PL) series in
order to compute quasi-periodic series representations of trajectories
approximating the motion on $q$--tori, up to a high order in a small
parameter. We give some details on the way by which appropriate
values of the torus frequencies are chosen in order for the PL
method to proceed. We also test numerically the convergence behavior of our
PL series. In particular we show that our series exhibit the phenomenon 
of cancellations between terms of a big size, leaving a small residual. 
Furthermore, the GALI indicator was used as an additional test for
verifying the dimension of our computed $q$--tori solutions.  

2) Using properties of the PL series construction, we study the
phenomenon of {\it energy localization} on $q$--tori. We present
a theory of propagation of initial `excitations' in the series terms.
Via this theory we predict theoretically the form and shape of energy
localization profiles for $q$--tori. Finally, we compare the latter
with the energy localization profiles of orbits lying in the
neighborhood of $q$--tori (called `FPU--trajectories').

3) Regarding the FPU--trajectories lying in the
neighborhood of $q$--tori solutions, we find that they exhibit
energy localization phenomena leading to an averaged normalized
energy spectrum tending to saturate to a form quite close to
that of a $q$--torus solution with similar initial excitation.
We provide numerical evidence that the so--called `first stage' 
in \cite{dresden} of energy transfer in FPU--trajectories, 
involves energy transfer among groups defined by the sets 
${\cal D}_k$, as in the $q$--tori case. A theoretical 
interpretation of this phenomenon remains an open question.

4) As a case of particular interest, we study low--frequency packet
excitations. In this case, we show that $q$--tori solutions predict
the appearance of exponential localization profiles, with a slope
depending logarithmically on the specific energy of the system 
and on the percentage of the modes excited. 
Thus, the normalized energy spectra are invariant with
respect to the re--scaled wavenumber $\kappa=q/N$, as long as the
fraction of modes excited $s/N$ and the specific energy are kept
constant. Via leading order estimates in the PL series, we derive
a law for the slope of the exponential energy profile, both in the
FPU--$\alpha $ and FPU--$\beta$ models. This turns out to be similar to
that of $q$--breather solutions in which the  `seed mode'
$q_0$ is allowed to vary proportionally to $N$. In fact, such laws
suggest the invariance of the $q$--tori's localization profiles
as we approach the thermodynamic limit, provided that
$s/N$ and specific energy $\varepsilon $ are kept constant.
However, the existence of $q$--tori as $N\rightarrow\infty$ is
still an open issue, as we have no proof of the convergence of
the PL series in a finite domain of initial conditions as we
approach this limit.

5) Excitations of packets of consecutive modes not in the low 
frequency part of the spectrum lead to non--trivial localization 
profiles that have not been extensively studied in the literature.
A study of the FPU--trajectories arising from such initial conditions, 
as well as of the times needed for such trajectories to reach 
equipartition, is an interesting open problem.

6) The computer--algebraic program of the PL series presently used 
for the determination $q$--tori was employed also in the case
of $q$--breathers, leading to high precision calculations 
corresponding to orders of the PL series higher than $k=60$. 
Using also our algorithm of systematic determination of 
frequencies (\ref{AppA}), we are able to locate $q$--breathers 
of quite high energy values. We find that, as the energy of the 
system increases, the distance between $q$--breathers and 
FPU--trajectories resulting from the same initial excitation 
also grows. Furthermore, the FPU--trajectories exhibit a dynamics 
consistent with an increasing number of incommensurable frequencies. 
Nevertheless, the spectra of the FPU--trajectories remain 
strongly localized in $q$--space. 

As a final remark, we note that the time for which `metastable
states' of FPU--trajectories started close to $q$--tori persist,
as well as its dependence on the system's parameters, is an
interesting open question that can be considered as complementary
to the study in \cite{benetal2011} (where the initially excited
packet is smaller than the natural one). This is proposed as a
subject for future study.

\paragraph{Acknowledgments}
We wish to thank A. Ponno and G. Benettin for their useful
discussions clarifying particular points of the paper. H.C.
gratefully acknowledges the hospitality of the Dipartimento di
Matematica Pura e Applicata, Universit\'{a} di Padova, during the
period May 2010 to May 2012, where this work was initiated and
completed.


\clearpage
\noindent
{\bf APPENDIX}\\

{\bf A. Algorithm of determination of frequency values for the PL
series construction \label{AppA}}

We give an iterative algorithm of determination of numerical
values of the frequencies $\omega_{q_i}$, $q_i\in{\cal D}_0$,
for a $q$--torus solution with initial excitation ${\cal D}_0$,
such that the solution constructed at every step corresponds
to a higher specific energy than the solution constructed in
the previous step. The algorithm consists of the following
steps:\\

{\bf Initialization.} i) We choose some value of `trial amplitudes'
$A_{q_i}^{trial}$ and define
\begin{equation}\label{linespe}
\epsilon_{l,trial} = {1\over N}\sum_{i=1}^s {1\over 2}
\Omega_{q_i}^2 A_{q_i,trial}^2,~~
q_i\in{\cal D}^{(0)}~~, i=1,\ldots s.
\end{equation}\\

ii) We compute `trial' frequencies by the lowest-order frequency
correction terms in the series (\ref{omeser}) corresponding
to the trial amplitudes $A_{q_i,trial}$.

In the FPU-$\alpha$ the frequencies $\omega_{q_i}$ appear in the
denominators of the lowest order terms, we implement a two-step
substitution-iteration procedure. Namely we set
\begin{eqnarray}\label{omamid}
\omega_{q,mid}^{(2)} &=&
- \frac{\Omega _q}{4} \sum_{\mathop{m \in
{\cal D}^{(0)} \cup {\cal D}^{(1)} }}
\Omega _m^2 \times
\Bigg[  \Omega _q^2 A_{q,trial}^2 B_{mqq}^2 \left( \frac{1}
{\Omega _m^2 - 4 \Omega _q^2} + \frac{ 2 }
{\Omega _m^2 }\right)\nonumber\\
&+& \sum_{\mathop{n \in {\cal D}^{(0)} }\limits_
{n \neq q,  j=1,...,4} }  \Omega _n^2 A_{n,trial}^2
\left( \frac{2 B_{qnm}^2}
{\Omega _m^2 -({\cal P}_j( \Omega _{q},\Omega _{n}))^2}
+\frac{B_{qqm} B_{mnn}}
{\Omega _m^2 -({\cal P}_j( \Omega _{n},\Omega _{n}))^2}
\right) \Bigg]~~~ \nonumber
\end{eqnarray}
for $q\in{\cal D}^{(0)}$, or
$$
\omega_{q,mid}^{(2)}=0
$$
for $q\notin{\cal D}^{(0)}$, and determine $\omega_{q,mid}=\Omega_q+
\mu^2\omega_{q,mid}^{(2)}$ for all $q=1,...N$. In the above formulae,
${\cal P}_1(x,y)=x+y$, ${\cal P}_2(x,y)=x-y$, ${\cal P}_3(x,y)=-x+y$,
${\cal P}_4(x,y)=-x-y$.  Then we compute
\begin{eqnarray}\label{omafin}
\omega _{q,trial}^{(2)} &=&
- \frac{\Omega _q}{4} \sum_{\mathop{m \in {\cal D}^{(0)}
\cup {\cal D}^{(1)} }}
\Omega _m^2 \times
\Bigg[  \Omega _q^2 A_{q,trial}^2 B_{mqq}^2 \left( \frac{1}
{\omega _{m,mid}^2 - 4 \omega_{q,mid}^2} +
\frac{ 2 }{\omega _m^2 }\right)\nonumber\\
&+& \sum_{\mathop{n \in {\cal D}^{(0)} }\limits_
{n \neq q,  j=1,...,4} }  \Omega _n^2 A_{n,trial}^2
\left( \frac{2 B_{qnm}^2}
{\omega_{m,mid}^2 -({\cal P}_j( \omega_{q,mid},\omega_{n,mid}))^2}
+\frac{B_{qqm} B_{mnn}}
{\omega_{m,mid}^2 -({\cal P}_j( \omega_{n,mid},\omega_{n,mid}))^2}
\right) \Bigg]~~. \nonumber
\end{eqnarray}
and set $\omega_{q,trial}=\Omega_q+
\mu^2\omega_{q,trial}^{(2)}$.

In the FPU--$\beta$, we simply have
\begin{equation}\label{omebguess}
\omega_{q,trial}= \Omega_q
+\mu \omega_q^{(1)}(A_{q_1,trial},\ldots,A_{q_s,trial})~~~.
\end{equation}\\

iii) We compute the PL series for the excitation ${\cal D_0}$,
using as frequency values $\omega_{q_i}=\omega_{q_i,trial}$.
We attempt to determine numerically a root of Eqs.(\ref{omeser})
for the amplitudes $A_{q_i}$, with a root-finding technique
starting by $A_{q_i,trial}$ as guess amplitudes. If this
fails, we return to substep (i), trying some lower value
for the amplitudes $A_{q_i,trial}$, until a successful
solution is found. We store the pairs $(\omega_{q_i},A_{q_i})$
of the latter.

iv) we repeat the process (i) to (iii) for some neighboring
trial amplitudes $A_{q_i,trial}'$ and store the pairs
$(\omega_{q_i}',A_{q_i}')$ for the corresponding solution.
We define $\Delta\epsilon_l=\epsilon_{l,trial}'
-\epsilon_{l,trial}$.\\

{\bf Iteration.} (i) We define
\begin{equation}\label{delome}
\Delta\omega = \left(\sum_{i=1}^s
(\omega_{q_i}-\omega_{q_i}')^2\right)^{1/2}~~,
\end{equation}
and denote by $(\omega_{q_i,0},A_{q_i,0})$ the original solution
$(\omega_{q_i},A_{q_i})$.

ii) We compute $s$ neighboring solutions corresponding to the set
of frequencies $\omega_{q_i,j}=\omega_{q_i}+\delta_{ij}\Delta\omega$.

iii) We compute the matrix $J_\omega=J_A^{-1}$, where $J_A$ is a
matrix defined by the finite differences
\begin{equation}\label{jacome}
J_{A,ij}={A_{q_i,j}^2-A_{q_i,0}^2\over\Delta\omega}~~.
\end{equation}

iv) Finally, we compute the next set of frequencies to be used
in PL series construction by
\begin{equation}\label{omeprime}
\omega_{q_i}' = \omega_{q_i}+{2N\Delta\epsilon_l\over s}
\sum_{j=1}^s {J_{\omega,ij}\over\Omega_{q_j}^2}~~.
\end{equation}

This completes one full step of the iterative algorithm of
determination of frequencies. We note that a change of the
frequencies as in Eq.(\ref{omeprime}) leads to an increment
of the total energy corresponding to each successive step
by an amount of order $\Delta\epsilon_l$.\\

\noindent

{\bf B.  Proof of the proposition of subsection \ref{sequence} \label{AppB}}


We give the proof of the proposition of subsection \ref{sequence} in the
case of the FPU--$\alpha$ model  (see \cite{chrietal2010} for
the proof in the case of the FPU--$\beta$ model).

The proof follows by induction. Let $M_k$ be the set defined in
Eq.(\ref{app}), which corresponds to the set of all modes
for which the r.h.s. of Eq.(\ref{moda}) is non--zero at the
$k$--th order. According to Definition 1, the set of modes
excited at the $k$--th order is given by
${\cal D}_k =M_k \setminus  \bigcup_{0 \leq j \leq k-1}
{\cal D}_{j}$, where $r(k)=k+1$.
For $k=1$, the r.h.s. of Eq.(\ref{moda}) is
non--zero if $B_{qq_1 q_2}\neq 0$, implying  that for all modes
$q \in M_k$ we have that $q$ is either of the form
$q = \mid\sigma^{(2)}q ^{(2)}\mid $, or of the form
$q=\big|2N-\mid\sigma^{(2)}q ^{(2)}\mid\big|$,
where the allowable combinations of values of
$q^{(2)}=(q_1,q_2) \in {\cal D}_0^2$ and of
$\sigma^{(2)}=(\sigma _1,\sigma _2)\in\Sigma ^2$, are
those leading to $1\leq q\leq N-1$.

Assuming, now, the proposition to be true at order
$k-1$, one finds that, at the $k$--th order, the r.h.s.
of Eq.(\ref{moda}) is non--zero if $B_{qlm}\neq 0$,
$Q_{l}^{(n_1)}\neq 0$ and $Q_{m}^{(n_2)}\neq 0$, where
$n_1+n_2=k-1$. For the modes $l \in M_{n_1}$ and
$m \in M_{n_2}$ we then have:
\begin{eqnarray} \label{lmn}
 l = \Bigg| 2 \nu _1 N - \mid  \sigma ^{(n_1+1)}
 q^{(n_1+1)} \mid \Bigg| ~~ , ~~
 m = \Bigg| 2 \nu _2 N - \mid  \sigma ^{(n_2+1)}
 q^{(n_2+1)} \mid \Bigg| ~~, \nonumber
 \end{eqnarray}
where $\nu _i = \left[  \frac{\mid  \sigma ^{(n_i+1)}
q^{(n_i+1)} \mid +N-1}{2N} \right] $.
However, the condition $B_{qlm}\neq 0$ implies that $q$
is necessarily of the form $q = \mid  l \pm m \mid $ or
$q = \big| 2N - \mid   l \pm m \mid  \big|$. Provided
that $1\leq q\leq N-1$, the two latter equations can
be written in a combined form as:
\begin{eqnarray} \label{wg}
q = \big| 2Ng - \mid   l \pm m \mid  \big|
\end{eqnarray}
where $g=0\mbox{ or }1$. Then, Eq.(\ref{wg}) takes
the form
\begin{eqnarray}
q &=& \big| 2Ng \pm 2 \nu _1 N  \pm  2 \nu _2 N
\mp \mid  \sigma ^{(n_1+1)} q^{(n_1+1)} \mid  \mp
\mid  \sigma ^{(n_2+1)} q^{(n_2+1)} \mid \big|  \nonumber\\
&=&\big| 2N (\underbrace{ \mid  g \pm \nu _1 \pm
\nu _2\mid }_{\nu }) - \mid  \sigma ^{(k+1)}
q^{(k+1)} \mid \big|, ~~
\end{eqnarray}
after a possible sign reversal within $|\cdot|$ (not affecting
the absolute value) and with $\nu $ a positive integer.

However, by the restriction $1\leq q\leq N-1$, one necessarily
has that $\nu=\left[  ( \mid  \sigma ^{(k+1)} q^{(k+1)}
\mid +N-1)/2N \right]  $. This concludes the proof of the
proposition.

\noindent

{\bf C. Explicit expressions for the leading order terms
of the PL series \label{leading}}


We give by the following Lemma an expression for the
leading order terms of the solution $Q_q(t)$, $q \in {\cal D}_k$ 
(see Definition 1). This will be used in section \ref{numqtor} in order to
predict the forms of energy localization profiles of $q$--tori.

Let ${\cal D}_0=\{q_1,q_2,\ldots,q_s\}$ be a zero order
excitation set, ${n^{(r)}\in{\cal D}_0^r}$ an $r$--vector in ${\cal D}_0^r$
and $\omega_{n}^{(r)}\equiv\{\omega_{n_1}, \dots,\omega_{n_{r}}\}$,
$\phi_{n}^{(r)} \equiv(\phi_{n_1},\dots,\phi_{n_r})$  its associated frequency and
phase  vectors respectively. Some basic features of the PL
construction arising in accordance with the Definitions 1 -- 4 of
subsection \ref{sequence} are:

i) If $q \in {\cal D}_k$, then $Q_q^{(k)}(t)$ is a leading order term
of the series (\ref{qser}).

ii) The only modes that admit frequency corrections are those 
in ${\cal D}_0$, for the rest holds $\omega _q = \Omega _q$, 
$q \notin {\cal D}_0$.

iii) At the $k$--th order, the expressions for $Q_q^{(k)}(t)$
contain divisors which are the products of $k$ factors of the form
$ \Omega_q^2-(\sigma ^{(r(m))}\omega^{(r(m))}_{n})^2$, 
 $m\leq k$. 
 
For convenience, in subsequent formulae we use exponential rather than trigonometric
expressions. In both the $\alpha$ and $\beta$ models, for
$q\in{\cal D}_k$ we find that:

{\bf Lemma:} {\it The leading order terms $Q_q^{(k)}(t)$  of Eqs.(\ref{moda}) and (\ref{modb}), 
starting by the zero order solution $Q_q^{(0)}(t)$ of Eq.(\ref{qq0}) read
\begin{eqnarray}\label{Qqkr}
Q_q^{(k)} (t) = \sum_{\mathop{n^{(r)} \in {\cal D}_0^r }\limits_
{\sigma ^{(r)} \in \Sigma ^{r} } }
\mathcal{R}_q^{(k)} (n^{(r)}) \mathcal{K}_q^{(k)} (n^{(r)})
e^{i \sigma ^{(r)} ( \omega _{n}^{(r)}t + \phi_{n}^{(r)})}
\end{eqnarray}
where $r(k)=k+1$ for the FPU--$\alpha$ and $r(k)=2k+1$ for the
FPU--$\beta$. In the above expression,  the factor $\mathcal{R}_q^{(k)}$
is given by
\begin{eqnarray}\label{Rqr}
\mathcal{R}_q^{(k)} (n^{(r)}) &=& \frac{ (-1)^k }{2^{r}}
\cdot\frac{\Omega_q\Omega_{n_1}\ldots\Omega_{n_{r}} A_{n_1}\ldots
A_{n_{r}}}{\Omega_q^2-(\sigma ^{(r)}\omega^{(r)}_{n})^2}
\end{eqnarray}
and $\mathcal{K}_q^{(k)}$  by the
recursive relation
\begin{eqnarray}\label{Kkga}
\mathcal{K}_{q ;\alpha}^{(k)}(n^{(r)}) =
\sum_{\mathop{l_{1,2}=0}\limits_
{l_1+l_2=k-1}}^{k-1}  \sum_{\mathop{ m_{i}\in
{\cal D}_{l_i}} \limits_
{i=1,2}} \mathfrak{L}_{m_1}^{(l_1)} (n^{(r(l_1))})
\mathfrak{L}_{m_2}^{(l_2)} (n^{(r(l_2))})
\mathcal{K}_{m_1}^{(l_1)} (n^{(r(l_1))})
\mathcal{K}_{m_2}^{(l_2)} (n^{(r(l_2))})
B_{q m_1m_2}
\end{eqnarray}
with
\begin{eqnarray}\label{nvecta}
n^{(r(k))}=(\underbrace{ n_1,\ldots ,n_{r(l_1)}}_{n^{(r(l_1))}}
,\underbrace{n_{r(l_1)+1}, \ldots ,n_{r(k)} }_{n^{(r(l_2))}} )\nonumber
\end{eqnarray}
in the $\alpha $ case and
\begin{eqnarray}\label{Kk}
\mathcal{K}_{q ;\beta } ^{(k)} (n^{(r)})
= \sum_{\mathop{l_{1,2,3}=0}\limits_
{l_1+l_2+l_3=k-1}}^{k-1} \sum_{\mathop{ m_{i}\in
{\cal D}_{l_i}} \limits_
{i=1,2,3}} \mathfrak{L}_{m_1}^{(l_1)} (n^{(r(l_1))})
\mathfrak{L}_{m_2}^{(l_2)} (n^{(r(l_2))})
\mathfrak{L}_{m_3}^{(l_3)} (n^{(r(l_3))} ) \nonumber\\
\mathcal{K}_{m_1}^{(l_1)} (n^{(r(l_1))})
\mathcal{K}_{m_2}^{(l_2)} (n^{(r(l_2))})
\mathcal{K}_{m_3}^{(l_3)} (n^{(r(l_3))})
C_{q m_1m_2m_3} ~~~,
\end{eqnarray}
with
\begin{eqnarray}\label{nvectb}
n^{(r(k))} =(\underbrace{ n_1,\ldots,n_{r(l_1)}}_{n^{(r(l_1))}},
\underbrace{n_{r(l_1)+1}, \ldots ,n_{r(l_1)+r(l_2)}}_{n^{(r(l_2))}},
\underbrace{ n_{r(l_1)+r(l_2)+1}, \ldots , n_{r(k)} }_{n^{(r(l_3))}})\nonumber
\end{eqnarray}
in the $\beta $ case, setting, in both cases, $\mathcal{K}_q^{(0)}=1$
at $k=0$. The terms $\mathfrak{L}_m^{(l)}$, $m \in {\cal D}_{l}$
entering the expressions (\ref{Kkga}) and (\ref{Kk}) are 
\begin{eqnarray}\label{Lm}
\mathfrak{L}_m^{(l)} (n^{(r(l))})=
\frac{\Omega _m^2}{\Omega_m^2-(\sigma^{(r(l))}
\omega^{(r(l))}_{n})^2}~~
\end{eqnarray}
for $l>0$, or $\mathfrak{L}_q^{(0)}=1$ at $l=0$. }
{\bf proof}\\
We prove by induction that the leading order terms $Q_q^{(k)}$ for
the modes $q \in {\cal D}_{k}$ are given by Eqs. (\ref{Qqkr})  
with the quantities $\mathcal{R}_q^{(k)}$, $\mathcal{K}_{q} ^{(k)}$ given by (\ref{Rqr}),
(\ref{Kkga}) and (\ref{Kk}). We focus again on the
FPU--$\alpha$ model.

For $k=1$, the solutions of Eqs.(\ref{moda}) read
\begin{eqnarray}\label{Qq1cora}
Q_q^{(1)} (t)&=& - \frac{\Omega_q}{4} \sum_{\mathop{n^{(2)}
\in {\cal D}^{(2)}_0}\limits_
{\sigma ^{2} \in \Sigma^2  }}
\Omega_{n_1} \Omega_{n_2} A_{n_1} A_{n_2} B_{qn_1n_2}
\frac{  e^{i(\sigma_1 \omega _{n_1} + \sigma _2 \omega_{n_2} )t}}
{\Omega_q^2 - (\sigma_1 \omega _{n_1} + \sigma _2 \omega_{n_2} )^2 }\nonumber\\
\end{eqnarray}
so, $Q_q^{(1)}$ satisfies Eq.(\ref{Qqkr}).

Assume now that Eq.(\ref{Qqkr}) holds true for the solution
$Q_q^{(k)}(t)$ at the order $k-1$. For simplicity, we use the
notation $\mathcal{R}_l^{(n_i)} = \mathcal{R}_l^{(n_i)}
(n^{(r(n_i))})$ and $\mathcal{K}_l^{(n_i)}
=\mathcal{K}_l^{(n_i)} (n^{(r(n_i))}) $, $i=1,\ldots,r(k)$.
At order $k$,  Eq.(\ref{moda}) takes the form
\begin{eqnarray}
\ddot {Q}_q^{(k)}+\Omega_q^2Q_{q}^{(k)}=
~~~~~~~~~~~~~~~~~~~~~~~~~~~~~~~~~~~~~~~~~~~~~~~~~~~~~~~~~~~~~\\
-\Omega_{q} \sum_{m_1,m_2=1}^{N-1}\Omega _{m_1}\Omega _{m_2}
B_{qm_1m_2}\sum_{\mathop{l_{1,2}=0}\limits_
{l_1+l_2=k-1}}^{k-1}\sum_{\mathop{n^{(r(k))} \in
{\cal D}_0^{r(k)} }\limits_
{\sigma ^{(r(k))} \in \Sigma ^{r(k)} } }
\mathcal{R}_{m_1}^{(l_1)} \mathcal{R}_{m_2}^{(l_2)}
\mathcal{K}_{m_1}^{(l_1)} \mathcal{K}_{m_2}^{(l_2)}
e^{i \sigma ^{(r(k))}  \omega _{n}^{(r(k))}t}  .\nonumber
\end{eqnarray}
By replacing the term
\begin{eqnarray}
\mathcal{R}_{m_1}^{(l_1)} \mathcal{R}_{m_2}^{(l_2)}
&=&{(-1)^{l_1+l_2} \over 2^{r(l_1)+r(l_2)}} \cdot
\frac{\mathfrak{L}_{m_1}^{(l_1)} \mathfrak{L}_{m_2}^{(l_2)}}
{\Omega _{m_1} \Omega _{m_2}}
\cdot \Omega_{n_1}\ldots\Omega_{n_{r(k)}} A_{n_1}\ldots
A_{n_{r(k)}} \nonumber\\
&=& -  \frac{\mathfrak{L}_{m_1}^{(l_1)} \mathfrak{L}_{m_2}^{(l_2)}  }
{\Omega _{m_1} \Omega _{m_2} }
\cdot \frac{\Omega_q^2 - (\sigma ^{(r(k))}  \omega _{n}^{(r(k))})^2}
{\Omega _{q}} \cdot \mathcal{R}_{q}^{(k)}
\end{eqnarray}
into the above equation, one has that
\begin{eqnarray}
\ddot {Q}_q^{(k)}+\Omega_q^2Q_{q}^{(k)}=
(\Omega_q^2 - (\sigma ^{(r(k))}  \omega _{n}^{(r(k))})^2)
~~~~~~~~~~~~~~~~~~~~~~~~~~~~~~~~~~~~~~~~~~~~~~~~
\nonumber\\
\times \sum_{\mathop{n^{(r(k))} \in {\cal D}_0^{r(k)} }\limits_
{\sigma ^{(r(k))} \in \Sigma ^{r(k)} } } \mathcal{R}_{q}^{(k)}
\left( \sum_{\mathop{l_{1,2}=0}\limits_
{l_1+l_2=k-1}}^{k-1} \sum_{m_1,m_2=1}^{N-1}
 \mathfrak{L}_{m_1}^{(l_1)} \mathfrak{L}_{m_2}^{(l_2)}
\mathcal{K}_{m_1}^{(l_1)} \mathcal{K}_{m_2}^{(l_2)} B_{qm_1m_2}
\right) e^{i \sigma ^{(r(k))}  \omega _{n}^{(r(k))}t}  \nonumber\\
=
(\Omega_q^2 - (\sigma ^{(r(k))}  \omega _{n}^{(r(k))})^2)
\sum_{\mathop{n^{(r(k))} \in {\cal D}_0^{r(k)} }\limits_
{\sigma ^{(r(k))} \in \Sigma ^{r(k)} } } \mathcal{R}_{q}^{(k)}
\mathcal{K}_{q}^{(k)}
e^{i \sigma ^{(r(k))}  \omega _{n}^{(r(k))}t}~~.~~~~~~~~~~~~~~~~~
\end{eqnarray}
However, the solution of the latter equation is
\begin{eqnarray}
Q_{q}^{(k)}(t)&=&
\sum_{\mathop{n^{(r(k))} \in {\cal D}_0^{r(k)} }\limits_
{\sigma ^{(r(k))} \in \Sigma ^{r(k)} } } \mathcal{R}_{q}^{(k)}
\mathcal{K}_{q}^{(k)}
e^{i \sigma ^{(r(k))}  \omega _{n}^{(r(k))}t}~~.
\end{eqnarray}
Thus, Eq.(\ref{Qqkr}) holds true at the order $k$, with the expressions
$\mathcal{R}_q^{(k)}$, $\mathcal{K}_{q} ^{(k)}$ given by (\ref{Rqr}),
(\ref{Kkga}) and (\ref{Kk}). \\

Both quantities $\mathcal{K}_{q ;\alpha}^{(k)}$,
and $\mathcal{K}_{q ;\beta } ^{(k)} $ are polynomials of degree $k-1$ in the terms
$\mathfrak{L}_m^{(l)}$, $l=0,1,\ldots$ and can be
computed iteratively. Explicit expressions for the
first few orders of the mapping
(\ref{Kkga}) are given in  \ref{AppC}.

\noindent

{\bf D. The mapping $\mathcal{K}_{q} ^{(k)}$ for FPU--$\alpha $ \label{AppC}}

We give the expressions of the quantities appearing in the mapping
 (\ref{Kkga}), up to the order $k=3$. We have
\begin{eqnarray}\label{Kqalpha}
&&\mathcal{K}_{q_1 ; \alpha} ^{(1)} = B_{q_1 q_0 q_0}  \nonumber\\
&&\mathcal{K}_{q_2 ; \alpha} ^{(2)} = 2  \mathfrak{L}_{q_0}^{(0)}
\mathfrak{L}_{q_1}^{(1)} \mathcal{K}_{q_0}^{(0)} \mathcal{K}_{q_1}^{(1)}
B_{q_2 q_1 q_0} = 2 \frac{\Omega _{q_1}^2}{\Omega_{q_1}^2 -
( \sigma _1 \omega _{q_0}+ \sigma _2 \omega _{q_0} )^2 } B_{q_1 q_0 q_0}
B_{q_2 q_1 q_0} \nonumber\\
&&\mathcal{K}_{q_3 ; \alpha} ^{(3)} = 2  \mathfrak{L}_{q_0}^{(0)}
\mathfrak{L}_{q_2}^{(2)} \mathcal{K}_{q_0}^{(0)} \mathcal{K}_{q_2}^{(2)}
B_{q_3 q_2 q_0} + [ \mathfrak{L}_{q_1}^{(1)} \mathcal{K}_{q_1}^{(1)} ]^2
B_{q_3 q_1 q_1} = \nonumber\\
&& 4 \frac{\Omega _{q_1}^2 \Omega _{q_2}^2  B_{q_1 q_0 q_0}
B_{q_2 q_1 q_0} B_{q_3 q_2 q_0}}{[\Omega_{q_1}^2 -
( \sigma _1 \omega _{q_0}+ \sigma _2 \omega _{q_0} )^2][\Omega_{q_2}^2
- ( \sigma _1 \omega _{q_0}+\sigma _2\omega _{q_0}+\sigma _3\omega _{q_0} )^2] }
\nonumber\\
&&+ \left[ \frac{\Omega _{q_1}^2 B_{q_1 q_0 q_0}}{\Omega_{q_1}^2 -
(\sigma _1 \omega _{q_0}+\sigma _2\omega _{q_0})^2 } \right]^2
B_{q_3 q_1 q_1},\nonumber
\end{eqnarray}
where $q_0$, $q_1$, $q_2$ and $q_3$ represent any mode belonging
in the sets ${\cal D}_0$, ${\cal D}_{1}$, ${\cal D}_{2}$ and
${\cal D}_{3}$ respectively.

\noindent

{\bf E. Localization profiles of $q$--tori estimated by leading
order terms in the PL series \label{AppE}}

We derive estimates for the form of the energy localization profiles 
for $q$--tori solutions corresponding to an initial excitation of the 
modes $1\leq q\leq s$, with $s$ varying proportionally to $N$. 

We first make the following estimates:\\

i) For any mode $q \in {\cal D}_{k}$ we use the approximation
$q\simeq c_k s$, where $c_k s$ is the mid mode of $ {\cal D}_{k}$,
i.e. $q=ks+[s/2]$ ($c_k\simeq k+1/2$) in FPU--$\alpha $ and
$q=2ks$ ($c_k=2k$) in FPU--$\beta $.

ii) For the unperturbed frequencies we use the approximation
$\Omega _q \simeq \pi q / N$.

iii) For  $m \in {\cal D}_{k}$ and for `almost resonant terms', 
for which $m-\sigma^{(r(k))}n=0$, we use the approximation 
$\big|\Omega_m-\sigma^{(r(k))}\omega^{(r(k))}_{n} \big| \simeq 
{\pi ^3 m^3}/ ({24 N^3})$ (See Appendix B of \cite{chrietal2010} 
for its derivation).

iv) For the quantities $\mathfrak{L}_m^{(k)}$ of Eq.(\ref{Lm})
we set $$\big| \mathfrak{L}_m^{(k)} (n^{(r(k))}) \big| \simeq
\frac{\Omega _m}{2[\Omega_m-\sigma^{(r(k))}
\omega^{(r(k))}_{n}]} \sim \frac{12 N^2 } {\pi ^2 m^2} $$

v) Finally, for the total energy of the system we use the
approximation $E \simeq \sum_{n=1,\ldots,s } E_n $, where
$E_n \simeq 1/2 A_n^2 \Omega _n^2$.

We will now derive an estimate for the quantity
$\big| Q_q^{(k)}\big| $ of Eq.(\ref{Qqkr}). We first write
some approximations for the terms of Eqs.(\ref{Rqr}), (\ref{Kkga})
and (\ref{Kk}). Recalling that $r(k)=k+1$ in the $\alpha $ model,
and $r(k)=2k+1$ in $\beta $, and taking into account the
approximations (i)--(v), we find
\begin{eqnarray}\label{Rqrapp}
\big| \mathcal{R}_q^{(k)} \big|  \simeq
\frac{\Omega_q }{\Omega_q^2 - (\sigma ^{(r)}\omega^{(r)}_{n})^2}
\cdot  \Omega_{n_1}\ldots\Omega_{n_{r}} A_{n_1}\ldots
A_{n_{r}} \sim \frac{12 N^3 } {\pi^3  q^3 } \cdot
\left( \frac{2 E}{s} \right)^{r(k)/2}.
\end{eqnarray}
The quantities $\mathcal{K}_{q ;\alpha }^{(k)}$  and
$ \mathcal{K}_{q ;\beta } ^{(k)} $ of  (\ref{Kkga}) and
(\ref{Kk}) cannot be evaluated analytically. However, as already
mentioned in section \ref{leading}, their form is a polynomial
of degree $k-1$ in the terms $\mathfrak{L}_m^{(l)}$, $l=0,1,
\ldots$ and a polynomial of degree $k$ in $B_{q,l,m}$
(or $C_{q,l,m,n}$). We denote here by ${\cal P}(q,n^{(r)},m^{(k-1)})$  the product of
$k$ factors of the coefficients $B_{q,l,m}$ (or $C_{q,l,m,n}$) in $\mathcal{K}_{q ;\alpha }^{(k)}$ 
(or $ \mathcal{K}_{q ;\beta } ^{(k)} $).  

The size of the mid
mode is then
\begin{eqnarray} \label{dddd}
&& A^{(k)}_q \sim \big| Q_q^{(k)} \big| \sim \sum_{\mathop{n^{(r)}
\in {\cal D}_0^r }\limits_
{\sigma ^{(r)} \in \Sigma ^{r} } }  \big|
\mathcal{R}_q^{(k)}  \big| \cdot   \big| \mathcal{K}_q^{(k)}  \big|
\nonumber\\ && \sim  \big|
\mathcal{R}_q^{(k)}  \big|  \cdot C_k \sum_{\mathop{n^{(r)}
\in {\cal D}_0^r }\limits_
{\sigma ^{(r)} \in \Sigma ^{r} } }
 \sum_{\mathop{ m_{i}\in
{\cal D}_{l_i}} \limits_
{i=1,\ldots,k-1}} \big| \mathfrak{L}_{m_1}^{(l_1)} \big|
\big| \mathfrak{L}_{m_2}^{(l_2)} \big| \ldots
\big| \mathfrak{L}_{m_{k-1}}^{(l_{k-1})} \big|
{\cal P}(q,n^{(r)},m^{(k-1)})\nonumber\\
&& \sim  \big|
\mathcal{R}_q^{(k)}  \big| \cdot C_k  \cdot
\left( \frac{12 N^2}{\pi ^2} \right)^{k-1} \sum_{\mathop{n^{(r)}
\in {\cal D}_0^r }\limits_
{\sigma ^{(r)} \in \Sigma ^{r} } }
 \sum_{\mathop{ m_{i}\in
{\cal D}_{l_i}} \limits_
{i=1,\ldots ,k-1}}  \frac{1}{\left( m_1  \ldots m_{k-1} \right)^2}
{\cal P}(q,n^{(r)},m^{(k-1)}) ~~, \nonumber\\
\end{eqnarray}
where by $C_k$ we note constants in powers of $k$. 
Due to the term ${\cal P}(q,n^{(r)},m^{(k-1)})$,
 the sums over $n^{(r)} $ and $m^{(k-1)}$ in (\ref{dddd})
give rise to a factor $s^{(r(k)+1)/2}$,
i.e. $s^{k/2+1}$ in $\alpha $ and to $s^{k+1}$ in $\beta $
\footnote{
These factors are found by recursively solving $\sum_{\mathop{n^{(r)}
\in {\cal D}_0^r }\limits_{\sigma ^{(r)} \in \Sigma ^{r} } }  
{\cal P}(q,n^{(r)},m^{(k-1)}) $, for those $q$ and $m^{(k-1)}$ that maximize
the results. In $\beta$ one finds for $k=1$ the factor $2s(s-1)$,
for $k=2$ the factor $4s^2(s-1)$, etc.}.
Replacing  $m_i$, $i=1,\ldots,k-1$ and $q$ in (\ref{dddd}) by their
mid mode expression $c_k s$, one has
\begin{eqnarray} \label{ddda}
A^{(k)}_q &\simeq &C_k ^{'} \frac{12 N^3 } {\pi^3  q^3 } \cdot
\left( \frac{2 E}{s} \right)^{r(k)/2} s^{(r(k)+1)/2}
\left( \frac{12 N^2}{\pi ^2 s^2} \right)^{k-1} \nonumber\\
&&= C_k ^{''}  \frac{N^{2k+1} } {\pi^{2k+1} s^{2k+1} } E^{r(k)/2}~~.
\end{eqnarray}
The energy in each group of modes is then estimated as
\begin{eqnarray}
E^{(k)} &\simeq &{1\over 2} \mu ^{2k}  \Omega _{q}^2
A_q^2 \simeq {1\over 2} \mu ^{2k}
\left( \frac{ \pi c_k s } {N} \right)^2
\cdot  \left( C_k ^{''}  \frac{N } {\pi s }
\right)^{2(2k+1)} E^{r(k)} \nonumber\\
&& \simeq C_k ^{'''} \cdot  \left(  \frac{ \mu N^2 }
{\pi^2 s^2 }  \right)^{2k} E^{r(k)} ~~.
\end{eqnarray}
Finally, replacing $r(k)$ and $\mu $ by $k+1$ and $\alpha /\sqrt{2N}$ for FPU--$\alpha$,
or $2k+1$ and $\beta /(2N)$ for FPU--$\beta$ respectively, we arrive at the exponential
laws
\begin{eqnarray}
E_{\alpha }^{(k)} &\sim & C_k ^{'''} \cdot
\left(  \frac{ \alpha^2 E  N^3 } {\pi^4 s^4 }  \right)^{k}
\sim \left(  \frac{ \alpha^2 \varepsilon   N^4 }
{\pi^4 s^4 }  \right)^{k} \nonumber\\
 E_{\beta }^{(k)} &\sim & C_k ^{'''} \cdot
 \left(  \frac{  \beta^2 E^2 N^2 } {\pi^4 s^4 }  \right)^{k}
 \sim\left(  \frac{  \beta^2 \varepsilon ^2 N^4 }
 {\pi^4 s^4 }  \right)^{k}  ~~.
\end{eqnarray}

{\bf F. Precise values of frequencies and amplitudes in 
all paper's numerical examples \label{AppF}}

\begin{equation} \label{Table1}
\small{
\begin{array}{lllll}
\bf{Example} & \bf{Frequencies} & \bf{Amplitudes} \nonumber\\
& & & \nonumber\\
 \text{Fig.1},~~ {\cal D}_0 = \{ 1,2,3,4\} &
\begin{array} {ll}\omega _1 = 0.09813690483108113 \\
 \omega _2 = 0.19603499627351076  \\
 \omega _3 = 0.29346131418681830 \\
 \omega _4 =0.39018079936090050
 \end{array}
&
\begin{array} {ll}A _1 = 0.098097447281306230 \\
 A _2 =  0.048959912346992665 \\
 A _3 = 0.032444239210574590 \\
 A _4 = 0.023892612612360370
 \end{array} \\
& & & \\
\text{Fig.4a},~~ {\cal D}_0 = \{1,11,21,31\} &
\begin{array} {l}\omega _{1} = 0.09814720863898445 \\
 \omega _{11} = 1.0282023436963477  \\
 \omega _{21} = 1.7154505429645952 \\
 \omega _{31} =1.9975829222954775
 \end{array} &
 \begin{array} {ll}A _1 = 0.2699420271816366000 \\
 A _{11} =  0.027437337334704704 \\
 A _{21} = 0.015901830855243902\\
 A _{31} = 0.014550407721151376
 \end{array} \\
 & & & \\
 \text{Fig.4b},~~ {\cal D}_0 = \{ 60,61,62,63\} &
\begin{array} {l}\omega _{60} = 1.9903455323483594 \\
 \omega _{61} = 1.9945569202421924  \\
 \omega _{62} = 1.9975668544869103 \\
 \omega _{63} =1.9993735255838023
 \end{array} &
 \begin{array} {ll}A _{60} = 0.010626655362545707 \\
 A _{61} =  0.010540492500677328 \\
 A _{62} = 0.010486090307194270\\
 A _{63} = 0.010462240436802113
 \end{array} \\
 & & & \\
 \text{Fig.4c},~~ {\cal D}_0 = \{ 63,64,65\} &
 \begin{array} {ll}
 \omega _{63} = 1.3967520180075357  \\
 \omega _{64} = 1.4142130013133483 \\
 \omega _{65} =1.4314611496716776
 \end{array} &
 \begin{array} {ll}
 A _{63} =  0.005128073559888574 \\
 A _{64} = 0.005045052412361571\\
 A _{65} = 0.004963639646819611
 \end{array} \\
 & & & \\
  \text{Fig.4d},~~ {\cal D}_0 = \{ 94,\ldots ,98 \} &
\begin{array} {l}\omega _{94} =1.8284112766464593 \\
 \omega _{95} = 1.8382194182395468  \\
 \omega _{96} = 1.8477507307792000 \\
 \omega _{97} =1.8570037790005570\\
  \omega _{98} =1.8659771693147402
 \end{array} &
 \begin{array} {ll}A _{94} = 0.008845316872292220 \\
 A _{95} =  0.008752766020034742\\
 A _{96} = 0.008700391793859069\\
 A _{97} = 0.008662026577864702 \\
  A _{98} =0.008661383820872627
 \end{array} \\
 & & & \\
  \text{Fig.6},~~ {\cal D}_0 = \{ 1 \} &
   \omega _{1} =0.09814109959448596  & A_1 =0.18574105489382606\\
    & & & \\
  \text{Fig.7},~~ {\cal D}_0 = \{ 25 \} &
   \omega _{25} =1.8830741847088537  & A_{25} =0.0512142216322969\\
    & & & \\
  \text{Fig.8},~~ {\cal D}_0 = \{ 1 \} &
   \omega _{1} =0.09844127049688513 & A_1 =1.3009149083500795\\
 \end{array}}
\end{equation}

\bibliographystyle{elsarticle-num}

\begin{thebibliography}{00}
\expandafter\ifx\csname natexlab\endcsname\relax\def\natexlab#1{#1}\fi
\expandafter\ifx\csname bibnamefont\endcsname\relax
  \def\bibnamefont#1{#1}\fi
\expandafter\ifx\csname bibfnamefont\endcsname\relax
  \def\bibfnamefont#1{#1}\fi
\expandafter\ifx\csname citenamefont\endcsname\relax
  \def\citenamefont#1{#1}\fi
\expandafter\ifx\csname url\endcsname\relax
  \def\url#1{\texttt{#1}}\fi
\expandafter\ifx\csname urlprefix\endcsname\relax\def\urlprefix{URL }\fi
\providecommand{\bibinfo}[2]{#2}
\providecommand{\eprint}[2][]{\url{#2}}


\bibitem{bampon2006}
\bibinfo{author}{\bibfnamefont{D.}~\bibnamefont{Bambusi}}
\bibnamefont{and}
\bibinfo{author}{\bibfnamefont{A.}~\bibnamefont{Ponno}},
\bibinfo{journal}{Comm. Math. Phys.} \textbf{\bibinfo{volume}{264}} (\bibinfo{year}{2006})
\bibinfo{pages}{539}.

\bibitem{benetal2011}
\bibinfo{author}{\bibfnamefont{G.}~\bibnamefont{Benettin}}
\bibnamefont{and}
\bibinfo{author}{\bibfnamefont{A.}~\bibnamefont{Ponno}},
\bibinfo{journal}{J. Stat. Phys.} \textbf{\bibinfo{volume}{144}} (\bibinfo{year}{2011})
\bibinfo{pages}{793}.

\bibitem{beretal2004}
\bibinfo{author}{\bibfnamefont{L.}~\bibnamefont{Berchialla}},
\bibinfo{author}{\bibfnamefont{A.}~\bibnamefont{Giorgilli}},
\bibnamefont{and}
\bibinfo{author}{\bibfnamefont{S.}~\bibnamefont{Paleari}},
\bibinfo{journal}{Phys. Lett. A} \textbf{\bibinfo{volume}{321}} (\bibinfo{year}{2004})
\bibinfo{pages}{167}.

\bibitem{beretal2005}
\bibinfo{author}{\bibfnamefont{L.}~\bibnamefont{Berchialla}},
\bibinfo{author}{\bibfnamefont{L.}~\bibnamefont{Galgani}},
\bibnamefont{and}
\bibinfo{author}{\bibfnamefont{A.}~\bibnamefont{Giorgilli}},
\bibinfo{journal}{Discr. Cont. Dyn. Sys. A} \textbf{\bibinfo{volume}{11}} (\bibinfo{year}{2005})
\bibinfo{pages}{855}.

\bibitem{Pasta73}
\bibinfo{author}{\bibfnamefont{R. L.}~\bibnamefont{Bivins}},
\bibinfo{author}{\bibfnamefont{N.}~\bibnamefont{Metropolis}},
\bibnamefont{and}
\bibinfo{author}{\bibfnamefont{J.}~\bibnamefont{Pasta}},
\bibinfo{journal}{J. Comp. Phys.}
\textbf{\bibinfo{volume}{12}} (\bibinfo{year}{1973})
\bibinfo{pages}{65}.

\bibitem{chechin2005}
\bibinfo{author}{\bibfnamefont{G. M.}~\bibnamefont{Chechin}},
\bibinfo{author}{\bibfnamefont{D. S.}~\bibnamefont{Ryabov}},
\bibnamefont{and}
\bibinfo{author}{\bibfnamefont{K. G.}~\bibnamefont{Zhukov}},
\bibinfo{journal}{Physica D} \textbf{\bibinfo{volume}{203}} (\bibinfo{year}{2005})
\bibinfo{pages}{121}.


\bibitem{chri06}
\bibinfo{author}{\bibfnamefont{H.}~\bibnamefont{Christodoulidi}}
\bibnamefont{and}
\bibinfo{author}{\bibfnamefont{T.}~\bibnamefont{Bountis}},
\bibinfo{journal}{Romai J.} \textbf{\bibinfo{volume}{2}} (\bibinfo{year}{2006})
\bibinfo{pages}{37}.

\bibitem{chrietal2010}
\bibinfo{author}{\bibfnamefont{H.}~\bibnamefont{Christodoulidi}},
\bibinfo{author}{\bibfnamefont{C.}~\bibnamefont{Efthymiopoulos}}
\bibnamefont{and} \bibinfo{author}{\bibfnamefont{T.}~\bibnamefont{Bountis}},
\bibinfo{journal}{Phys. Rev. E} \textbf{\bibinfo{volume}{81}} (\bibinfo{year}{2010})
\bibinfo{pages}{016210}.


\bibitem{eli1997}
\bibinfo{author}{\bibfnamefont{L.}~\bibnamefont{Eliasson}},
\bibinfo{journal}{Math. Phys. Electron. J.} \textbf{\bibinfo{volume}{2}} (\bibinfo{year}{1997})
\bibinfo{pages}{1}.


\bibitem{feretal1955}
\bibinfo{author}{\bibfnamefont{E.}~\bibnamefont{Fermi}},
\bibinfo{author}{\bibfnamefont{J.}~\bibnamefont{Pasta}},
\bibnamefont{and}
\bibinfo{author}{\bibfnamefont{S.}~\bibnamefont{Ulam}},
\bibinfo{journal}{Los Alamos report No LA-1940}
(\bibinfo{year}{1955}) \bibinfo{pages}{977}.



\bibitem{flaetal2005}
\bibinfo{author}{\bibfnamefont{S.}~\bibnamefont{Flach}},
\bibinfo{author}{\bibfnamefont{M.~V.} \bibnamefont{Ivanchenko}},
\bibnamefont{and} \bibinfo{author}{\bibfnamefont{O.~I.}
\bibnamefont{Kanakov}}, \bibinfo{journal}{Phys. Rev. Lett.}
\textbf{\bibinfo{volume}{95}} (\bibinfo{year}{2005}) \bibinfo{pages}{064102}.

\bibitem{flaetal2006}
\bibinfo{author}{\bibfnamefont{S.}~\bibnamefont{Flach}},
\bibinfo{author}{\bibfnamefont{M.~V.} \bibnamefont{Ivanchenko}},
\bibnamefont{and} \bibinfo{author}{\bibfnamefont{O.~I.}
\bibnamefont{Kanakov}}, \bibinfo{journal}{Phys. Rev. E}
\textbf{\bibinfo{volume}{73}} (\bibinfo{year}{2006}) \bibinfo{pages}{036618}.



\bibitem{flapon2008}
\bibinfo{author}{\bibfnamefont{S.}~\bibnamefont{Flach}}
\bibnamefont{and}
\bibinfo{author}{\bibfnamefont{A.}~\bibnamefont{Ponno}},
\bibinfo{journal}{Physica D} \textbf{\bibinfo{volume}{237}} (\bibinfo{year}{2008})
\bibinfo{pages}{908}.

\bibitem{flaetal2007}
\bibinfo{author}{\bibfnamefont{S.}~\bibnamefont{Flach}},
\bibinfo{author}{\bibfnamefont{O.}~\bibnamefont{Kanakov}},
\bibinfo{author}{\bibfnamefont{M.}~\bibnamefont{Ivanchenko}},
\bibnamefont{and} \bibinfo{author}{\bibfnamefont{K.}~\bibnamefont{Mishagin}},
\bibinfo{journal}{Int. J. Mod. Phys. B} \textbf{\bibinfo{volume}{21}} (\bibinfo{year}{2007})
\bibinfo{pages}{3925}.

\bibitem{flachtiz}
\bibinfo{author}{\bibfnamefont{S.}~\bibnamefont{Flach}}
\bibnamefont{and} \bibinfo{author}{\bibfnamefont{T.}~\bibnamefont{Penati}},
\bibinfo{journal}{Chaos} \textbf{\bibinfo{volume}{17}} (\bibinfo{year}{2007})
\bibinfo{pages}{023102}.


\bibitem{fucetal1982}
\bibinfo{author}{\bibfnamefont{F.}~\bibnamefont{Fucito}},
\bibinfo{author}{\bibfnamefont{F.}~\bibnamefont{Marchesoni}},
\bibinfo{author}{\bibfnamefont{E.}~\bibnamefont{Marinari}},
\bibinfo{author}{\bibfnamefont{G.}~\bibnamefont{Parisi}},
\bibinfo{author}{\bibfnamefont{L.}~\bibnamefont{Politi}},
\bibinfo{author}{\bibfnamefont{S.}~\bibnamefont{Ruffo}},
\bibnamefont{and}
\bibinfo{author}{\bibfnamefont{A.}~\bibnamefont{Vulpiani}},
\bibinfo{journal}{J. Physique} \textbf{\bibinfo{volume}{43}} (\bibinfo{year}{1982})
\bibinfo{pages}{707}.


\bibitem{gal1994}
\bibinfo{author}{\bibfnamefont{G.}~\bibnamefont{Gallavotti}},
\bibinfo{journal}{Commun. Math. Phys.} \textbf{\bibinfo{volume}{164}} (\bibinfo{year}{1994})
\bibinfo{pages}{145}.


\bibitem{gentaetal}
\bibinfo{author}{\bibfnamefont{T.}~\bibnamefont{Genta}},
\bibinfo{author}{\bibfnamefont{A.}~\bibnamefont{Giorgilli}},
\bibinfo{author}{\bibfnamefont{S.}~\bibnamefont{Paleari}},
\bibnamefont{and}
\bibinfo{author}{\bibfnamefont{T.}~\bibnamefont{Penati}},
\bibinfo{journal}{Phys. Lett. A} \textbf{\bibinfo{volume}{376}} (\bibinfo{year}{2012})
\bibinfo{pages}{2038}.

\bibitem{giomur2006}
\bibinfo{author}{\bibfnamefont{A.}~\bibnamefont{Giorgilli}},
\bibnamefont{and}
\bibinfo{author}{\bibfnamefont{D.}~\bibnamefont{Muraro}},
\bibinfo{journal}{Boll. Unione Mat. Ital. B} \textbf{\bibinfo{volume}{9}} (\bibinfo{year}{2006})
\bibinfo{pages}{1}.

\bibitem{hem1959}
\bibinfo{author}{\bibfnamefont{P.}~\bibnamefont{Hemmer}},
\bibinfo{journal}{Dynamic and stochastic type of motion by the linear chain.
Det Physiske Seminar i Trondheim} \textbf{\bibinfo{volume}{2}} (\bibinfo{year}{1959})
\bibinfo{pages}{66}.

\bibitem{flaetal2006b}
\bibinfo{author}{\bibfnamefont{M.~V.} \bibnamefont{Ivanchenko}},
\bibinfo{author}{\bibfnamefont{O.~I.} \bibnamefont{Kanakov}},
\bibinfo{author}{\bibfnamefont{K.~G.} \bibnamefont{Mishagin}},
\bibnamefont{and}
\bibinfo{author}{\bibfnamefont{S.}~\bibnamefont{Flach}},
\bibinfo{journal}{Phys. Rev. Lett.}
\textbf{\bibinfo{volume}{97}} (\bibinfo{year}{2006}) \bibinfo{pages}{025505}.


\bibitem{kanetal2007}
\bibinfo{author}{\bibfnamefont{O.}~\bibnamefont{Kanakov}},
\bibinfo{author}{\bibfnamefont{S.}~\bibnamefont{Flach}},
\bibinfo{author}{\bibfnamefont{M.}~\bibnamefont{Ivanchenko}},
\bibnamefont{and} \bibinfo{author}{\bibfnamefont{K.}~\bibnamefont{Mishagin}},
\bibinfo{journal}{Phys. Lett. A} \textbf{\bibinfo{volume}{365}} (\bibinfo{year}{2007})
\bibinfo{pages}{416}.

\bibitem{lichetal2008}
\bibinfo{author}{\bibfnamefont{A.}~\bibnamefont{Lichtenberg}},
\bibinfo{author}{\bibfnamefont{R.}~\bibnamefont{Livi}},
\bibinfo{author}{\bibfnamefont{M.}~\bibnamefont{Pettini}},
\bibnamefont{and}
\bibinfo{author}{\bibfnamefont{S.}~\bibnamefont{Ruffo}},
\bibinfo{journal}{Lect. Notes Phys.} \textbf{\bibinfo{volume}{728}} (\bibinfo{year}{2008})
\bibinfo{pages}{21}.

\bibitem{livetal1985}
\bibinfo{author}{\bibfnamefont{R.}~\bibnamefont{Livi}},
\bibinfo{author}{\bibfnamefont{M.}~\bibnamefont{Pettini}},
\bibinfo{author}{\bibfnamefont{S.}~\bibnamefont{Ruffo}},
\bibnamefont{and}
\bibinfo{author}{\bibfnamefont{A.}~\bibnamefont{Vulpiani}},
\bibinfo{journal}{Phys. Rev. A} \textbf{\bibinfo{volume}{31}} (\bibinfo{year}{1985})
\bibinfo{pages}{2740}.

\bibitem{delucetal1995}
\bibinfo{author}{\bibfnamefont{J.}~\bibnamefont{De Luca}},
\bibinfo{author}{\bibfnamefont{A.J.}~\bibnamefont{Lichtenberg}},
\bibnamefont{and}
\bibinfo{author}{\bibfnamefont{M.A.}~\bibnamefont{Lieberman}},
\bibinfo{journal}{Chaos} \textbf{\bibinfo{volume}{5}} (\bibinfo{year}{1995})
\bibinfo{pages}{283}.

\bibitem{deletal1999}
\bibinfo{author}{\bibfnamefont{J.~D.} \bibnamefont{Luca}},
\bibinfo{author}{\bibfnamefont{A.~J.} \bibnamefont{Lichtenberg}},
\bibnamefont{and} \bibinfo{author}{\bibfnamefont{S.}~\bibnamefont{Ruffo}},
\bibinfo{journal}{Phys. Rev. E} \textbf{\bibinfo{volume}{60}} (\bibinfo{year}{1999})
\bibinfo{pages}{3781}.

\bibitem{poggi1997}
\bibinfo{author}{\bibfnamefont{P.}~\bibnamefont{Poggi}}
\bibnamefont{and}
\bibinfo{author}{\bibfnamefont{S.}~\bibnamefont{Ruffo}},
\bibinfo{journal}{Physica D} \textbf{\bibinfo{volume}{103}} (\bibinfo{year}{1997})
\bibinfo{pages}{251}.

\bibitem{ponbam2005}
\bibinfo{author}{\bibfnamefont{A.}~\bibnamefont{Ponno}}
\bibnamefont{and}
\bibinfo{author}{\bibfnamefont{D.}~\bibnamefont{Bambusi}},
\bibinfo{journal}{Chaos} \textbf{\bibinfo{volume}{15}} (\bibinfo{year}{2005})
\bibinfo{pages}{015107}.

\bibitem{dresden}
\bibinfo{author}{\bibfnamefont{A.}~\bibnamefont{Ponno}},
\bibinfo{author}{\bibfnamefont{H.}~\bibnamefont{Christodoulidi}},
\bibinfo{author}{\bibfnamefont{Ch.}~\bibnamefont{Skokos}}
\bibnamefont{and}
\bibinfo{author}{\bibfnamefont{S.}~\bibnamefont{Flach}},
\bibinfo{journal}{Chaos} \textbf{\bibinfo{volume}{21}} (\bibinfo{year}{2011})
\bibinfo{pages}{043127}.

\bibitem{rink2003}
\bibinfo{author}{\bibfnamefont{B.}~\bibnamefont{Rink}},
\bibinfo{journal}{Physica D}
\textbf{\bibinfo{volume}{175}} (\bibinfo{year}{2003}) \bibinfo{pages}{31}.


\bibitem{sansetal2011}
\bibinfo{author}{\bibfnamefont{M.}~\bibnamefont{Sansottera}},
\bibinfo{author}{\bibfnamefont{U.}~\bibnamefont{Locatelli}},
\bibnamefont{and}
\bibinfo{author}{\bibfnamefont{A.}~\bibnamefont{Giorgilli}},
\bibinfo{journal}{Cel. Mech. Dyn. Astron.} \textbf{\bibinfo{volume}{111}} (\bibinfo{year}{2011})
\bibinfo{pages}{337}.

\bibitem{skoetal2007}
\bibinfo{author}{\bibfnamefont{C.}~\bibnamefont{Skokos}},
\bibinfo{author}{\bibfnamefont{T.}~\bibnamefont{Bountis}},
\bibnamefont{and}
\bibinfo{author}{\bibfnamefont{C.}~\bibnamefont{Antonopoulos}},
\bibinfo{journal}{Physica D} \textbf{\bibinfo{volume}{231}} (\bibinfo{year}{2007})
\bibinfo{pages}{30}.

\bibitem{skoetal2008}
\bibinfo{author}{\bibfnamefont{C.}~\bibnamefont{Skokos}},
\bibinfo{author}{\bibfnamefont{T.}~\bibnamefont{Bountis}},
\bibnamefont{and}
\bibinfo{author}{\bibfnamefont{C.}~\bibnamefont{Antonopoulos}},
\bibinfo{journal}{Eur. Phys. J. Special Topics} \textbf{\bibinfo{volume}{165}} (\bibinfo{year}{2008})
\bibinfo{pages}{5}.

 \end{thebibliography}

\end{document}